\documentclass[11pt,a4paper]{article}
\pdfoutput=1
\usepackage{jheppub}
\usepackage{amsthm,amsbsy,amsfonts,mathrsfs,enumerate,float,wrapfig,amsmath}
\usepackage[utf8]{inputenc}
\DeclareUnicodeCharacter{2212}{-}
\usepackage{subfigure}

\newcommand{\be}{\begin{equation}}
\newcommand{\ee}{\end{equation}}
\newcommand{\bea}{\begin{eqnarray}}
\newcommand{\eea}{\end{eqnarray}}
\newcommand{\ba}{\begin{align}}
\newcommand{\ea}{\end{align}}

\usepackage{longtable,caption,multirow}
\usepackage{amsmath}
\usepackage{amssymb}
\usepackage{MnSymbol}
\DeclareMathOperator{\PE}{PE}
\DeclareMathOperator{\HWG}{HWG}
\DeclareMathOperator{\HS}{HS}
\DeclareMathOperator{\MQ}{MQ}
\DeclareMathOperator{\EQ}{EQ}

\usepackage{array}
\newcolumntype{L}[1]{>{\raggedright\let\newline\\\arraybackslash\hspace{0pt}}m{#1}}
\newcolumntype{C}[1]{>{\centering\let\newline\\\arraybackslash\hspace{0pt}}m{#1}}
\newcolumntype{R}[1]{>{\raggedleft\let\newline\\\arraybackslash\hspace{0pt}}m{#1}}

\usepackage{tikz}
\usetikzlibrary{decorations.pathmorphing}
\usetikzlibrary{decorations.pathreplacing}
\usetikzlibrary{shapes, shapes.geometric, shapes.symbols, shapes.arrows, shapes.multipart, shapes.callouts, shapes.misc}
\tikzset{snake it/.style={decorate, decoration=snake}}
\tikzset{7brane/.style={circle, draw=black, fill=black,ultra thick,inner sep=1.5 pt, minimum size=1 pt,}, c/.default={4pt}}

\tikzset{big7brane/.style={circle, draw=black, fill=black,ultra thick,inner sep=2.5 pt, minimum size=1 pt,}, c/.default={4pt}}
\tikzset{u/.style={circle, draw=black, fill=white, thick,inner sep=2 pt, minimum size=2 pt,},f/.style={square, draw=black, fill=white,ultra thick,inner sep=4 pt, minimum size=2 pt,}}

\tikzset{so/.style={circle, draw=black, fill=red, thick,inner sep=2 pt, minimum size=2 pt,},f/.style={square, draw=black, fill=white,ultra thick,inner sep=4 pt, minimum size=2 pt,}}

\tikzset{sp/.style={circle, draw=black, fill=blue,thick,inner sep=2 pt, minimum size=2 pt,},f/.style={square, draw=black, fill=white,ultra thick,inner sep=4 pt, minimum size=2 pt,}}

\tikzset{uf/.style={rectangle, draw=black, fill=white, thick,inner sep=2.5 pt, minimum size=4 pt,}}

\tikzset{spf/.style={rectangle, draw=black, fill=blue, thick,inner sep=2.5 pt, minimum size=4 pt,}}

\tikzset{sof/.style={rectangle, draw=black, fill=red, thick,inner sep=2.5 pt, minimum size=4 pt,}}
\usetikzlibrary{positioning}
\usetikzlibrary{arrows}
\title{Factorised 3d $\mathcal{N}=4$ orthosymplectic quivers}
\author[a]{Mohammad Akhond,}
\author[b]{Federico Carta,}
\author[c]{Siddharth Dwivedi,}
\author[d]{Hirotaka Hayashi,}
\author[e]{Sung-Soo Kim,}
\author[f]{and Futoshi Yagi}
\affiliation[a]{Department of Physics, Swansea University, \\
 Singleton Park, Swansea, SA2 8PP, U.K.}
\affiliation[b]{Department of Mathematical Sciences, Durham University,\\
	Durham, DH$1$ $3$LE, United Kingdom}
\affiliation[c]{Center for Theoretical Physics, College of Physical Science and Technology, Sichuan University, \\Chengdu, 610064, China}
\affiliation[d]{Department of Physics, School of Science, Tokai University,\\ 4-1-1 Kitakaname, Hiratsuka-shi, Kanagawa 259-1292, Japan}
\affiliation[e]{School of Physics, University of Electronic Science and Technology of China, \\
No.2006, Xiyuan Ave, West Hi-Tech Zone, 
Chengdu, Sichuan 611731, China}
\affiliation[f]{School of Mathematics, Southwest Jiaotong University,\\ 
West zone, High-tech district, Chengdu, Sichuan 611756, China}
\emailAdd{akhondmohammad@gmail.com}
\emailAdd{federico.carta@durham.ac.uk}
\emailAdd{sdwivedi@scu.edu.cn}
\emailAdd{h.hayashi@tokai.ac.jp}
\emailAdd{sungsoo.kim@uestc.edu.cn}
\emailAdd{futoshi\_yagi@swjtu.edu.cn}

\abstract{We study the moduli space of 3d $\mathcal{N}=4$ quiver gauge theories with unitary, orthogonal and symplectic gauge nodes, that fall into exceptional sequences. We find that both the Higgs and Coulomb branches of the moduli space factorise into decoupled sectors.
Each decoupled sector is described by a single quiver gauge theory with only unitary gauge nodes. The orthosymplectic quivers serve as magnetic quivers for 5d $\mathcal{N}=1$ superconformal field theories which can be engineered in type IIB string theories both with and without an O5 plane. We use this point of view to postulate the dual pairs of unitary and orthosymplectic quivers by deriving them as magnetic quivers of the 5d theory. We use this correspondence to conjecture exact highest weight generating functions for the Coulomb branch Hilbert series of the orthosymplectic quivers, and provide tests of these results by directly computing the Hilbert series for the orthosymplectic quivers in a series expansion.
}
\begin{document}
\preprint{CTP-SCU/2021017}
\maketitle
\section{Introduction and summary of results}\label{sec:intro}
Gauge theories in three spacetime dimensions are strongly coupled in the IR, determining their low energy dynamics is therefore generically difficult. One arena in which one can overcome this difficulty is the realm of 3d $\mathcal{N}=4$ gauge theories. Their relevance to string theory was highlighted very early after the D-brane revolution in a landmark paper by Hanany and Witten \cite{Hanany:1996ie}, which facilitated further explorations of the subject. A more recent development is to use 3d $\mathcal{N}=4$ theories as a probe to study higher dimensional superconformal field theories (SCFTs) as well as gauge theories \cite{Akhond:2020vhc,Bourget:2019rtl, Bourget:2020asf,Bourget:2020gzi,Bourget:2020xdz,Closset:2020scj,vanBeest:2020kou, vanBeest:2020civ,Bourget:2020mez,Cabrera:2018jxt,Cabrera:2019izd, Eckhard:2020jyr, Closset:2020afy}. The idea is to relate the Higgs branch of these higher dimensional theories to the Coulomb branch of the 3d theory, the latter of which is dubbed magnetic quiver (MQ). In addition to their significance to string theory or higher dimensional theories, 3d $\mathcal{N}=4$ theories possess rich dynamics, making them interesting objects in their own right.
Notable among their many rich properties is mirror symmetry~\cite{Intriligator:1996ex, Hanany:1996ie, Gaiotto:2008ak}, a duality which relates pairs of 3d $\mathcal{N}=4$ gauge theories where the role of masses and Fayet-Illiopolous terms are exchanged. The moduli space of vacua of a 3d $\mathcal{N}=4$ gauge theory is generically comprised of a Higgs branch and a Coulomb branch,\footnote{One can also consider mixed branches, but we will not explore that in this work.} which are exchanged under mirror symmetry. In certain cases, it can be shown that the magnetic quiver of a given theory, is the 3d mirror of the torus compactification of that theory to 3d \cite{Benini:2009gi,Benini:2010uu,Collinucci:2020kdm,Closset:2020scj}. Another interesting aspect of 3d $\mathcal{N}=4$ theories is the enhancement of their global symmetry in the infrared (IR) limit. Gauge theories in 3d possess a topological (or magnetic) symmetry, which is valued in the centre of the Langland dual of the gauge group $G$. Classically, this is an abelian symmetry, but in the IR this is typically enhanced to a non-abelian global symmetry. 

It will be convenient for us to make a distinction between quivers which are made entirely of unitary gauge nodes, and those which can also have orthogonal and/or symplectic gauge nodes in addition. We will refer to the former as unitary and the latter as orthosymplectic (OSp) quivers respectively. The primary focus of this paper is orthosymplectic magnetic quivers. 

A convenient tool to study the moduli space of 3d $\mathcal{N}=4$ theories is the Hilbert series, which enumerates gauge invariant operators graded by their conformal dimension. The Coulomb branch Hilbert series can be computed using the monopole formula \cite{Cremonesi:2013lqa}, while the Higgs branch Hilbert series can be evaluated using the Molien-Weyl formula \cite{Gray:2008yu}. The Coulomb branch Hilbert series is sensitive to the pattern of symmetry enhancement discussed above, so long as one finds a way to refine the computation. This is a longstanding challenge in the case of OSp quivers due to a current lack of understanding of such computations.\footnote{This difficulty is related to the notion of hidden FI parameters in OSp quivers, see e.g. \cite{Feng:2000eq}.} One of the main results of this paper is the refined Hilbert series, and therefore the enhanced magnetic symmetry of the OSp quivers under our study. Together with the other tools of the Plethystic programme \cite{Benvenuti:2006qr,Feng:2007ur}, the Hilbert series can be used to give an algebraic description of the moduli space as a variety. We will in particular make use of the notion of highest weight generators (HWGs) developed in \cite{Hanany:2014dia} in order to write down closed form expressions for the Coulomb branch Hilbert series of the OSp magnetic quivers under our consideration.   

In this paper we uncover an interesting phenomenon which is common to all models under our consideration; the moduli space of the OSp quivers that we study generically factorizes into two decoupled sectors, each of which has an alternative description, in terms of the moduli space of a single connected unitary quiver. An upshot of this result is that we can write exact highest weight generating functions (HWGs) encoding the Coulomb branch Hilbert series of several families of orthosymplectic quivers using known results for the individual factors. This result ultimately follows from fact that the OSp quivers that we study serve as magnetic quivers to 5d $\mathcal{N}=1$ SCFTs which are the UV fixed point of a 5d IR gauge theory whose gauge group is a product of SO(4) factors, and with matter representations transforming as either spinors or conjugate spinors of each SO(4) factor. Since SO(4) is locally isomorphic to SU(2)$\times$SU(2), and a spinor and conjugate spinor transform under different SU(2) factors, each such theory can be reformulated as a product of two decoupled theories, each of which has a gauge group that is a product of SU(2)s. The theories containing SO(4) factors can be engineered using a single type IIB brane web with the inclusion of O5-planes, which can then be used to obtain an OSp magnetic quiver \cite{Akhond:2020vhc, Bourget:2020gzi}. On the other hand the formulation in terms of the product of theories with SU(2) factors is engineered by two independent brane webs, giving rise to two magnetic quivers, which will be unitary by construction \cite{Cabrera:2018jxt}.   

We note that an analogous factorization phenomenon happens for 4d N=2 theories of class-S of D-type. In this context, it is well known that there are cases in which a single three-punctured sphere describes the direct sum of 2 SCFTS, each of which also admits a realization in A-type class-S \cite{Chacaltana:2011ze,Distler:2017xba,Distler:2018gbc,Ergun:2020fnm}. Indeed, we identify some 4d N=2 theories of D-type class-S which exhibit this factorization, and for which the orthosymplectic 3d mirror theories correspond to the magnetic quiver derived from the 5-brane webs with O5-plane. Likewise, the 3d mirror of the two A-type factors also corresponds to the unitary magnetic quiver derived from the 5-brane web without O5-plane.

For ease of presentation, we tabulate a list of all the orthosymplectic quivers appearing in this work,  along with their Coulomb branch symmetry and the refined highest weight generating functions of the Coulomb branch in table \ref{TableHWGOSpQuivers}.

The organization of this paper is as follows. In section \ref{plethystics} we review the relevant tools that we will need from the plethystic programme. We discuss the monopole formula for the Coulomb branch and the Molien-Weyl formula for the Higgs branch and review the notion of highest weight generating functions. Section \ref{Product sequences}, contains our main results. Here we will present the product exceptional sequences of OSp quivers, their 5d origin from brane webs with O5 planes and their unitary counterparts as well as their 5d origin from ordinary brane webs. In this section we also state the Hilbert series results for all the quivers. Finally, in section \ref{discussion} we discuss potential applications of our results and state open problems which we find deserve further investigation. Appendix \ref{appendixA} contains additional details for some of the Higgs branch Hilbert series computations. We collect the results of HWGs for the unitary quivers in the table \ref{TableHWGUnitaryQuivers} of appendix \ref{appendixB}. In appendix \ref{appendixC}, we give more details on the computations of the Coulomb branch Hilbert series for the OSp quivers. Appendix \ref{appendixD} contains a review of the class S technology that is used within the main sections.  
\captionsetup{width=15cm}
\begin{longtable}{|c|C{2.5cm}|C{6.5cm}|}
        \caption{Summary of the orthosymplectic quivers, their Coulomb branch symmetry and the associated HWG. The corresponding fugacities are denoted by subscripts in the symmetry groups. Note that for $N=1$, there is an enhancement in the symmetry as detailed in the later sections. The notations for the quivers are defined in the beginning of section \ref{Product sequences}.}
			\label{TableHWGOSpQuivers} \\ \hline 
					Quiver&Symmetry&PL[HWG]\\\hline 
      $\begin{array}{c}
             \begin{scriptsize}
             \begin{tikzpicture}
             \node[label=below:{1}][u](1){};
             \node (dots)[right of=1]{$\cdots$};
             \node[label=below:{$2N-1$}][u](2n-1)[right of=dots]{};
             \node[label=below:{$2N$}][sp](sp2n)[right of=2n-1]{};
             \node[label=above:{$4$}][sof](sof)[above of=sp2n]{};
             \draw(1)--(dots);
             \draw(dots)--(2n-1);
             \draw(2n-1)--(sp2n);
             \draw(sp2n)--(sof);
             \end{tikzpicture}
             \end{scriptsize}
        \end{array}$ & SU($2N$)$_{\mu}$ $\times$ SU($2N$)$_{\nu}$ & $\sum\limits_{k=1}^N(\mu_k\,\mu_{2N-k} + \nu_k\,\nu_{2N-k})\,t^{2k}$ \\\hline 
				$\begin{array}{c}
        \begin{scriptsize}
        \begin{tikzpicture}
            \node[label=below:{1}][u](1){};
            \node (dots)[right of=1]{$\cdots$};
            \node[label=below:{$2N-1$}][u](2N-1)[right of=dots]{};
            \node[label=below:{$2N$}][sp](sp2N)[right of=2N-1]{};
            \node[label=below:{$1$}][u](11)[right of=sp2N]{};
            \node[label=above:{$1$}][uf](1f)[above of=11]{};
            \draw(1)--(dots);
            \draw(dots)--(2N-1);
            \draw(2N-1)--(sp2N);
            \draw[double distance=2pt](sp2N)--(11);
             \path [draw,snake it](11)--(1f);
        \end{tikzpicture}
        \end{scriptsize}\end{array}$ & SU($2N$)$_\mu$ $\times$ SU($2N$)$_\nu$ $\times$ U(1)$_q$ & $\sum\limits_{k=1}^N(\mu_k\,\mu_{2N-k}+\nu_k\,\nu_{2N-k})\,t^{2k} + t^2+\left(q+q^{-1}\right)\nu_N\,t^{N+1} -\nu_N^2\,t^{2N+2} $ \\\hline 
		$\begin{array}{c}
        \begin{scriptsize}
        \begin{tikzpicture}
            \node[label=below:{1}][u](1){};
            \node (dots)[right of=1]{$\cdots$};
            \node[label=below:{$2N-1$}][u](2N-1)[right of=dots]{};
            \node[label=above:{$2N-2$}][sp](sp2N)[above of=2N-1]{};
            \node[label=below:{$1$}][u](11)[right of=2N-1]{};
            \node[label=above:{$2$}][uf](1f)[above of=11]{};
            \draw(1)--(dots);
            \draw(dots)--(2N-1);
            \draw(2N-1)--(sp2N);
            \draw[double distance=2pt](2N-1)--(11);
             \path [draw,snake it](11)--(1f);
        \end{tikzpicture}
        \end{scriptsize}
    \end{array}$ & SU($2N$)$_\mu$ $\times$ SU($2N$)$_\nu$ $\times$ U(1)$_q$ & $\sum\limits_{k=1}^N\mu_k\,\mu_{2N-k}\,t^{2k}+\sum\limits_{k=1}^{N-1}\nu_k\nu_{2N-k}\,t^{2k}  + t^2+\left(\nu_{N+1}q+\nu_{N-1}q^{-1}\right)\,t^{N+1} -\nu_{N+1}\,\nu_{N-1}\,t^{2N+2} $ \\\hline 
		$\begin{array}{c}\begin{scriptsize}
         \begin{tikzpicture}
             \node[label=below:{1}][u](1){};
             \node (dots)[right of=1]{$\cdots$};
             \node[label=below:{$2N-1$}][u](2N-1)[right of=dots]{};
             \node[label=below:{$2N$}][sp](sp2N)[right of=2N-1]{};
             \node[label=above:{$2$}][so](so2)[above right of=sp2N]{};
             \node[label=below:{$1$}][u](u1)[below right of=sp2N]{};
             \draw(1)--(dots);
             \draw(dots)--(2N-1);
             \draw(sp2N)--(2N-1);
             \draw(sp2N)--(so2);
             \draw(sp2N)--(u1);
             \draw(u1)--(so2);
         \end{tikzpicture}
         \end{scriptsize}\end{array}$ & SU($2N$)$_\mu$ $\times$ SU($2N$)$_\nu$ $\times$ U(1)$_{q_1}$ $\times$ U(1)$_{q_2}$ & $\sum\limits_{k=1}^{N}(\mu_k\,\mu_{2N-k}+\nu_k\,\nu_{2N-k})\,t^{2k}+2t^2+\left(q_1+q_1^{-1}\right)\mu_N\,t^{N+1}-\mu_N^2\,t^{2N+2}+\left(q_2+q_2^{-1}\right)\nu_N\,t^{N+1}-\nu_N^2\,t^{2N+2}$ \\ \hline
				$\begin{array}{c}\begin{scriptsize}
         \begin{tikzpicture}
                    \node[label=below:{1}][u](1){};
             \node (dots)[right of=1]{$\cdots$};
             \node[label=below:{$2N-2$}][u](2N-2)[right of=dots]{};
             \node[label=below:{$2N-2$}][u](2N-2')[right of=2N-2]{};
             \node[label=below:{$2N-2$}][sp](sp2N-2)[right of=2N-2']{};
             \node[label=above:{$2$}][so](so2)[above of=sp2N-2]{};
             \node[label=above:{$1$}][u](u1)[above of=2N-2]{};
             \draw(1)--(dots);
             \draw(dots)--(2N-2);
             \draw(2N-2)--(2N-2');
             \draw(2N-2')--(sp2N-2);
             \draw(sp2N-2)--(so2);
             \draw[double distance=2pt](so2)--(u1);
             \draw(u1)--(2N-2);
         \end{tikzpicture}
         \end{scriptsize}\end{array}$ & SU($2N$)$_\mu$ $\times$ SU($2N$)$_\nu$ $\times$ U(1)$_{q_1}$ $\times$ U(1)$_{q_2}$ & $\sum\limits_{k=1}^{N-1}(\mu_k\,\mu_{2N-k}+\nu_k\,\nu_{2N-k})\,t^{2k}+2t^2+\left(\mu_{N+1}q_1+\mu_{N-1}q_1^{-1}\right)\,t^{N+1}-\mu_{N+1}\,\mu_{N-1}\,t^{2N+2}+\left(\nu_{N+1}q_2+\nu_{N-1}q_2^{-1}\right)\,t^{N+1}-\nu_{N+1}\,\nu_{N-1}\,t^{2N+2}
			$\\ \hline 
		$\begin{array}{c}
         \begin{scriptsize}
         \begin{tikzpicture}
               \node[label=below:{1}][u](1){};
             \node (dots)[right of=1]{$\cdots$};
             \node[label=below:{$2N-1$}][u](2N-1)[right of=dots]{};
             \node[label=below:{$2N-2$}][sp](sp2N-2)[right of=2N-1]{};
           \node[label=below:{1}][u](u1)[below right of=sp2N-2]{};
           \node[label=above:{1}][u](u11)[above right of=sp2N-2]{};
           \draw(1)--(dots);
           \draw(dots)--(2N-1);
           \draw(2N-1)--(sp2N-2);
           \draw(u1)--(2N-1);
           \draw(2N-1)--(u11);
            \draw[dashed, double distance=2 pt] (u1)--(u11);
           \draw(u1)to[out =30, in=-30](u11);
         \end{tikzpicture}
         \end{scriptsize}\end{array}$ & SU($2N$)$_\mu$ $\times$ SU($2N$)$_\nu$ $\times$ U(1)$_{q_1}$ $\times$ U(1)$_{q_2}$ & $\sum\limits_{k=1}^{N}\mu_k\,\mu_{2N-k}\,t^{2k}+\sum\limits_{k=1}^{N-1}\nu_k\,\nu_{2N-k}\,t^{2k}+2t^2+\left(q_1+q_1^{-1}\right)\mu_N\,t^{N+1}-\mu_N^2\,t^{2N+2}+\left(\nu_{N+1}q_2+\nu_{N-1}q_2^{-1}\right)\,t^{N+1}-\nu_{N+1}\,\nu_{N-1}\,t^{2N+2}
		 $ \\ \hline
		$\begin{array}{c}
         \begin{scriptsize}
         \begin{tikzpicture}
             \node[label=below:{1}][u](1){};
             \node (dots) [right of=1]{$\cdots$};
             \node[label=below:{$2N$}][u](2N)[right of=dots]{};
             \node[label=below:{$2N$}][sp](sp2N)[right of=2N]{};
             \node[label=above:{$2$}][sof](sof)[above of=sp2N]{};
             \node[label=above:{$1$}][uf](uf)[above of=2N]{};
             \draw(1)--(dots);
             \draw(dots)--(2N);
             \draw(2N)--(sp2N);
             \draw(sp2N)--(sof);
             \draw(2N)--(uf);
         \end{tikzpicture}
         \end{scriptsize}
    \end{array}$ & SU($2N+1$)$_\mu$ $\times$ SU($2N+1$)$_\nu$ & $\sum\limits_{k=1}^N\left(\mu_k\mu_{2N+1-k}+\nu_k\nu_{2N+1-k}\right)t^{2k}$ \\ \hline
				$\begin{array}{c}\begin{scriptsize}
     \begin{tikzpicture}
     \node[label=left:{1}][u](1){};
     \node (dots)[above of=1]{$\vdots$};
     \node[label=left:{$2N-2$}][u](2N-2)[above of=dots]{};
     \node[label=below:{$2N-2$}][u](2N-2')[right of=2N-2]{};
     \node[label=below:{$2N-2$}][u](2N-2'')[right of=2N-2']{};
     \node[label=below:{$2N-2$}][sp](sp2N-2)[right of=2N-2'']{};
     \node[label=above:{1}][uf](uf)[above of=2N-2]{};
     \node[label=above:{2}][sof](sof)[above of=sp2N-2]{};
     \draw(1)--(dots);
     \draw(dots)--(2N-2);
     \draw(2N-2)--(2N-2');
     \draw(2N-2')--(2N-2'');
     \draw(2N-2'')--(sp2N-2);
     \draw(sp2N-2)--(sof);
     \draw(2N-2)--(uf);
     \end{tikzpicture}
     \end{scriptsize}
     \end{array}$ & SU($2N+1$)$_\mu$ $\times$ SU($2N+1$)$_\nu$ & $\sum\limits_{k=1}^{N-1}\left(\mu_k\mu_{2N+1-k}+\nu_k\nu_{2N+1-k}\right)t^{2k}$ \\ \hline
		$\begin{array}{c}
         \begin{scriptsize}
         \begin{tikzpicture}
         \node[label=left:{1}][u](1){};
         \node (dots) [above of=1]{$\vdots$};
         \node[label=above:{$2N-2$}][u](2N-2)[above of=dots]{};
         \node[label=below:{$2N-1$}][u](2N-1)[right of=2N-2]{};
         \node[label=below:{$2N-1$}][u](2N-1')[right of=2N-1]{};
         \node[label=below:{$2N-2$}][sp](sp2N-2)[right of=2N-1']{};
         \node[label=above:{1}][uf](uf)[above of=2N-1]{};
         \node[label=above:{1}][uf](uf')[above of=2N-1']{};
         \draw(1)--(dots);
         \draw(dots)--(2N-2);
         \draw(2N-2)--(2N-1);
         \draw(2N-1)--(2N-1');
         \draw(2N-1')--(sp2N-2);
         \draw(2N-1)--(uf);
         \draw(2N-1')--(uf');
         \end{tikzpicture}
         \end{scriptsize}
    \end{array} $ & SU($2N+1$)$_\mu$ $\times$ SU($2N+1$)$_\nu$ & $\sum\limits_{k=1}^N\mu_k\mu_{2N+1-k}t^{2k}+\sum\limits_{j=1}^{N-1}\nu_j\nu_{2N+1-j}t^{2j}$ \\ \hline
		$\begin{array}{c}
         \begin{scriptsize}
         \begin{tikzpicture}
             \node[label=right:{1}][u](1){};
             \node (dots)[below of=1]{$\vdots$};
             \node[label=right:{$2N$}][u](2N-2)[below of=dots]{};
             \node[label=below:{$2N$}][sp](sp2N-2)[below of=2N-2]{};
             \node[label=below:{1}][u](u1)[left of=sp2N-2]{};
             \node[label=below:{2}][so](so2)[left of=u1]{};
             \draw(1)--(dots);
             \draw(dots)--(2N-2);
             \draw(2N-2)--(sp2N-2);
             \draw(sp2N-2)--(u1);
             \draw(u1)--(so2);
             \draw(u1)--(2N-2);
         \end{tikzpicture}
         \end{scriptsize}
    \end{array}$ & SU($2N+1$)$_\mu$ $\times$ SU($2N+1$)$_\rho$ $\times$ SU(2)$_\nu$ $\times$ U(1)$_q$ & $\sum\limits_{i=1}^N(\mu_i\mu_{2N+1-i}+\rho_i\rho_{2N+1-i})t^{2i}+(\nu^2+1)t^2+\nu(\mu_Nq+\mu_{N+1}q^{-1})t^{N+1}-\nu^2\mu_{N}\mu_{N+1}t^{2N+2}$ \\ \hline
		$\begin{array}{c}
         \begin{scriptsize}
         \begin{tikzpicture}
             \node[label=right:{1}][u](1){};
             \node (dots)[below of=1]{$\vdots$};
             \node[label=right:{$2N-1$}][u](2N-1)[below of=dots]{};
             \node[label=right:{$2N-1$}][u](2N-1')[below of=2N-1]{};
             \node[label=right:{$2N-2$}][sp](sp2N-2)[below of=2N-1']{};
             \node[label=below:{$1$}][u](1')[left of=2N-1]{};
             \node[label=below:{$2$}][sp](sp2)[left of=1']{};
             \node[label=below:{$2$}][so](so2)[left of=sp2]{};
                          \draw(1)--(dots);
             \draw(dots)--(2N-1);
             \draw(2N-1)--(2N-1');
             \draw(2N-1')--(sp2N-2);
             \draw[dashed](2N-1')--(1');
             \draw (1')--(2N-1);
             \draw[double distance=2pt](1')--(sp2);
             \draw(sp2)--(so2);
         \end{tikzpicture}
         \end{scriptsize}
    \end{array}$ & SU($2N+1$)$_\mu$ $\times$ SU($2N+1$)$_\lambda$ $\times$ SU(2)$_\nu$ $\times$ SU(2)$_\eta$ $\times$ U(1)$_q$ & $\sum\limits_{i=1}^N(\mu_i\mu_{2N+1-i}+\lambda_i\lambda_{2N+1-i})t^{2i}+(\nu^2+1)t^2+\nu(\mu_Nq+\mu_{N+1}q^{-1})t^{N+1}-\nu^2\mu_{N}\mu_{N+1}t^{2N+2}+\eta^2t^2$ \\ \hline
		$\begin{array}{c}
             \begin{scriptsize}
             \begin{tikzpicture}
             \node[label=below:{2}][so](so2){};
             \node[label=below:{2}][sp](sp2)[right of=so2]{};
             \node[label=below:{2}][so](so2')[below right of=sp2]{};
             \node[label=above:{1}][u](u11)[above right of=sp2]{};
             \node[label=below:{$2N$}][sp](sp2'')[right of=so2']{};
             \node[label=above:{$2N$}][u](u2)[right of=u11]{};
             \node (dots)[right of=u2]{$\cdots$};
             \node[label=above:{1}][u](u1)[right of=dots]{};
             \draw(so2)--(sp2);
             \draw(sp2)--(so2');
             \draw(so2')--(sp2'');
             \draw(sp2)--(u11);
             \draw(u11)--(u2);
             \draw(dots)--(u2);
             \draw(dots)--(u1);
             \draw(u2)--(sp2'');
             \end{tikzpicture}
             \end{scriptsize}
        \end{array}$ & SU($2N+1$)$_\mu$ $\times$ SU($2N+1$)$_\eta$ $\times$ SU(2)$_\nu$ $\times$ SU(2)$_\lambda$ $\times$ U(1)$_q$ $\times$ U(1)$_r$ & $\sum\limits_{i=1}^N(\mu_i\mu_{2N+1-i}+\eta_i\eta_{2N+1-i})t^{2i}+(\nu^2+\lambda^2+2)t^2+\nu(\mu_Nq+\mu_{N+1}q^{-1})t^{N+1} +\lambda(\eta_Nr+\eta_{N+1}r^{-1})t^{N+1}-\nu^2\mu_{N}\mu_{N+1}t^{2N+2}- \lambda^2\eta_{N}\eta_{N+1}t^{2N+2}$ \\ \hline
				$\begin{array}{c}
             \begin{scriptsize}
             \begin{tikzpicture}
                 \node[label=right:{1}][u](1){};
                 \node (dots)[below of=1]{$\vdots$};
                 \node[label=right:{$2N+1$}][u](2N-1)[below of=dots]{};
                 \node[label=right:{$2N+2$}][sp](sp2N)[below of=2N-1]{};
                 \node[label=below:{4}][so](so4)[below of=sp2N]{};
                 \node[label=below:{2}][sp](sp2)[right of=so4]{};
                 \node[label=below:{2}][sp](sp2')[left of=so4]{};
                 \node[label=below:{2}][so](so2)[right of=sp2]{};
                 \node[label=below:{2}][so](so2')[left of=sp2']{};
                 \draw(1)--(dots);
                 \draw(dots)--(2N-1);
                 \draw(2N-1)--(sp2N);
                 \draw(sp2N)--(so4);
                 \draw(so4)--(sp2);
                 \draw(so4)--(sp2');
                 \draw(sp2)--(so2);
                 \draw(so2')--(sp2');
             \end{tikzpicture}
             \end{scriptsize}
        \end{array}$ & SU($2N+2$)$_\mu$ $\times$ SU($2N+2$)$_\lambda$ $\times$ SU(2)$_{\nu_1}$ $\times$ SU(2)$_{\nu_2}$ $\times$ SU(2)$_{\rho_1}$ $\times$ SU(2)$_{\rho_2}$ & $\sum\limits_{i=1}^{N+1}(\mu_i\mu_{2N+2-i}+\lambda_i\lambda_{2N+2-i})t^{2i}+(\nu_1^2+\nu_2^2+\rho_1^2+\rho_2^2)t^2+2t^4+\nu_1\nu_2\mu_{N+1}(t^{N+1}+t^{N+3})+\rho_1\rho_2\lambda_{N+1}(t^{N+1}+t^{N+3})-\nu_1^2\nu_2^2\mu_{N+1}^2t^{2N+6}-\rho_1^2\rho_2^2\lambda_{N+1}^2t^{2N+6}$ \\ \hline
				$\begin{array}{c}
         \begin{scriptsize}
         \begin{tikzpicture}
         \node[label=below:{1}][so](o1){};
         \node[label=left:{$2N$}][sp](sp4)[above of=o1]{};
         \node[label=left:{$4N+1$}][so](so5)[above of=sp4]{};
         \node[label=below:{$4N$}][sp](sp4')[right of=so5]{};
         \node[label=below:{$4N+1$}][so](so5')[right of=sp4']{};
         \node[label=right:{$4N$}][sp](sp4'')[right of=so5']{};
         \node[label=right:{$4N$}][so](so4)[below of=sp4'']{};
         \node (dots)[below of=so4]{$\vdots$};
         \node[label=right:{2}][sp](sp2)[below of=dots]{};
         \node[label=right:{2}][so](so2)[below of=sp2]{};
         \node[label=above:{1}][sof](so1)[above of=sp4'']{};
         \node[label=right:{$2N$}][sp](sp2')[above of=so5]{};
         \node[label=above:{1}][sof](so3f)[above of=sp2']{};
         \draw(o1)--(sp4);
         \draw(sp4)--(so5);
         \draw(so5)--(sp2');
         \draw(sp2')--(so3f);
         \draw(so5)--(sp4');
         \draw(sp4')--(so5');
         \draw(so5')--(sp4'');
         \draw(sp4'')--(so1);
         \draw(sp4'')--(so4);
         \draw(so4)--(dots);
         \draw(dots)--(sp2);
         \draw(sp2)--(so2);
         \end{tikzpicture}
         \end{scriptsize}
    \end{array}$ & SO($4N+6$)$_\mu$ $\times$ SO($4N+6$)$_\nu$ & $\sum\limits_{k=1}^N\left(\mu_{2k}+\nu_{2k}\right)t^{2k}$ \\ \hline
		$\begin{array}{c}
         \begin{scriptsize}
         \begin{tikzpicture}
             \node[label=right:{2}][so](so2){};
             \node[label=right:{2}][sp](sp2)[above of=so2]{};
             \node (dots)[above of=sp2]{$\vdots$};
             \node[label=above:{$4N+2$}][so](so4N+2)[above of=dots]{};
             \node[label=below:{$4N+2$}][sp](sp4N+2)[left of=so4N+2]{};
             \node[label=left:{$4N+2$}][so](so4N+2')[left of=sp4N+2]{};
             \node[label=above:{$2N$}][sp](sp2N)[above of=so4N+2']{};
             \node[label=below:{$2N$}][sp](sp2N')[below of=so4N+2']{};
             \node[label=above:{$1$}][u](u1)[above of=sp4N+2]{};
             \draw(so2)--(sp2);
             \draw(sp2)--(dots);
             \draw(dots)--(so4N+2);
             \draw(so4N+2)--(sp4N+2);
             \draw(sp4N+2)--(so4N+2');
             \draw(so4N+2')--(sp2N);
             \draw(so4N+2')--(sp2N');
             \draw(sp4N+2)--(u1);
         \end{tikzpicture}
         \end{scriptsize}
    \end{array}$ & SO($4N+6$)$_\mu$ $\times$ SO($4N+6$)$_\nu$ $\times$ U(1)$_q$ & $\sum\limits_{i=1}^{N}\left(\mu_{2i}+\nu_{2i}\right)t^{2i}+t^2+\left(\mu_{2N+2}q+\mu_{2N+3}q^{-1}\right)t^{N+1}$ \\ \hline
		$\begin{array}{c}
         \begin{scriptsize}
         \begin{tikzpicture}
         \node[label=below:{2}][so](so2){};
         \node[label=right:{$2N+2$}][sp](sp4)[above of=so2]{};
         \node[label=left:{$4N+4$}][so](so8)[above of=sp4]{};
         \node[label=below:{$4N+2$}][sp](sp6)[right of=so8]{};
         \node[label=above:{$4N+2$}][so](so6)[right of=sp6]{};
         \node(dots)[below of=so6]{$\vdots$};
         \node[label=right:{2}][sp](sp2)[below of=dots]{};
         \node[label=right:{2}][so](so2')[below of=sp2]{};
         \node[label=right:{$2N+2$}][sp](sp4'')[above of=so8]{};
         \node[label=right:{1}][u](u1)[above of=sp4'']{};
         \draw(so2)--(sp4);
         \draw(sp4)--(so8);
         \draw(so8)--(sp6);
         \draw(sp6)--(so6);
         \draw(sp2)--(dots);
         \draw(so6)--(dots);
         \draw(sp2)--(so2');
         \draw(so8)--(sp4'');
         \draw(sp4'')--(u1);
         \end{tikzpicture}
         \end{scriptsize}
    \end{array}$ & SO($4N+6$)$_\mu$ $\times$ SO($4N+6$)$_\nu$ $\times$ U(1)$_{q_1}$ $\times$ U(1)$_{q_2}$ & $\sum\limits_{i=1}^{N}(\mu_{2i}+\nu_{2i})t^{2i}+2t^2+\left(\mu_{2N+2}q_1+\mu_{2N+3}q_1^{-1}\right)t^{N+1}   +\left(\nu_{2N+2}q_2+\nu_{2N+3}q_2^{-1}\right)t^{N+1}$ \\ \hline
		$\begin{array}{c}
         \begin{scriptsize}
         \begin{tikzpicture}
         \node[label=below:{2}][so](so2){};
         \node[label=left:{2}][sp](sp2)[above of=so2]{};
         \node[label=left:{4}][so](so4)[above of=sp2]{};
         \node[label=left:{$2N+4$}][sp](sp6)[above of=so4]{};
         \node[label=left:{$4N+6$}][so](so10)[above of=sp6]{};
         \node[label=below:{$4N+4$}][sp](sp8)[right of=so10]{};
         \node[label=above:{$4N+4$}][so](so8)[right of=sp8]{};
         \node (dots)[below of=so8]{$\vdots$};
         \node[label=right:{2}][sp](sp2')[below of=dots]{};
         \node[label=right:{2}][so](so2')[below of=sp2']{};
         \node[label=above:{$2N+2$}][sp](sp4)[above of=so10]{};
         \draw(so2)--(sp2);
         \draw(sp2)--(so4);
         \draw(so4)--(sp6);
         \draw(sp6)--(so10);
         \draw(so10)--(sp4);
         \draw(so10)--(sp8);
         \draw(sp8)--(so8);
         \draw(so8)--(dots);
         \draw(dots)--(sp2');
         \draw(sp2')--(so2');
         \end{tikzpicture}
         \end{scriptsize}
    \end{array}$ & SO($4N+8$)$_\mu$ $\times$ SO($4N+8$)$_\lambda$ $\times$ SU(2)$_\nu$ $\times$ SU(2)$_\rho$ & $\sum\limits_{i=1}^{N+1}(\mu_{2i}+\lambda_{2i})t^{2i}+2t^4+(\nu^2+\rho^2)t^2+\nu\mu_{2N+4}(t^{N+1}+t^{N+3})+\rho\lambda_{2N+4}(t^{N+1}+t^{N+3})+(\mu_{2N+4}^2+\lambda_{2N+4}^2)t^{2N+4}-\nu^2\mu_{2N+4}t^{2N+6}-\rho^2\lambda_{2N+4}t^{2N+6}$ \\\hline
				    \end{longtable}

\section{Preliminaries: tools from the plethystic programme}\label{plethystics}

In this section we review the material we need for the computation of the Hilbert series. The discussion will be minimal and will cover only those aspects necessary for the subsequent section. For more details the reader can consult the original papers. The literature on this material is vast, but we will mostly follow \cite{Gray:2008yu,Benvenuti:2006qr,Hanany:2014dia,Cremonesi:2013lqa,Bourget:2020xdz}. In the following subsections we consider a 3d $\mathcal{N}=4$ theory with gauge group $G$ of rank $r$ and $n_H$ hypermultiplets transforming under the representation $R_i$ of $G$ ($i=1,\cdots,n_H$).
\subsection{Coulomb branches}
The bosonic fields in a 3d $\mathcal{N}=4$ vector multiplet consist of a gauge field and 3 real scalars. Upon dualising the gauge field to a scalar we have 4 scalars at our disposal, which can be pairwise complexified. These are the coordinates on the Coulomb branch at large vevs. Each of these complex scalars forms the scalar component of an $\mathcal{N}=2$ chiral superfield. For small vevs one needs to replace the complex scalar containing the dual photon by a BPS monopole operator \cite{Gaiotto:2008ak}, which can again be thought of as the bottom component of an $\mathcal{N}=2$ chiral multiplet. The quantum coordinates of the Coulomb branch are therefore the BPS monopole $V_m$ of magnetic charge $m$ and the adjoint scalar $\phi$. The Hilbert series (HS) for the Coulomb branch of a good or ugly (in the sense of \cite{Gaiotto:2008ak}) 3d $\mathcal{N}=4$ theory is then computed by the monopole formula \cite{Cremonesi:2013lqa}
\begin{equation}
    \HS(t,z)=\sum_{m\in\Lambda_m^{G^\vee}/\mathcal{W}_{G^\vee}}z^{J(m)}t^{2\Delta(m)}P_G(t,m)\;,
\end{equation}
where the sum is over the magnetic lattice of the gauge group $G$, we refer to \cite{Bourget:2020xdz} for a detailed account. $\Delta(m)$ is the conformal dimension of the monopole operator with magnetic charge $m$, and is given by
\begin{equation}
    \Delta(m)=-\sum_{\alpha\in\Delta^+}|\alpha(m)|+\frac{1}{2}\sum_{i=1}^{n_H}\sum_{\rho_i\in R_i}|\rho_i(m)|\;,
\end{equation}
where $\Delta^+$ is the set of positive roots and $\rho_i$ are the weights of the representation $R_i$.
The classical dressing factor $P_G(t,m)$ counts invariants built out of the adjoint scalars $\phi$  
\begin{equation}
    P_G(t,m)=\prod_{i=1}^r\frac{1}{1-t^{2d(i)}}\;,
\end{equation}
where $d(i)$ are the degrees of the Casimir invariants of the gauge symmetry $H\subset G$ which is the unbroken part of the original gauge symmetry in the presence of a monopole operator $V_m$ of charge $m$. Finally $z$ denote the fugacities of the topological symmetry whose exponent $J(m)$ denotes the topological current.
\subsection{Higgs branches}
The Higgs branch of a 3d $\mathcal{N}=4$ theory is parametrised by vevs of the scalar components of the hypermultiplets. When dealing with a gauge theory, each such operator will be in an irrep of $G$. The Higgs branch operators are therefore those constructed by considering all symmetrised tensor powers of these irreps. The symmetrisation is done in order to be consistent with Pauli statistics. To avoid overcounting, the relations among these scalar operators due to the superpotential need to be imposed. One will then need to project onto the gauge singlet states in order for the resulting operators to be well defined gauge invariant operators. The Higgs branch Hilbert series is therefore computed using the Molien-Weyl formula \cite{Gray:2008yu}
\begin{equation}\label{molien-weyl formula}
  \HS^{\mathcal{H}}(t)=  \int_Gd\mu_G\frac{\PE\left[\sum_{i=1}^{n_H}\chi_{R_i}(x)t\right]}{\PE\left[\chi_{\textbf{Adj}}(x)t^2\right]}\;,
\end{equation}
where $\chi_{R_i}(x)$ is the character of the representation $R_i$ of $G$ under which the scalars in the $i$-th hypermultiplet transform, $\chi_{\textbf{Adj}}$ is the character of the adjoint representation of $G$, the representation carried by the relations. The function $\PE\left[\cdot\right]$ is the plethystic exponential, defined via
\begin{equation}
    \PE\left[f\left(z_1,\cdots,z_r\right)\right]=\exp\left(\sum_{k=1}^\infty \frac{1}{k}f\left(z_1^k,\cdots,z_r^k\right)\right)\;,
\end{equation}
it is a symmetrising function that generates the characters of the symmetrised tensor powers of $\chi_{R_i}$. Finally the projection onto gauge invariant operators is achieved by integrating over the group manifold using the Haar measure.
In a suitable basis, the Haar measure can be taken to be
\begin{equation}
    \int_Gd\mu_G=\frac{1}{\left(2\pi \text{i}\right)^r}\prod_{i=1}^r\oint_{|x_i|=1}\frac{dx_i}{x_i}\prod_{\alpha\in\Delta^+}\left(1-\prod_{k=1}^rx_k^{\alpha_k}\right)\;.
\end{equation}
\subsection{Highest weight generating functions}
The refined Hilbert series for a given theory can be generically expanded as Taylor series in $t$ such that the coefficients are sum of the characters of representations of global symmetry of the theory. In general, given a global symmetry of rank $r$, the refined Hilbert series can be expanded as:
\begin{equation}\label{refined HS generic}
    \sum_{n_1=0}^{\infty} \sum_{n_2=0}^{\infty} \cdots \sum_{n_r=0}^{\infty} \chi_{\left[f_1,f_2,\cdots,f_r\right]}\,t^{\eta}\;,
\end{equation}
where each of $f_1,\ldots,f_r$ and $\eta$ can be some polynomial function in variables $n_1,\ldots, n_r$. The notation $\chi_{\left[f_1,\cdots,f_r\right]}$ is the character of the irrep of the global symmetry whose highest weight is $f_1 \Lambda_1+\ldots + f_r \Lambda_r$, where $f_i$ are the Dynkin labels of the irrep and $\Lambda_i$ are the fundamental weights of the global symmetry group. A convenient way to repackage the same information is in terms of highest weight generating functions or HWGs \cite{Hanany:2014dia}. One introduces a set of fugacities $\{\mu_1,\ldots,\mu_{r}\}$, called highest weight fugacities, and one writes the characters in terms of $\mu_i$ according to the map 
\begin{equation}
    \chi_{\left[f_1,\ldots,f_r\right]} \leftrightarrow \mu_1^{f_1}\ldots \mu_r^{f_r}\;.
\end{equation}
With this map, the Hilbert series becomes a formal power series which can be resummed, the corresponding generating function is termed as its HWG:
 \begin{equation}
    \HWG = \sum_{n_1=0}^{\infty} \cdots \sum_{n_r=0}^{\infty} \mu_1^{f_1}\ldots \mu_r^{f_r}\,t^{\eta}\;.
\end{equation}

\section{Product sequences}\label{Product sequences}
In this section we present sequences of orthosymplectic magnetic quivers whose moduli space is the product of two decoupled sectors, each of which enjoys a description as the moduli space of a unitary quiver. Each sequence is parameterised by an integer $N$ which is the sequence number, and labelled $E_m\times E_n$, in accordance with the Coulomb branch isometry of the $N=1$ case. For $N=1$, some of the theories become particularly simple such that we can prove the equivalence of the orthosymplectic quiver with the two unitary quivers. The $E_m\times E_n$ orthosymplectic quiver with sequence number 1 corresponds to the magnetic quiver for infinite coupling limit of 5d $\mathcal{N}=1$ SO(4) gauge theory with $m-1$ hypermultiplets in the spinor representation denoted by $\textbf{s}$ and $n-1$ hypermultiplets in the conjugate spinor representation of SO(4) denoted by $\textbf{c}$. Correspondingly the dual unitary quivers with sequence number 1 correspond to magnetic quivers for infinite coupling limit of 5d $\mathcal{N}=1$ SU(2)$\times$SU(2) gauge theory with $m-1$ hypermultiplets in the (\textbf{F},\textbf{1}) and $n-1$ hypermultiplets in the (\textbf{1},\textbf{F}) representation of SU(2)$\times$SU(2), where we denote by $\mathbf{F}$ the fundamental representation of associated gauge group. One can engineer these theories using 5-brane webs with O5-planes \cite{Zafrir:2015ftn}, as well as using ordinary brane webs \cite{Aharony:1997bh}. This pattern generalizes for higher sequence numbers, namely one can provide an intuition for the reason that the orthosymplectic quivers factorise into two decoupled sectors by viewing them as magnetic quivers of a 5d theory. We will therefore employ this perspective in the following. We will use EQ$_{m,n}$ to denote the 5d OSp electric quivers for the $E_m\times E_n$ sequence, while we use the notation $\EQ_m$ to denote the 5d unitary electric quivers to which the former factorise. Similarly, we will denote the OSp magnetic quivers of the $E_m\times E_n$ sequence by $\MQ_{m,n}$, while we denote the unitary components to which they factorise by $\MQ_m$. Occasionally there will be more than one generalisation of a given sequence for higher sequence numbers, in which case we will distinguish the different sequences by a prime. 

Our conventions for the magnetic quivers are identical to those appearing in \cite{Akhond:2020vhc} which we briefly review in the following. We denote by a white, red and blue node, the groups U$(n)$, SO$(n)$ and USp($2n$) respectively. A circular node is to be understood as a gauge group, while a square node denotes a flavour group. We use solid lines to denote bifundamental hypermultiplets, in the case where the solid line connects an orthogonal to a symplectic node there is a reality condition which renders the link to be a half hypermultiplet. We use a dashed link between two unitary nodes to denote a fundamental-fundamental hypermultiplet and a jagged line between a unitary flavour and a gauge U$(1)$ node to denote a charge 2 hypermultiplet.

\subsection{The \texorpdfstring{$E_1\times E_1$}{TEXT} sequence}

Consider the 5-brane web constructed by collapsing $2N$ NS5 branes on top of an O5-plane that is asymptotically O5$^+$ as in figure \ref{web diagrams EQ11 and EQ1}. By resolving this web to go on the Coulomb branch one can identify the following low energy quiver description, 
\begin{equation}\label{E1xE1 electric OSp}
\EQ_{1,1}
=\begin{array}{c}
\begin{tikzpicture}
    \node{$\text{SO(4)}-\text{USp(0)}-\text{SO(4)}-\cdots-\text{USp(0)}-\text{SO(4)}$};
    \draw [thick,decorate,decoration={brace,amplitude=6pt},xshift=0pt,yshift=10pt]
(-3.75,0) -- (3.75,0)node [black,midway,xshift=0pt,yshift=15pt] {
$2N-1$};
    \end{tikzpicture}
    \end{array}\ .
\end{equation}
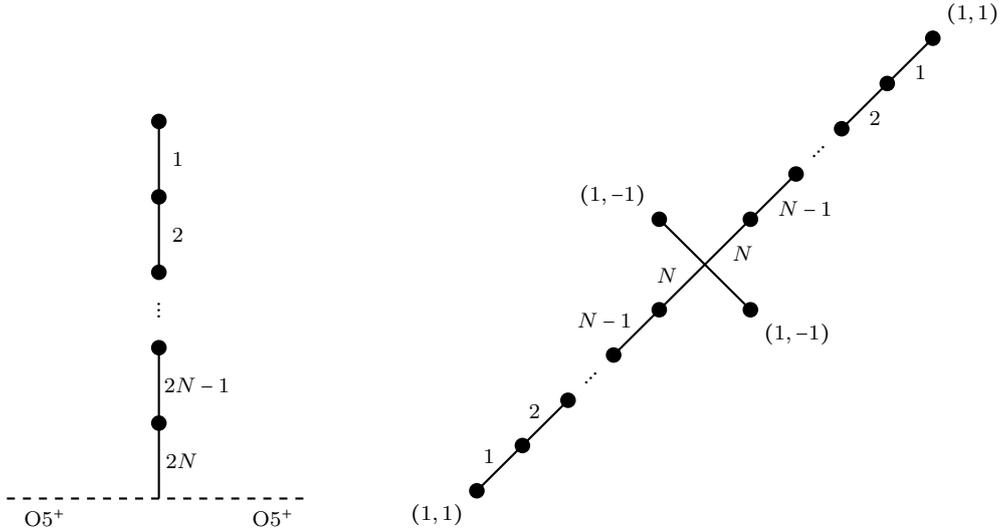
\begin{figure}[t]
    \centering
             \begin{scriptsize}
         \begin{tikzpicture}
    \draw[thick,dashed](0,0)--(4,0);
    \draw[thick](2,3)--(2,5);
    \node at (.5,-0.25){O5$^+$};
    \node at (3.5,-0.25){O5$^+$};
    \node at (2.25,4.5){$1$};
    \node at (2.25,3.5){$2$};
    \node at (2.5,1.5){$2N-1$};
    \node at (2.3,.5){$2N$};
    \draw[thick](2,0)--(2,2);
    \node[7brane]at(2,1){};
    \node[7brane]at(2,2){};
    \node at (2,2.5) {$\vdots$};
    \node[7brane]at(2,3){};
    \node[7brane]at(2,4){};
    \node[7brane]at(2,5){};
    \end{tikzpicture}
    \hspace{1cm}
    \begin{tikzpicture}[scale=.6]
    \draw[thick](-2,-2)--(2,2);
    \draw[thick](-3,-3)--(-5,-5);
    \draw[thick](3,3)--(5,5);
    \node[label=above right:{$(1,1)$}][7brane]at(5,5){};
    \node[7brane]at(4,4){};
    \node[7brane]at(3,3){};
    \node[7brane]at(2,2){};
    \node[7brane]at(1,1){};
        \node[label=below left:{$(1,1)$}][7brane]at(-5,-5){};
    \node[7brane]at(-4,-4){};
    \node[7brane]at(-3,-3){};
    \node[7brane]at(-2,-2){};
    \node[7brane]at(-1,-1){};
    \draw[thick](-1,1)--(1,-1);
    \node[label=above left:{$(1,-1)$}][7brane]at(-1,1){};
    \node[label=below right:{$(1,-1)$}][7brane]at(1,-1){};
    \node at (2.5,2.5){$\udots$};
    \node at (-2.5,-2.5){$\udots$};
    \node[label=below right:{1}] at(4.25,4.75){};
    \node[label=below right:{2}] at(3.25,3.75){};
    \node[label=below right:{$N-1$}] at(1.25,1.75){};
    \node[label=below right:{$N$}] at(.25,.75){};
    
    \node[label=above left:{1}] at(-4.25,-4.75){};
    \node[label=above left:{2}] at(-3.25,-3.75){};
    \node[label=above left:{$N-1$}] at(-1.25,-1.75){};
    \node[label=above left:{$N$}] at(-.25,-.75){};
    \end{tikzpicture}
         \end{scriptsize}
    \caption{Brane webs engineering EQ$_{1,1}$ (left) and EQ$_1$ (right). The numbers near each 5-brane denote the number of coincident 5-branes in the stack in that segment. Black dots denote 7-branes of charge $(p, q)$.}
    \label{web diagrams EQ11 and EQ1}
\end{figure} %
\noindent One can recast this theory, by using the accidental Lie algebra isomorphism $\mathfrak{so}(4)\cong\mathfrak{su}(2)\oplus \mathfrak{su}(2)$ as a direct sum of two decoupled 5d quiver gauge theories. 
For each SU(2) gauge node, we need to fix the discrete theta angles, $\theta=0$ or $\theta=\pi$. It is known that the former has a Higgs branch with dimension one while the latter does not have Higgs branch in the infinite coupling limit. Since the SO(4) gauge theory constructed as in the left of figure \ref{web diagrams EQ11 and EQ1} with $N=1$ has Higgs branch with dimension two in the infinite coupling limit, it should correspond to SU(2)$_0 \times$ SU(2)$_0$ gauge theory. Further support for this claim is given in appendix \ref{appendixE} by constructing the SU(2)$_{\theta_1} \times$ SU(2)$_{\theta_2}$ gauge theories with different discrete theta angles by using 5-brane web diagram with O5-planes. Since each SO(4) gauge node corresponds to SU(2)$_0 \times$ SU(2)$_0$, the generalization to generic $N$ should be\footnote{Analogous discussion on the discrete theta angle will apply to the later examples. In the following subsections, we suppress the label ``0'' of SU(2)$_0$ denoting its discrete theta angle.}
\begin{equation}\label{E1xE1 electric unitary}
    \EQ_{1,1}=(\EQ_1)^2=\left(\begin{array}{c}
\begin{tikzpicture}
    \node{$\text{SU(2)}_0-\text{SU(2)}_0-\text{SU(2)}_0-\cdots-\text{SU(2)}_0$};
    
    \draw [thick,decorate,decoration={brace,amplitude=6pt},xshift=0pt,yshift=10pt]
(-3,0) -- (3,0)node [black,midway,xshift=0pt,yshift=15pt] {
$N$};
    \end{tikzpicture}
    \end{array}\right)^2\ .
\end{equation}
Since this description involves only special unitary gauge groups, we should be able to engineer it using (two copies of) ordinary brane webs, i.e. without using O5-planes. It is constructed in such a way that it is decomposed into the copy of SU(2)$_0$ gauge theories in the region where the bi-fundamental masses are large enough \cite{Bergman:2013aca}. The result is given in the right of figure \ref{web diagrams EQ11 and EQ1}. This is also discussed in appendix \ref{appendixE}.


At this point we need to clarify which aspect of the two theories \eqref{E1xE1 electric OSp} and \eqref{E1xE1 electric unitary} are expected to be the same. This is because we used an isomorphism at the level of Lie algebra, ignoring any issues related to the global structure of the gauge group (with the exception of the choice of discrete theta angle mentioned above). In particular, any information related to local operators in the two theories is likely to agree, while questions about e.g. line and surface operators in general will be sensitive to the global structure of the gauge group. Our primary interest in these theories is in their Higgs branch, which is parameterised by local operators, and thus should agree regardless of any subtle differences such as those alluded to above.

Having constructed web diagrams for $\EQ_{1,1}$ and $\EQ_1$ in figure \ref{web diagrams EQ11 and EQ1}, we can now proceed to derive their magnetic quivers, following the rules introduced in \cite{Cabrera:2018jxt, Bourget:2020gzi, Akhond:2020vhc}. From the orientifold web in figure \ref{web diagrams EQ11 and EQ1} we obtain the OSp magnetic quiver \footnote{The flavor SO(4) node appearing in the quiver $\MQ_{1,1}$ was argued \cite{Akhond:2020vhc} to arise from the intersection between (0,1) 5-brane and O5$^+$-plane. This is based on the intuition that an O5$^+$ carries the same charge as an O5$^-$ and two (immobile) D5 branes.}
     \begin{equation}\label{MQ11}\MQ_{1,1}=
        \begin{array}{c}
             \begin{scriptsize}
             \begin{tikzpicture}
             \node[label=below:{1}][u](1){};
             \node (dots)[right of=1]{$\cdots$};
             \node[label=below:{$2N-1$}][u](2n-1)[right of=dots]{};
             \node[label=below:{$2N$}][sp](sp2n)[right of=2n-1]{};
             \node[label=above:{$4$}][sof](sof)[above of=sp2n]{};
             \draw(1)--(dots);
             \draw(dots)--(2n-1);
             \draw(2n-1)--(sp2n);
             \draw(sp2n)--(sof);
             \end{tikzpicture}
             \end{scriptsize}
        \end{array}\;,
    \end{equation}
    while taking two copies of the unitary web in figure \ref{web diagrams EQ11 and EQ1} leads us to conjecture
    
    \begin{equation}\label{MQ1^2}
   \MQ_{1,1}= \left(\MQ_{1}\right)^2=\left(\begin{array}{c}
             \begin{scriptsize}
             \begin{tikzpicture}
             \node[label=below:{1}][u](2){};
             \node (dots)[right of=2]{$\cdots$};
             \node[label=below:{$N$}][u](N)[right of=dots]{};
             \node (dots')[right of=N]{$\cdots$};
             \node[label=below:{1}][u](2')[right of=dots']{};
             \node[label=above:{1}][u](11)[above of=N]{};
             \draw(2)--(dots);
             \draw(dots)--(N);
             \draw[double distance=2pt](N)--(11);
             \draw(N)--(dots');
             \draw(dots')--(2');
             \end{tikzpicture}
             \end{scriptsize}
        \end{array}\right)^2    \;,
    \end{equation}
  where the first equality above is to be understood as an equality of moduli spaces. We will use this notation throughout.  
	Note that the set of balanced nodes in the unitary quiver implies an $\text{SU}(2N)\times \text{SU}(2N)$ symmetry, with each factor coming from the balanced nodes in one of the unitary quivers \eqref{MQ1^2}. This is consistent with the claim of Gaiotto and Witten \cite{Gaiotto:2008ak}, that whenever a chain of balanced unitary nodes terminate on a balanced symplectic node, the isometry of the Coulomb branch is doubled. A second consistency check, is that for $N=1$, the two theories are obviously identical, the OSp quiver in this case is the so called $T(\text{SO}(4))$, while the unitary side is two copies of $T(\text{SU}(2))$. In other words for $N=1$ we recover the statement
    \begin{equation}
        T(\text{SO}(4))\leftrightarrow T(\text{SU}(2)\times\text{SU}(2))\leftrightarrow T(\text{SU}(2))\times T(\text{SU}(2))\;.
    \end{equation}
    One upshot is that the HWG for the unitary quiver is straightforward to extract, given that its refined Hilbert series can be computed. Indeed the unitary quiver has already appeared in previous studies, for instance in \cite{Bourget:2019rtl}. Therefore our conjecture implies that the HWG of the OSp quiver \eqref{MQ11} is simply given by doubling the known result for \eqref{MQ1^2}, namely:  

\begin{equation}
\HWG_{1,1}=\PE\left[\sum_{k=1}^N\left(\mu_k\mu_{2N-k}+\nu_k\nu_{2N-k}\right)t^{2k}\right]    \;,
\end{equation}
where $\mu$ and $\nu$ are highest weight fugacities for $\text{SU}(2N)\times \text{SU}(2N)$. 
We can confirm this proposal by computing the unrefined Hilbert Series for the OSp quiver \eqref{MQ11} for small values of $N$. Some of the results are given in table \ref{TableHSOSp}.

Even more remarkable, is that the agreement between the quivers \eqref{MQ11} and \eqref{MQ1^2} is also valid on the Higgs branch. From the 5d perspective there is no a priori reason why this should be so, but it can be confirmed by an explicit calculation of the unrefined Hilbert series (see appendix \ref{appendixA} for a derivation of this formula)
\begin{equation}\label{MQ11 HB formula}
    \prod_{q=2}^{2N}(1-t^{2q})\int d\mu_{C_N}\PE\left[\chi_{\left[0,1,0,\cdots,0\right]_{C_N}}t^2+4\chi_{\left[1,0,\cdots,0\right]_{C_N}}t\right]\;.
\end{equation}
Let us evaluate this integral for $N=2$. The measure over the USp$(4)$ group can be taken to be
\begin{equation}
    \int d\mu_{C_2}=\oint_{|x_1|=1}\frac{d x_1}{2\pi \text{i} x_1}\oint_{|x_2|=1}\frac{d x_2}{2\pi \text{i} x_2}(1-x_1^2)(1-x_2)(1-\frac{x_1^2}{x_2})(1-\frac{x_2^2}{x_1^2})\;,
\end{equation}
while the characters for the fundamental and second rank antisymmetric representation of USp$(4)$ are given respectively by\footnote{These characters are computed as follows. For a weight $w = [w_1, w_2] \equiv w_1\Lambda_1 + w_2 \Lambda_2$ appearing in the weight system of a representation $R$ of $C_2$, the corresponding monomial in the character will be $x_1^{w_1}x_2^{w_2}$. For example, the weights appearing in the weight system of fundamental representation of $C_2$ are $\{[1,0],[-1,1],[1,-1],[-1,0]\}$ where each weight is written in the fundamental weight basis. Thus the character for fundamental representation will be simply $x_1^{1}x_2^{0}+x_1^{-1}x_2^{1}+x_1^{1}x_2^{-1}+x_1^{-1}x_2^{0}$. }
\begin{equation}
\begin{split}
    \chi_{\left[1,0\right]_{C_2}}&=x_1+\frac{x_2}{x_1}+\frac{x_1}{x_2}+\frac{1}{x_1}\;,\\
    \chi_{\left[0,1\right]_{C_2}}&=x_2+\frac{x_1^2}{x_2}+1+\frac{x_2}{x_1^2}+\frac{1}{x_2}\;,\\
\end{split}
\end{equation}
Thus the expression we need to evaluate is
\begin{equation}
\begin{gathered}
\prod_{q=2}^4(1-t^{2q})\oint_{|x_1|=1}\frac{d x_1}{2\pi \text{i} x_1}\oint_{|x_2|=1}\frac{d x_2}{2\pi \text{i} x_2}(1-x_1^2)(1-x_2)(1-\frac{x_1^2}{x_2})(1-\frac{x_2^2}{x_1^2})\times\\
\times\frac{1}{(1-x_2t^2)(1-\frac{x_1^2}{x_2}t^2)(1-t^2)(1-\frac{x_2}{x_1^2}t^2)(1-\frac{t^2}{x_2})(1-x_1t)^4(1-\frac{x_2}{x_1}t)^4(1-\frac{x_1}{x_2}t)^4(1-\frac{t}{x_1})^4}\;.
\end{gathered}
\end{equation}
This integral can now be evaluated using residues to arrive at the following Hilbert series
\begin{equation}\label{HB HS of MQ1^2}
    \HS_{1,1}^{\mathcal{H}}\rvert_{N=2}=\frac{(1-t^6)^2(1-t^8)^2}{(1-t^4)^6(1-t^2)^6}\;.
\end{equation}
We recognise this as the Coulomb branch Hilbert series of two copies of U$(2)$ with 4 fundamental hypermultiplets \cite{Cremonesi:2013lqa}, which is the mirror of the $N=2$ quiver of \eqref{MQ1^2}. Therefore we see that the agreement between the OSp quiver \eqref{MQ11} and unitary quiver \eqref{MQ1^2} holds also on the Higgs branch.
\subsection{The \texorpdfstring{$E_1\times E_3$}{TEXT} sequence}
In the previous subsection, we saw that the Higgs branch of the fixed point limit of 5$d$ SO(4) gauge theory, factorises to two copies of the Higgs branch of the 5$d$ pure SU(2) gauge theory. It is natural to ask whether this pattern holds if we include matter transforming under SO(4). The two matter representations which are straightforward to obtain from the brane web are the vector of SO(4) and the two spinor representations of opposite chirality. Since the vector of SO(4) corresponds to bifundamental of SU(2)$\times$SU(2), this will not lead to the desired factorised theory. However, each of the two spinor representations, denoted by \textbf{s} and \textbf{c} respectively, will only transform under one of the two SU(2) factors in SO(4). In this and subsequent subsections, we will exploit this well-known fact.

\begin{figure}[!htb]
    \centering
             \begin{scriptsize}
         \begin{tikzpicture}
    \draw[thick,dashed](0,0)--(4,0);
    \draw[thick](2,4)--(2,6);
    \node at (.5,-0.25){O5$^+$};
    \node at (3.5,-0.25){O5$^+$};
    \node[label=right:{$1$}] at (2,5.5){};
    \node[label=right:{$2$}] at (2,4.5){};
    \node[label=right:{$2N-1$}] at (2,2.5){};
    \node[label=right:{$2N$}] at (2,1.5){};
    \draw[thick](2,0)--(4,1);
    \draw[thick](2,0)--(0,1);
    \draw[thick](2,0)--(2,3);
    \node[label=above:{(2,1)}][7brane]at(4,1){};
    \node[label=above:{(2,-1)}][7brane]at(0,1){};
    \node[7brane]at(2,2){};
    \node[7brane]at(2,3){};
    \node at (2,3.5) {$\vdots$};
    \node[7brane]at(2,4){};
    \node[7brane]at(2,5){};
    \node[7brane]at(2,6){};
    \end{tikzpicture}
    \hspace{1cm}
    \begin{tikzpicture}[scale=.6]
    \draw[thick](-2.5,-2.5)--(2.5,2.5);
    \draw[thick](-3.5,-3.5)--(-5.5,-5.5);
    \draw[thick](3.5,3.5)--(5.5,5.5);
    \node[label=above right:{$(1,1)$}][7brane]at(5.5,5.5){};
    \node[7brane]at(4.5,4.5){};
    \node[7brane]at(3.5,3.5){};
    \node[7brane]at(2.5,2.5){};
    \node[7brane]at(1.5,1.5){};
        \node[label=below left:{$(1,1)$}][7brane]at(-5.5,-5.5){};
    \node[7brane]at(-4.5,-4.5){};
    \node[7brane]at(-3.5,-3.5){};
    \node[7brane]at(-2.5,-2.5){};
    \node[7brane]at(-1.5,-1.5){};
    \draw[thick](-2,0)--(2,0);
    \draw[thick](0,-2)--(0,2);
    \node[7brane]at(-2,0){};
    \node[7brane]at(2,0){};
    \node[7brane]at(0,-2){};
    \node[7brane]at(0,2){};
    \node at (3,3){$\udots$};
    \node at (-3,-3){$\udots$};
    \node[label=below right:{1}] at(4.75,5.25){};
    \node[label=below right:{2}] at(3.75,4.25){};
    \node[label=below right:{$N-1$}] at(1.75,2.25){};
    \node[label=below right:{$N$}] at(.75,1.25){};
    \node[label=above left:{1}] at(-4.75,-5.25){};
    \node[label=above left:{2}] at(-3.75,-4.25){};
    \node[label=above left:{$N-1$}] at(-1.75,-2.25){};
    \node[label=above left:{$N$}] at(-.75,-1.25){};
    \end{tikzpicture}
         \end{scriptsize}
    \caption{Brane webs engineering EQ$_{1,3}$ (left) and EQ$_3$ (right). For the unitary brane web engineering EQ$_1$, see figure \ref{web diagrams EQ11 and EQ1}.}
    \label{web diagrams EQ13 and EQ3}
\end{figure}
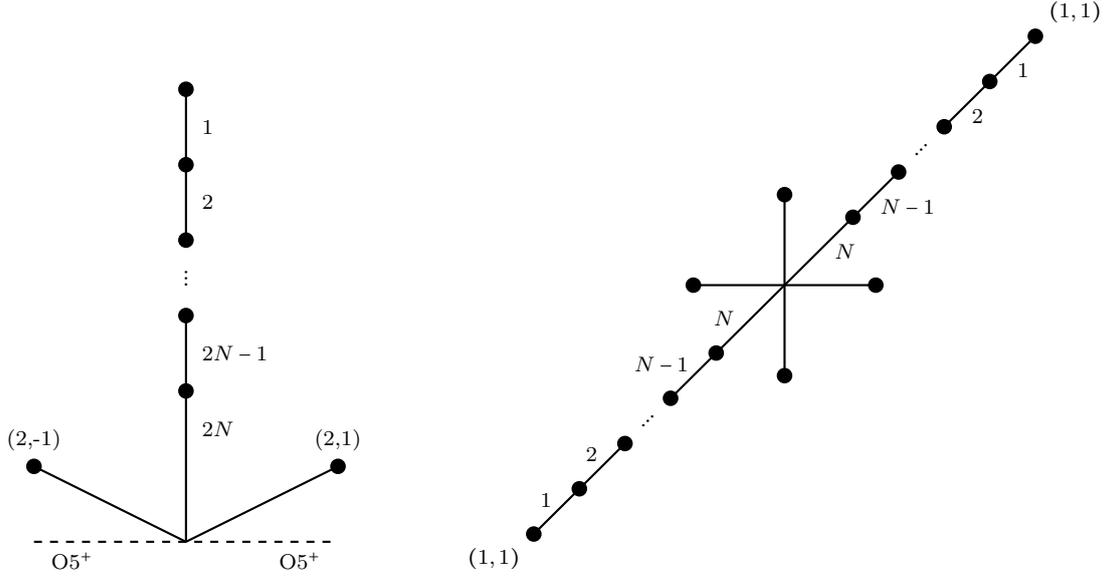

Consider the orientifold web diagram presented in figure \ref{web diagrams EQ13 and EQ3}. The corresponding IR quiver gauge theory description is given by the electric quiver
\begin{equation}\EQ_{1,3}=
\begin{array}{c}
\begin{tikzpicture}
    \node{$[1\;\textbf{s}]-\text{SO(4)}-\text{USp(0)}-\text{SO(4)}-\cdots-\text{USp(0)}-\text{SO(4)}-[1\;\textbf{s}]$};
    
    \draw [thick,decorate,decoration={brace,amplitude=6pt},xshift=0pt,yshift=10pt]
(-3.75,0) -- (3.75,0)node [black,midway,xshift=0pt,yshift=15pt] {
$2N-1$};
    \end{tikzpicture}
    \end{array}\;.
    \end{equation}
By using the isomorphism $\mathfrak{so}(4)\cong \mathfrak{su}(2)\times\mathfrak{su}(2)$, we can rewrite this theory as a product of the following electric quivers.
\begin{equation}\label{E1xE3 electric unitary}
    \EQ_{1,3}=\EQ_1\times \EQ_3=\begin{array}{c}
\begin{tikzpicture}
    \node{$\text{SU(2)}-\text{SU(2)}-\cdots-\text{SU(2)}$};
    
    \draw [thick,decorate,decoration={brace,amplitude=6pt},xshift=0pt,yshift=10pt]
(-2,0) -- (2,0)node [black,midway,xshift=0pt,yshift=15pt] {
$N$};
    \end{tikzpicture}
    \end{array}\times\begin{array}{c}
\begin{tikzpicture}
    \node{${\underset{\underset{\text{\large$\left[1\textbf{F}\right]$}}{\textstyle\vert}}{\text{SU}(2)}}-\text{SU(2)}-\cdots-{\underset{\underset{\text{\large$\left[1\textbf{F}\right]$}}{\textstyle\vert}}{\text{SU}(2)}}$};
    
    \draw [thick,decorate,decoration={brace,amplitude=6pt},xshift=0pt,yshift=10pt]
(-2.25,.5) -- (2.25,.5)node [black,midway,xshift=0pt,yshift=15pt] {
$N$};
    \end{tikzpicture}
    \end{array}
\end{equation}
The unitary brane web for $\EQ_1$ is given in figure \ref{web diagrams EQ11 and EQ1}, while the unitary brane web for $\EQ_3$ is presented in figure \ref{web diagrams EQ13 and EQ3}. The orientifold web in figure \ref{web diagrams EQ13 and EQ3} admits two maximal subdivisions. Accordingly the Higgs branch of this theory is the union of two cones, given by the two OSp magnetic quivers in table \ref{tab:MQ13}.
On the other hand, we expect these magnetic quivers to be equivalent to the product $\MQ_1\times\left(\MQ_3^{(\text{I})}\cup\MQ_3^{(\text{II})}\right)$, with the latter factor obtained from the unitary web for $\EQ_3$ in figure \ref{web diagrams EQ13 and EQ3}. As a further non-trivial check, we can compute the Coulomb branch Hilbert series of the unitary and OSp quivers. The unitary quivers $\MQ_1$ along with $\MQ_3^{(\text{I})}$ and $\MQ_3^{(\text{II})}$ have known HWGs \cite{Bourget:2019rtl}, which we have tabulated in table \ref{TableHWGUnitaryQuivers}.
We are now in a position to write down the HWGs for the OSp magnetic quivers in table \ref{tab:MQ13}. 
\begin{table}[!htb]
            \centering
            \begin{tabular}{|c|c|c|}\hline
            MS&OSp&Unitary\\\hline
                 $\MQ_{1,3}^{(\text{I})}$ & $\begin{array}{c}
        \begin{scriptsize}
        \begin{tikzpicture}
            \node[label=below:{1}][u](1){};
            \node (dots)[right of=1]{$\cdots$};
            \node[label=below:{$2N-1$}][u](2N-1)[right of=dots]{};
            \node[label=below:{$2N$}][sp](sp2N)[right of=2N-1]{};
            \node[label=below:{$1$}][u](11)[right of=sp2N]{};
            \node[label=above:{$1$}][uf](1f)[above of=11]{};
            \draw(1)--(dots);
            \draw(dots)--(2N-1);
            \draw(2N-1)--(sp2N);
            \draw[double distance=2pt](sp2N)--(11);
             \path [draw,snake it](11)--(1f);
        \end{tikzpicture}
        \end{scriptsize}\end{array}$ & $\begin{array}{cc}
             \begin{scriptsize}
             \begin{tikzpicture}
             \node[label=below:{1}][u](2){};
             \node (dots)[above right of=2]{$\udots$};
             \node[label=below:{$N$}][u](N)[above right of=dots]{};
             \node (dots')[below right of=N]{$\ddots$};
             \node[label=below:{1}][u](2')[below right of=dots']{};
             \node[label=above:{1}][u](11)[above of=N]{};
             \draw(2)--(dots);
             \draw(dots)--(N);
             \draw[double distance=2pt](N)--(11);
             \draw(N)--(dots');
             \draw(dots')--(2');
             \end{tikzpicture}
             \end{scriptsize}&
             \begin{scriptsize}
         \begin{tikzpicture}
             \node[label=below:{1}][u]{};
             \node (dots)[above right of=1]{$\udots$};
             \node[label=below:{$N$}][u](N)[above right of=dots]{};
             \node (dots')[below right of=N]{$\ddots$};
             \node[label=below:{1}][u](11)[below right of=dots']{};
             \node[label=above:{1}][u](u1)[above left of=N]{};
             \node[label=above:{1}][u](u11)[above right of=N]{};
             \draw(1)--(dots);
             \draw(dots)--(N);
             \draw(N)--(dots');
             \draw(dots')--(11);
             \draw(N)--(u1);
             \draw(N)--(u11);
             \draw(u1)--(u11);
         \end{tikzpicture}
         \end{scriptsize}
        \end{array}$\\
        \hline          $\MQ_{1,3}^{(\text{II})}$&$\begin{array}{c}
        \begin{scriptsize}
        \begin{tikzpicture}
            \node[label=below:{1}][u](1){};
            \node (dots)[right of=1]{$\cdots$};
            \node[label=below:{$2N-1$}][u](2N-1)[right of=dots]{};
            \node[label=above:{$2N-2$}][sp](sp2N)[above of=2N-1]{};
            \node[label=below:{$1$}][u](11)[right of=2N-1]{};
            \node[label=above:{$2$}][uf](1f)[above of=11]{};
            \draw(1)--(dots);
            \draw(dots)--(2N-1);
            \draw(2N-1)--(sp2N);
            \draw[double distance=2pt](2N-1)--(11);
             \path [draw,snake it](11)--(1f);
        \end{tikzpicture}
        \end{scriptsize}
    \end{array}$ & $\begin{array}{cc}
             \begin{scriptsize}
             \begin{tikzpicture}
             \node[label=below:{1}][u](2){};
             \node (dots)[above right of=2]{$\udots$};
             \node[label=below:{$N$}][u](N)[above right of=dots]{};
             \node (dots')[below right of=N]{$\ddots$};
             \node[label=below:{1}][u](2')[below right of=dots']{};
             \node[label=above:{1}][u](11)[above of=N]{};
             \draw(2)--(dots);
             \draw(dots)--(N);
             \draw[double distance=2pt](N)--(11);
             \draw(N)--(dots');
             \draw(dots')--(2');
             \end{tikzpicture}
             \end{scriptsize}&
             \begin{scriptsize}
              \begin{tikzpicture}
                  \node[label=below:{1}][u](1){};
                  \node(dots)[above of=1]{$\vdots$};
                  \node[label=left:{$N-1$}][u](N-1)[above of=dots]{};
                  \node[label=below:{$N-1$}][u](N-1')[right of=N-1]{};
                  \node[label=right:{$N-1$}][u](N-1'')[right of=N-1']{};
                  \node (dots')[below of=N-1'']{$\vdots$};
                  \node[label=below:{1}][u](1')[below of=dots']{};
                  \node[label=above:{1}][u](u1)[above of=N-1]{};
                  \node[label=above:{1}][u](u11)[above of=N-1'']{};
                  \draw(1)--(dots);
                  \draw(dots)--(N-1);
                  \draw(N-1)--(N-1');
                  \draw(N-1')--(N-1'');
                  \draw(N-1'')--(dots');
                  \draw(dots')--(1');
                  \draw(N-1)--(u1);
                  \draw(N-1'')--(u11);
                  \draw[double distance=2pt](u1)--(u11);
              \end{tikzpicture}
              \end{scriptsize}
             \end{array}$\\\hline
            \end{tabular}
            \caption{OSp and unitary representation of the two cones on the Higgs branch of $\EQ_{1,3}$. The unitary quivers appearing in the extreme right of the two rows of the table are respectively $\MQ_3^{(\text{I})}$ and $\MQ_3^{(\text{II})}$.}
            \label{tab:MQ13}
        \end{table}
The final result for the first cone reads
\begin{equation}
    \HWG_{1,3}^{\text{(I)}}=\PE\left[\sum_{k=1}^N\mu_k\mu_{2N-k}t^{2k}\right] \PE\left[t^2+\left(q+q^{-1}\right)\nu_Nt^{N+1}+\sum_{k=1}^{N}\nu_k\nu_{2N-k}t^{2k}-\nu_N^2t^{2N+2}\right]\;,
\end{equation}
where $\mu$ and $\nu$ are the highest weight fugacities for SU$(2N)\times$SU$(2N)$ while $q$ keeps track of the U(1) charge. The HWG for the second cone reads
\begin{align}
    \HWG_{1,3}^{\text{(II)}} &= \PE\left[\sum_{k=1}^N\mu_k\mu_{2N-k}t^{2k}\right] \nonumber \\
    &\times \PE\left[t^2+\left(\nu_{N+1}q+\nu_{N-1}q^{-1}\right)t^{N+1}+\sum_{k=1}^{N-1}\nu_k\nu_{2N-k}t^{2k}-\nu_{N+1}\nu_{N-1}t^{2N+2}\right] ~.
\end{align}
This can be verified upon comparison with the result of an unrefined Hilbert series computation on the OSp side. The results for low values of $N$ are given in table \ref{TableHSOSp}. 

We can also compute the Higgs branch HS for MQ$_{1,3}^{(\text{I})}$ for $N=2$ exactly. The computation is very similar to that of the Higgs branch of $\MQ_{1,1}$ discussed around \eqref{HB MQ11 initial integral}. We need to evaluate the following integral
\begin{equation}
    \prod_{q=1}^4(1-t^{2q})\int d\mu_{C_2}\int d\mu_{U(1)}\PE\left[\chi_{\left[0,1\right]}^{C_2}t^2+2\chi_{\left[1,0\right]}^{C_2}(q+q^{-1})t+(q^2+q^{-2})t\right]
\end{equation}
Evaluating this integral by finding the residues one arrives at the following 
\begin{equation}\label{HB MQ13(I)}
    \HS_{1,3,(\text{I})}^\mathcal{H}=\frac{(1 - t + t^2) (1 + t^4) (1 + t^3 + t^4 + t^5 + t^6 + t^7 + 
   t^{10})}{(1 - t)^8 (1 + t)^6 (1 + t^2)^3 (1 + t + t^2 + t^3 + t^4)}\;.
\end{equation}
This is to be compared with the product of the Higgs branch HS of the two unitary quivers appearing in the first row of table \ref{tab:MQ13}. We already know the result for one of these, which is identical to $\MQ_1$ of \eqref{MQ1^2}. Its Higgs branch HS was discussed in the previous section and is given by the square root of the expression in \eqref{HB HS of MQ1^2}. The Higgs branch HS of the other quiver in the first row of table, \ref{tab:MQ13}, which we dub $\MQ_3^{(\text{I})}$ is straightforward to compute. Specialising to the case $N=2$, we need to evaluate the following integral
\begin{equation}
\begin{gathered}
\int d\mu_{\text{U}(2)}(x,q_x)\int d\mu_{\text{U}(1)}(u)H^2_{T\left[\text{SU}(2)\right]}(x)H_{\text{glue}}^{\text{U}(2)}(x,q_x)\times\\\times H_{[2]-[1]}(x,q_x)H_{[2]-[1]}(x,q_x,u)H_{[1]-[1]}(u)H_\text{glue}^{\text{U}(1)}(u) \\
    =\oint_{|x|=1}\frac{dx}{2\pi\text{i}x}(1-x^2)\oint_{|q_x|=1}\frac{dq_x}{2\pi\text{i}q_x}\oint_{|u|=1}\frac{du}{2\pi\text{i}u}(1-t^2)^2(1-t^4)^2\times\\
    \times \PE\left[(x^2+1+x^{-2})t^2+(x+x^{-1})(q_x+q_x^{-1})(u+u^{-1})t+(u+u^{-1})t+(x+x^{-1})(q_x+q_x^{-1})t\right]\;.
    \end{gathered}
\end{equation}
Evaluating this integral by computing its residues results in
\begin{equation}
    \HS_{3,(\text{I})}^\mathcal{H}=\frac{(1 + t^3 + t^4 + t^5 + t^6 + t^7 + t^{10})}{(1 - t)^4 (1 + t)^2 (1 + 
   t^2) (1 + t + t^2) (1 + t + t^2 + t^3 + t^4)}\;.
\end{equation}
Together with the result for the HB of MQ$_1$, this precisely reproduces the computation on the OSp side \eqref{HB MQ13(I)}.
\subsection{The \texorpdfstring{$E_3\times E_3$}{TEXT} sequence}
\begin{figure}[!htb]
    \centering
           \begin{scriptsize}
         \begin{tikzpicture}
         
    \draw[thick](-1,0)--(5,0);
    \draw[thick, dashed](-1,0)--(-2,0);
    \draw[thick, dashed](5,0)--(6,0);
    \draw[thick](2,4)--(2,6);
    \node at (-1.5,-0.25){O5$^-$};
    \node at (5.5,-0.25){O5$^-$};
    \node[label=right:{$1$}] at (2,5.5){};
    \node[label=right:{$2$}] at (2,4.5){};
    \node[label=right:{$2N-1$}] at (2,2.5){};
    \node[label=right:{$2N$}] at (2,1.5){};
    \draw[thick](2,0)--(3,1);
    \draw[thick](2,0)--(1,1);
    \draw[thick](2,0)--(2,3);
    \node[label=above right:{(1,1)}][7brane]at(3,1){};
    \node[label=above left:{(1,-1)}][7brane]at(1,1){};
    \node[7brane]at(2,2){};
    \node[7brane]at(2,3){};
    \node at (2,3.5) {$\vdots$};
    \node[7brane]at(2,4){};
    \node[7brane]at(2,5){};
    \node[7brane]at(2,6){};
    \node[label=below:{$\frac{1}{2}$}]at(4.5,0){};
    \node[label=below:{$1$}]at(3.5,0){};
    \node[label=below:{$\frac{1}{2}$}]at(-.5,0){};
    \node[label=below:{$1$}]at(.5,0){};
    \node[7brane]at(-1,0){};
    \node[7brane]at(0,0){};
    \node[7brane]at(5,0){};
    \node[7brane]at(4,0){};
    
    \end{tikzpicture}
    \end{scriptsize}
    \caption{Orientifold web for EQ$_{3,3}$.}
    \label{EQ33 orientifold web}
\end{figure}
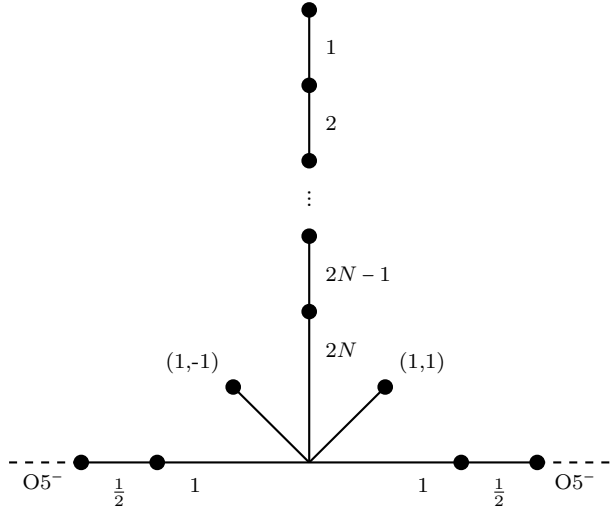
The $E_3\times E_3$ sequence corresponds to the fixed point limit of the electric quiver
   \begin{equation}\label{E3xE3 electric OSp}
   \EQ_{3,3}=\left(\EQ_3\right)^2=
    \begin{array}{c}
\begin{tikzpicture}
    \node{${\underset{\underset{\text{\large$\left[1\textbf{s}+1\textbf{c}\right]$}}{\textstyle\vert}}{\text{SO}(4)}}-\text{USp(0)}-\text{SO(4)}-\cdots-\text{USp(0)}-{\underset{\underset{\text{\large$\left[1\textbf{s}+1\textbf{c}\right]$}}{\textstyle\vert}}{\text{SO}(4)}}$};
    
    \draw [thick,decorate,decoration={brace,amplitude=6pt},xshift=0pt,yshift=10pt]
(-3.75,0.5) -- (3.75,0.5)node [black,midway,xshift=0pt,yshift=15pt] {
$2N-1$};
    \end{tikzpicture}
    \end{array}\;.
\end{equation}
The orientifold web which engineers this theory is given in figure \ref{EQ33 orientifold web}. This brane web admits three maximal subdivisions leading to the three OSp magnetic quivers in table \ref{tab:E3xE3 magnetic quivers}. It can be understood as a limiting case of the $Y_N^{1,1}$ theory in \cite{Akhond:2020vhc}.
One can provide a purely unitary description of this theory in terms of the following electric quiver:
\begin{equation}\label{E3xE3 electric unitary}
    \EQ_{3,3}=\EQ_3^2=\left(\begin{array}{c}
\begin{tikzpicture}
    \node{${\underset{\underset{\text{\large$\left[1\textbf{F}\right]$}}{\textstyle\vert}}{\text{SU}(2)}}-\text{SU(2)}-\cdots-{\underset{\underset{\text{\large$\left[1\textbf{F}\right]$}}{\textstyle\vert}}{\text{SU}(2)}}$};
    
    \draw [thick,decorate,decoration={brace,amplitude=6pt},xshift=0pt,yshift=10pt]
(-2.25,.5) -- (2.25,.5)node [black,midway,xshift=0pt,yshift=15pt] {
$N$};
    \end{tikzpicture}
    \end{array}\right)^2 ~,
\end{equation}
which can be engineered by taking two copies of the unitary web shown in figure \ref{web diagrams EQ13 and EQ3}. This brane web admits two maximal subdivisions whose magnetic quivers were discussed in the previous subsection. Since we are taking two copies, a third cone arises when we take a different maximal subdivision for each web diagram. This leads us to the unitary magnetic quivers in table \ref{tab:E3xE3 magnetic quivers}. 

\begin{table}[!htb]
    \centering
        \begin{tabular}{|c|c|c|}
    \hline MS&OSp&Unitary\\\hline 
      $\MQ_{3,3}^{(\text{I})}$ &$  \begin{array}{c}\begin{scriptsize}
         \begin{tikzpicture}
             \node[label=below:{1}][u](1){};
             \node (dots)[right of=1]{$\cdots$};
             \node[label=below:{$2N-1$}][u](2N-1)[right of=dots]{};
             \node[label=below:{$2N$}][sp](sp2N)[right of=2N-1]{};
             \node[label=above:{$2$}][so](so2)[above right of=sp2N]{};
             \node[label=below:{$1$}][u](u1)[below right of=sp2N]{};
             \draw(1)--(dots);
             \draw(dots)--(2N-1);
             \draw(sp2N)--(2N-1);
             \draw(sp2N)--(so2);
             \draw(sp2N)--(u1);
             \draw(u1)--(so2);
         \end{tikzpicture}
         \end{scriptsize}\end{array}$& $\left(\begin{array}{c}\begin{scriptsize}
         \begin{tikzpicture}
             \node[label=below:{1}][u]{};
             \node (dots)[right of=1]{$\cdots$};
             \node[label=below:{$N$}][u](N)[right of=dots]{};
             \node (dots')[right of=N]{$\cdots$};
             \node[label=below:{1}][u](11)[right of=dots']{};
             \node[label=above:{1}][u](u1)[above left of=N]{};
             \node[label=above:{1}][u](u11)[above right of=N]{};
             \draw(1)--(dots);
             \draw(dots)--(N);
             \draw(N)--(dots');
             \draw(dots')--(11);
             \draw(N)--(u1);
             \draw(N)--(u11);
             \draw(u1)--(u11);
         \end{tikzpicture}
         \end{scriptsize}\end{array}\right)^2$ \\\hline 
				$\MQ_{3,3}^{(\text{II})}$&
        $ \begin{array}{c}\begin{scriptsize}
         \begin{tikzpicture}
                    \node[label=below:{1}][u](1){};
             \node (dots)[right of=1]{$\cdots$};
             \node[label=below:{$2N-2$}][u](2N-2)[right of=dots]{};
             \node[label=below:{$2N-2$}][u](2N-2')[right of=2N-2]{};
             \node[label=below:{$2N-2$}][sp](sp2N-2)[right of=2N-2']{};
             \node[label=above:{$2$}][so](so2)[above of=sp2N-2]{};
             \node[label=above:{$1$}][u](u1)[above of=2N-2]{};
             \draw(1)--(dots);
             \draw(dots)--(2N-2);
             \draw(2N-2)--(2N-2');
             \draw(2N-2')--(sp2N-2);
             \draw(sp2N-2)--(so2);
             \draw[double distance=2pt](so2)--(u1);
             \draw(u1)--(2N-2);
         \end{tikzpicture}
         \end{scriptsize}\end{array}$&$\left(\begin{array}{c}
              \begin{scriptsize}
              \begin{tikzpicture}
                  \node[label=below:{1}][u](1){};
                  \node(dots)[right of=1]{$\cdots$};
                  \node[label=below:{$N-1$}][u](N-1)[right of=dots]{};
                  \node[label=below:{$N-1$}][u](N-1')[right of=N-1]{};
                  \node[label=below:{$N-1$}][u](N-1'')[right of=N-1']{};
                  \node (dots')[right of=N-1'']{$\cdots$};
                  \node[label=below:{1}][u](1')[right of=dots']{};
                  \node[label=above:{1}][u](u1)[above of=N-1]{};
                  \node[label=above:{1}][u](u11)[above of=N-1'']{};
                  \draw(1)--(dots);
                  \draw(dots)--(N-1);
                  \draw(N-1)--(N-1');
                  \draw(N-1')--(N-1'');
                  \draw(N-1'')--(dots');
                  \draw(dots')--(1');
                  \draw(N-1)--(u1);
                  \draw(N-1'')--(u11);
                  \draw[double distance=2pt](u1)--(u11);
              \end{tikzpicture}
              \end{scriptsize}
         \end{array}\right)^2$\\\hline $\MQ_{3,3}^{(\text{III})}$&$\begin{array}{c}
         \begin{scriptsize}
         \begin{tikzpicture}
               \node[label=below:{1}][u](1){};
             \node (dots)[right of=1]{$\cdots$};
             \node[label=below:{$2N-1$}][u](2N-1)[right of=dots]{};
             \node[label=below:{$2N-2$}][sp](sp2N-2)[right of=2N-1]{};
           \node[label=below:{1}][u](u1)[below right of=sp2N-2]{};
           \node[label=above:{1}][u](u11)[above right of=sp2N-2]{};
           \draw(1)--(dots);
           \draw(dots)--(2N-1);
           \draw(2N-1)--(sp2N-2);
           \draw(u1)--(2N-1);
           \draw(2N-1)--(u11);
            \draw[dashed, double distance=2 pt] (u1)--(u11);
           \draw(u1)to[out =30, in=-30](u11);
         \end{tikzpicture}
         \end{scriptsize}\end{array}$& 
         $\begin{array}{cc}
             \begin{scriptsize}
         \begin{tikzpicture}
             \node[label=below:{1}][u]{};
             \node (dots)[above right of=1]{$\udots$};
             \node[label=below:{$N$}][u](N)[above right of=dots]{};
             \node (dots')[below right of=N]{$\ddots$};
             \node[label=below:{1}][u](11)[below right of=dots']{};
             \node[label=above:{1}][u](u1)[above left of=N]{};
             \node[label=above:{1}][u](u11)[above right of=N]{};
             \draw(1)--(dots);
             \draw(dots)--(N);
             \draw(N)--(dots');
             \draw(dots')--(11);
             \draw(N)--(u1);
             \draw(N)--(u11);
             \draw(u1)--(u11);
         \end{tikzpicture}
         \end{scriptsize} &   \begin{scriptsize}
              \begin{tikzpicture}
                  \node[label=below:{1}][u](1){};
                  \node(dots)[above of=1]{$\vdots$};
                  \node[label=left:{$N-1$}][u](N-1)[above of=dots]{};
                  \node[label=below:{$N-1$}][u](N-1')[right of=N-1]{};
                  \node[label=right:{$N-1$}][u](N-1'')[right of=N-1']{};
                  \node (dots')[below of=N-1'']{$\vdots$};
                  \node[label=below:{1}][u](1')[below of=dots']{};
                  \node[label=above:{1}][u](u1)[above of=N-1]{};
                  \node[label=above:{1}][u](u11)[above of=N-1'']{};
                  \draw(1)--(dots);
                  \draw(dots)--(N-1);
                  \draw(N-1)--(N-1');
                  \draw(N-1')--(N-1'');
                  \draw(N-1'')--(dots');
                  \draw(dots')--(1');
                  \draw(N-1)--(u1);
                  \draw(N-1'')--(u11);
                  \draw[double distance=2pt](u1)--(u11);
              \end{tikzpicture}
              \end{scriptsize}
         \end{array}$\\\hline
    \end{tabular}
    \caption{Unitary and OSp magnetic quivers for the $E_3\times E_3$ sequence.}
    \label{tab:E3xE3 magnetic quivers}
\end{table}
We can now infer the HWG for the OSp quivers in table \ref{tab:E3xE3 magnetic quivers} by taking those of the corresponding unitary magnetic quivers as building blocks. This reasoning leads us to conjecture the following HWG for the three cones in table \ref{tab:E3xE3 magnetic quivers}
\begin{align}\label{HWG MQ33}
    \HWG_{3,3}^{\text{(I)}}(t^2)&= \PE\left[t^2 + (\mu_N q_1 + \mu_N q_1^{-1})t^{N+1} + \sum_{k=1}^{N}(\mu_k\mu_{2N-k}t^{2k}) - \mu_N^2 t^{2N+2} \right]
		 \nonumber \\ &\times\PE\left[ t^2 + (\nu_N q_2 + \nu_N q_2^{-1})t^{N+1} + \sum_{k=1}^{N}(\nu_k\nu_{2N-k}t^{2k}) - \nu_N^2 t^{2N+2} \right] \nonumber \\
     \HWG_{3,3}^{\text{(II)}}(t^2)&=\PE\left[t^2+\left(\mu_{N+1}p+\mu_{N-1}p^{-1}\right)t^{N+1}+\sum_{k=1}^{N-1}\mu_k\mu_{2N-k}t^{2k}-\mu_{N+1}\mu_{N-1}t^{2N+2}\right]\nonumber \\&\times\PE\left[t^2+\left(\nu_{N+1}q+\nu_{N-1}q^{-1}\right)t^{N+1}+\sum_{k=1}^{N-1}\nu_k\nu_{2N-k}t^{2k}-\nu_{N+1}\nu_{N-1}t^{2N+2}\right]\nonumber \\
     \HWG_{3,3}^{\text{(III)}}(t^2)&=\PE\left[t^2+\left(\mu_{N+1}p+\mu_{N-1}p^{-1}\right)t^{N+1}+\sum_{k=1}^{N-1}\mu_k\mu_{2N-k}t^{2k}-\mu_{N+1}\mu_{N-1}t^{2N+2}\right]\nonumber\\&\times\PE\left[t^2+\left(q+q^{-1}\right)\nu_Nt^{N+1}+\sum_{k=1}^{N}\nu_k\nu_{2N-k}t^{2k}-\nu_N^2t^{2N+2}\right] ~.
     \end{align}
Here $\mu$ and $\nu$ are the fuagicites for the two SU($2N$) groups and $p$ and $q$ are the U(1) charges. We have verified this result by an explicit unrefined Hilbert series computation of the Coulomb branch of the OSp quivers which are presented in table \ref{TableHSOSp} for low values of $N$. 
\subsection{The \texorpdfstring{$E'_3\times E'_3$}{TEXT} sequence}
\begin{figure}[!htb]
    \centering
   \begin{scriptsize}
         \begin{tikzpicture}[scale=.75]
    \draw[thick](2,0)--(6,0);
    \draw[thick, dashed](2,0)--(0,0);
    \draw[thick, dashed](7,0)--(6,0);
    \draw[thick](2,4)--(2,6);
    \node[label=below:{O5$^+$}] at (.5,0){};
    \node[label=below:{O5$^-$}] at (6.5,0){};
    \node[label=right:{$1$}] at (2,5.5){};
    \node[label=right:{$2$}] at (2,4.5){};
    \node[label=right:{$2N$}] at (2,2.5){};
    \node[label=right:{$2N+1$}] at (2,1.5){};
    \draw[thick](2,0)--(2,3);
    \node[7brane]at(2,2){};
    \node[7brane]at(2,3){};
    \node at (2,3.5) {$\vdots$};
    \node[7brane]at(2,4){};
    \node[7brane]at(2,5){};
    \node[7brane]at(2,6){};
    \node[label=below:{$\frac{1}{2}$}]at(5.5,0){};
    \node[label=below:{$1$}]at(4.5,0){};
    \node[label=below:{$\frac{3}{2}$}]at(3.5,0){};
    \node[label=below:{$2$}]at(2.5,0){};
    \node[7brane]at(5,0){};
    \node[7brane]at(4,0){};
    \node[7brane]at(6,0){};
    \node[7brane]at(3,0){};
    
    \end{tikzpicture}
    \end{scriptsize}
    \begin{scriptsize}
    \begin{tikzpicture}[scale=.55]
    \draw[thick](-2.5,-2.5)--(2.5,2.5);
    \draw[thick](-3.5,-3.5)--(-5.5,-5.5);
    \draw[thick](3.5,3.5)--(5.5,5.5);
    \node[label=above right:{$(1,1)$}][7brane]at(5.5,5.5){};
    \node[7brane]at(4.5,4.5){};
    \node[7brane]at(3.5,3.5){};
    \node[7brane]at(2.5,2.5){};
    \node[7brane]at(1.5,1.5){};
        \node[label=below left:{$(1,1)$}][7brane]at(-5.5,-5.5){};
    \node[7brane]at(-4.5,-4.5){};
    \node[7brane]at(-3.5,-3.5){};
    \node[7brane]at(-2.5,-2.5){};
    \node[7brane]at(-1.5,-1.5){};
    \draw[thick](0,0)--(4,0);
    \draw[label=above:{(1,-1)}][thick](0,0)--(-1.5,1.5);
    \node at (-1.75,1){1};
    \node[7brane]at(-1.5,1.5){};
    \node[7brane]at(2,0){};
    \node[7brane]at(4,0){};
    \node at (3,3){$\udots$};
    \node at (-3,-3){$\udots$};
    \node[label=above:{1}] at(4.75,4.75){};
    \node[label=above:{2}] at(3.75,3.75){};
    \node[label=above:{$N-1$}] at(1.55,1.75){};
    \node[label=above:{$N$}] at(.6,.7){};
    \node[label=above left:{1}] at(-4.75,-5.25){};
    \node[label=above left:{2}] at(-3.75,-4.25){};
    \node[label=above left:{$N$}] at(-1.75,-2.25){};
    \node[label=above left:{$N+1$}] at(-.75,-1.25){};
    \node[label=below:{1}]at(3,0){};
    \node[label=below:{2}]at(1,0){};
    \end{tikzpicture}
         \end{scriptsize}
    \caption{Brane webs for EQ$_{3',3'}$ (left) and EQ$_{3'}$ (right).}
    \label{fig:E3'xE3' brane webs}
\end{figure}
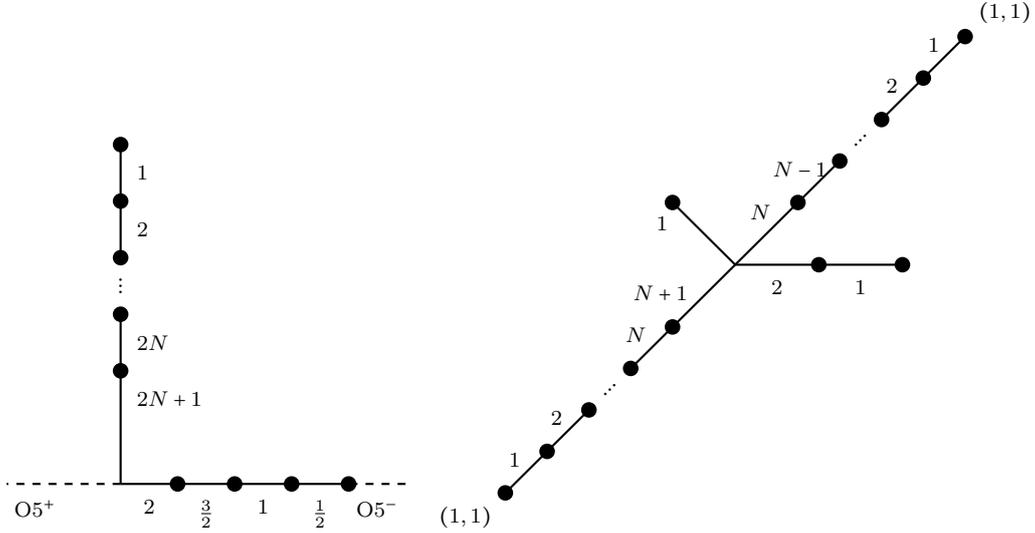
There is another sequence whose first member has an $E_3\times E_3$ symmetry. We will refer to this as the $E_3'\times E_3'$ sequence. In figure \ref{fig:E3'xE3' brane webs} we present the orientifold web that engineers the 5d electric quivers in this sequence, the IR quiver description is given by
   \begin{equation}\label{E3'xE3' electric OSp}
   \EQ_{3',3'}=
    \begin{array}{c}
\begin{tikzpicture}
    \node{$\text{SO(4)}-\text{USp(0)}-\text{SO(4)}-\cdots-\text{USp(0)}-{\underset{\underset{\text{\large$\left[2\textbf{s}+2\textbf{c}\right]$}}{\textstyle\vert}}{\text{SO}(4)}}$};
    
    \draw [thick,decorate,decoration={brace,amplitude=6pt},xshift=0pt,yshift=10pt]
(-3.75,0.5) -- (3.75,0.5)node [black,midway,xshift=0pt,yshift=15pt] {
$2N-1$};
    \end{tikzpicture}
    \end{array}
\end{equation}
It can also be given a unitary web description, once rewritten as a quiver theory with SU(2) nodes:
\begin{equation}
\EQ_{3',3'}=\EQ_{3'}^2=\left(\begin{array}{c}
\begin{tikzpicture}
    \node{$\text{SU(2)}-\text{SU(2)}-\cdots-{\underset{\underset{\text{\large$\left[2\textbf{F}\right]$}}{\textstyle\vert}}{\text{SU}(2)}}$};
    
    \draw [thick,decorate,decoration={brace,amplitude=6pt},xshift=0pt,yshift=10pt]
(-2.25,.5) -- (2.25,.5)node [black,midway,xshift=0pt,yshift=15pt] {
$N$};
    \end{tikzpicture}
    \end{array}\right)^2\;.
\end{equation}
We present the unitary web engineering each EQ$_3'$ factor in figure \ref{fig:E3'xE3' brane webs}. Now we can use the unitary and orientifold webs in figure \ref{fig:E3'xE3' brane webs} to obtain unitary and OSp magnetic quivers for the $E_3'\times E_3'$ sequence, which appear in table \ref{tab:E3'xE3' sequence}. These MQs can be obtained by considering the maximal subdivisions appearing in figure \ref{fig:maximal subdivisions E3'xE3'}. The first subdivision was already discussed in \cite{Akhond:2020vhc}, while the second and third were overlooked. It was noticed in \cite{vanBeest:2020civ} that there should be 2 additional OSp cones to match the analysis on the unitary side. We claim that the missing cones in this case correspond to the two new maximal subdivisions appearing in figure \ref{fig:maximal subdivisions E3'xE3'}. 

\begin{figure}[!htb]
    \centering
    \begin{scriptsize}
    \begin{tikzpicture}[scale=.7]
        \draw[thick,blue](2,-.05)--(3,-.05);
        \draw[thick,brown](4,-.05)--(3,-.05);
        \draw[thick,green](5,-.05)--(4,-.05);
        \draw[thick,yellow](6,-.05)--(5,-.05);
    \draw[thick, dashed,blue](2,-0.05)--(1,-0.05);
    \draw[thick, dashed](7,0)--(6,0);
    \node[label=below:{O5$^+$}] at (1.5,0){};
    \node[label=below:{O5$^-$}] at (6.5,0){};
    \node[label=left:{$2N$}] at (2,2.5){};
    \node[label=left:{$2N$}] at (2,1.5){};
    \node[label=right:{$1$}] at (2,1.5){};
    \draw[thick](1.95,0)--(1.95,2);
    \draw[thick,cyan](2,2)--(2,3);
    \draw[thick,red](2.05,0.05)--(2.05,2);
    \draw[thick,red](2.05,.05)--(3,.05);
    \draw[thick,red](3,.05)--(4,.05);
    \draw[thick,red](5,.05)--(4,.05);
    \node[7brane]at(2,2){};
    \node[7brane]at(2,3){};
    \node at (2,3.5) {$\vdots$};
    \node[label=below:{$\frac{1}{2}$}]at(5.5,0){};
    \node[label=below:{$\frac{1}{2}$}]at(4.5,0){};
    \node[label=above:{$\frac{1}{2}$}]at(4.5,0){};
    \node[label=below:{$\frac{1}{2}$}]at(3.5,0){};
    \node[label=above:{$1$}]at(3.5,0){};
    \node[label=below:{$\frac{1}{2}$}]at(2.5,0){};
    \node[label=above:{$\frac{3}{2}$}]at(2.5,0){};
    \node[7brane]at(5,0){};
    \node[7brane]at(4,0){};
    \node[7brane]at(6,0){};
    \node[7brane]at(3,0){};
    \end{tikzpicture}
    \end{scriptsize}
    \hspace{1 pt}
    \begin{scriptsize}
    \begin{tikzpicture}[scale=.7]
        \draw[thick,blue](2,-.05)--(3,-.05);
        \draw[thick,brown](4,-.05)--(3,-.05);
        \draw[thick,purple](5,-.05)--(4,-.05);
        \draw[thick,orange](6,-.05)--(5,-.05);
    \draw[thick, dashed,blue](2,-0.05)--(1,-0.05);
    \draw[thick, dashed](7,0)--(6,0);
    \node[label=below:{O5$^+$}] at (1.5,0){};
    \node[label=below:{O5$^-$}] at (6.5,0){};
    \node[label=left:{$2N-2$}] at (2,2.5){};
    \node[label=left:{$2N-2$}] at (2,3.5){};
    \node[label=left:{$2N-2$}] at (2,4.5){};
    \node[label=right:{$2$}] at (2,2.5){};
    \node[label=right:{$1$}] at (2,3.5){};
    \node[label=left:{$2N-2$}] at (2,1.5){};
    \node[label=right:{$3$}] at (2,1.5){};
    \draw[thick](1.95,0)--(1.95,2);
    \draw[thick,cyan](1.95,2)--(1.95,3);
    \draw[thick,red](2.05,2)--(2.05,3);
    \draw[thick,red](2.05,4)--(2.05,3);
    \draw[thick,yellow](2,4)--(2,5);
    \draw[thick,green](1.95,4)--(1.95,3);
    \draw[thick,red](2.05,0.05)--(2.05,2);
    \draw[thick,red](2.05,.05)--(3,.05);
    \node[7brane]at(2,2){};
    \node[7brane]at(2,3){};
    \node[7brane]at(2,4){};
    \node[7brane]at(2,5){};
    \node at (2,5.5) {$\vdots$};
    \node[label=below:{$\frac{1}{2}$}]at(5.5,0){};
    \node[label=below:{$1$}]at(4.5,0){};
    \node[label=below:{$\frac{3}{2}$}]at(3.5,0){};
    \node[label=below:{$\frac{1}{2}$}]at(2.5,0){};
    \node[label=above:{$\frac{3}{2}$}]at(2.5,0){};
    \node[7brane]at(5,0){};
    \node[7brane]at(4,0){};
    \node[7brane]at(6,0){};
    \node[7brane]at(3,0){};
    \end{tikzpicture}
    \end{scriptsize}
    \hspace{1 pt}
    \begin{scriptsize}
    \begin{tikzpicture}[scale=.7]
        \draw[thick,blue](2,-.05)--(3,-.05);
        \draw[thick,brown](4,-.05)--(3,-.05);
        \draw[thick,purple](5,-.05)--(4,-.05);
        \draw[thick,orange](6,-.05)--(5,-.05);
    \draw[thick, dashed,blue](2,-0.05)--(1,-0.05);
    \draw[thick, dashed](7,0)--(6,0);
    \node[label=below:{O5$^+$}] at (1.5,0){};
    \node[label=below:{O5$^-$}] at (6.5,0){};
    \node[label=left:{$2N-1$}] at (2,2.5){};
    \node[label=left:{$2N-1$}] at (2,3.5){};
    \node[label=left:{$2N-2$}] at (2,4.5){};
    \node[label=right:{$1$}] at (2,2.5){};
    \node[label=left:{$2N-2$}] at (2,1.5){};
    \node[label=right:{$3$}] at (2,1.5){};
    \draw[thick](1.95,0)--(1.95,2);
    \draw[thick,cyan](1.95,2)--(1.95,3);
    \draw[thick,red](2.05,2)--(2.05,3);
    \draw[thick,yellow](2,4)--(2,5);
    \draw[thick,green](1.95,4)--(1.95,3);
    \draw[thick,red](2.05,0.05)--(2.05,2);
    \draw[thick,red](2.05,.05)--(3,.05);
    \draw[thick,red](4.05,.05)--(3,.05);
    \node[7brane]at(2,2){};
    \node[7brane]at(2,3){};
    \node[7brane]at(2,4){};
    \node[7brane]at(2,5){};
    \node at (2,5.5) {$\vdots$};
    \node[label=below:{$\frac{1}{2}$}]at(5.5,0){};
    \node[label=below:{$1$}]at(4.5,0){};
    \node[label=below:{$\frac{1}{2}$}]at(3.5,0){};
    \node[label=above:{$1$}]at(3.5,0){};
    \node[label=below:{$\frac{1}{2}$}]at(2.5,0){};
    \node[label=above:{$\frac{3}{2}$}]at(2.5,0){};
    \node[7brane]at(5,0){};
    \node[7brane]at(4,0){};
    \node[7brane]at(6,0){};
    \node[7brane]at(3,0){};
    \end{tikzpicture}
    \end{scriptsize}
    \caption{Maximal subdivisions of the Higgs branch of EQ$_{3',3'}$ at the superconformal limit. The subweb coloured in red is frozen and contributes only as flavour nodes to the magnetic quivers in table \ref{tab:E3'xE3' sequence}.}
    \label{fig:maximal subdivisions E3'xE3'}
\end{figure}

\begin{table}[!htb]
    \centering
    \begin{tabular}{|c|c|c|}
    \hline
         MS&OSp & Unitary\\\hline
         $\MQ_{3',3'}^{(\text{I})}$&$\begin{array}{c}
         \begin{scriptsize}
         \begin{tikzpicture}
             \node[label=below:{1}][u](1){};
             \node (dots) [right of=1]{$\cdots$};
             \node[label=below:{$2N$}][u](2N)[right of=dots]{};
             \node[label=below:{$2N$}][sp](sp2N)[right of=2N]{};
             \node[label=above:{$2$}][sof](sof)[above of=sp2N]{};
             \node[label=above:{$1$}][uf](uf)[above of=2N]{};
             \draw(1)--(dots);
             \draw(dots)--(2N);
             \draw(2N)--(sp2N);
             \draw(sp2N)--(sof);
             \draw(2N)--(uf);
         \end{tikzpicture}
         \end{scriptsize}
    \end{array}$ & $\left(\begin{array}{c}
        \begin{scriptsize}
             \begin{tikzpicture}
                 \node[label=above:{1}][u](1){};
                 \node[label=below:{$N$}][u](3)[below left of=1]{};
                 \node[label=below:{$N$}][u](4)[below right of=1]{};
                  \node (dots)[below right of=4]{$\ddots$};
                   \node (dots')[below left of=3]{$\udots$};
                  \node[label=below:{1}][u](u1)[below left of=dots']{};
                  \node[label=below:{1}][u](u11)[below right of=dots]{};
                  \draw(3)--(dots');
                  \draw(4)--(dots);
                  \draw(dots)--(u11);
                  \draw(dots')--(u1);
                 \draw(1)--(4);
                 \draw(1)--(3);
                 \draw(3)--(4);
             \end{tikzpicture}
    \end{scriptsize}\end{array}\right)^2$\\\hline
     $\MQ_{3',3'}^{(\text{II})}$&$\begin{array}{c}\begin{scriptsize}
     \begin{tikzpicture}
     \node[label=below:{1}][u](1){};
     \node (dots)[right of=1]{$\cdots$};
     \node[label=below:{$2N-2$}][u](2N-2)[right of=dots]{};
     \node[label=below:{$2N-2$}][u](2N-2')[right of=2N-2]{};
     \node[label=below:{$2N-2$}][u](2N-2'')[right of=2N-2']{};
     \node[label=below:{$2N-2$}][sp](sp2N-2)[right of=2N-2'']{};
     \node[label=above:{1}][uf](uf)[above of=2N-2]{};
     \node[label=above:{2}][sof](sof)[above of=sp2N-2]{};
     \draw(1)--(dots);
     \draw(dots)--(2N-2);
     \draw(2N-2)--(2N-2');
     \draw(2N-2')--(2N-2'');
     \draw(2N-2'')--(sp2N-2);
     \draw(sp2N-2)--(sof);
     \draw(2N-2)--(uf);
     \end{tikzpicture}
     \end{scriptsize}
     \end{array}\begin{array}{c}
     \begin{scriptsize}
     \begin{tikzpicture}
     \node[label=right:{2}][so](so2){};
     \node[label=right:{2}][sp](sp2)[above of=so2]{};
     \node[label=right:{4}][sof](sof)[above of=sp2]{};
     \draw(so2)--(sp2);
     \draw(sp2)--(sof);
     \end{tikzpicture}
     \end{scriptsize}
     \end{array}$ & $\left(\begin{array}{c}
         \begin{scriptsize}
         \begin{tikzpicture}
         \node[label=above:{1}][u](u1){};
         \node[label=left:{$N-1$}][u](N-1)[above left of=u1]{};
         \node[label=right:{$N-1$}][u](N-1')[above right of=u1]{};
         \node[label=above:{$N-1$}][u](N-1'')[above of=N-1]{};
         \node[label=above:{$N-1$}][u](N-1''')[above of=N-1']{};
         \node[label=below:{1}][u](u11)[below of=u1]{};
         \node (dots)[below left of=N-1]{$\udots$};
         \node[label=below:{1}][u](1)[below left of=dots]{};
         \node (dots')[below right of=N-1']{$\ddots$};
         \node[label=below:{1}][u](1')[below right of=dots']{};
         \draw(1)--(dots);
         \draw(dots)--(N-1);
         \draw(N-1''')--(N-1');
         \draw(N-1)--(N-1'');
         \draw(N-1')--(u1);
         \draw(N-1)--(u1);
         \draw(N-1'')--(N-1''');
         \draw(N-1')--(dots');
         \draw(dots')--(1');
         \draw[double distance=2pt](u1)--(u11);
         \end{tikzpicture}
         \end{scriptsize}
    \end{array}\right)^2$ \\\hline
    $\MQ_{3',3'}^{(\text{III})}$& $\begin{array}{c}
         \begin{scriptsize}
         \begin{tikzpicture}
         \node[label=below:{1}][u](1){};
         \node (dots) [right of=1]{$\cdots$};
         \node[label=below:{$2N-2$}][u](2N-2)[right of=dots]{};
         \node[label=below:{$2N-1$}][u](2N-1)[right of=2N-2]{};
         \node[label=below:{$2N-1$}][u](2N-1')[right of=2N-1]{};
         \node[label=below:{$2N-2$}][sp](sp2N-2)[right of=2N-1']{};
         \node[label=above:{1}][uf](uf)[above of=2N-1]{};
         \node[label=above:{1}][uf](uf')[above of=2N-1']{};
         \draw(1)--(dots);
         \draw(dots)--(2N-2);
         \draw(2N-2)--(2N-1);
         \draw(2N-1)--(2N-1');
         \draw(2N-1')--(sp2N-2);
         \draw(2N-1)--(uf);
         \draw(2N-1')--(uf');
         \end{tikzpicture}
         \end{scriptsize}
    \end{array} \begin{array}{c}
         \begin{scriptsize}
         \begin{tikzpicture}
         \node[label=below:{2}][so](so2){};
         \node[label=above:{2}][spf](spf)[above of=so2]{};
         \draw(so2)--(spf);
         \end{tikzpicture}
         \end{scriptsize}
    \end{array}$ &$\begin{array}{c}
        \begin{scriptsize}
             \begin{tikzpicture}
                 \node[label=above:{1}][u](1){};
                 \node[label=left:{$N$}][u](3)[below left of=1]{};
                 \node[label=right:{$N$}][u](4)[below right of=1]{};
                  \node (dots)[below of=4]{$\vdots$};
                   \node (dots')[below of=3]{$\vdots$};
                  \node[label=below:{1}][u](u1)[below of=dots']{};
                  \node[label=below:{1}][u](u11)[below of=dots]{};
                  \draw(3)--(dots');
                  \draw(4)--(dots);
                  \draw(dots)--(u11);
                  \draw(dots')--(u1);
                 \draw(1)--(4);
                 \draw(1)--(3);
                 \draw(3)--(4);
             \end{tikzpicture}
    \end{scriptsize}\end{array} \begin{array}{c}
         \begin{scriptsize}
         \begin{tikzpicture}
         \node[label=above:{1}][u](u1){};
         \node[label=left:{$N-1$}][u](N-1)[above left of=u1]{};
         \node[label=right:{$N-1$}][u](N-1')[above right of=u1]{};
         \node[label=above:{$N-1$}][u](N-1'')[above of=N-1]{};
         \node[label=above:{$N-1$}][u](N-1''')[above of=N-1']{};
         \node[label=below:{1}][u](u11)[below of=u1]{};
         \node (dots)[below of=N-1]{$\vdots$};
         \node[label=below:{1}][u](1)[below of=dots]{};
         \node (dots')[below of=N-1']{$\vdots$};
         \node[label=below:{1}][u](1')[below of=dots']{};
         \draw(1)--(dots);
         \draw(dots)--(N-1);
         \draw(N-1''')--(N-1');
         \draw(N-1)--(N-1'');
         \draw(N-1')--(u1);
         \draw(N-1)--(u1);
         \draw(N-1'')--(N-1''');
         \draw(N-1')--(dots');
         \draw(dots')--(1');
         \draw[double distance=2pt](u1)--(u11);
         \end{tikzpicture}
         \end{scriptsize}
    \end{array} $ \\\hline
    \end{tabular}
    \caption{Magnetic quivers for the $E_3'\times E_3'$ sequence}
    \label{tab:E3'xE3' sequence}
\end{table}
Notice that for $N=1$, the relation between the unitary and OSp quivers in the first row of table \ref{tab:E3'xE3' sequence} was already suggested in \cite{Beratto:2020wmn}, our result generalises this to higher $N$. The HWG for the unitary quiver is known and appears in \cite{Bourget:2019rtl}. Given the correspondence between the unitary and OSp magnetic quivers in table \ref{tab:E3'xE3' sequence}, we can use the results for the HWGs of the unitary quivers to obtain the HWGs for the OSp quivers. In order to do this, let us point out the following useful fact; one of the unitary quivers appearing in the second and third row of table \ref{tab:MQ34}, is itself a product of two unitary quivers:
\begin{equation}
     \begin{array}{c}
         \begin{scriptsize}
         \begin{tikzpicture}
         \node[label=above:{1}][u](u1){};
         \node[label=left:{$N-1$}][u](N-1)[above left of=u1]{};
         \node[label=right:{$N-1$}][u](N-1')[above right of=u1]{};
         \node[label=above:{$N-1$}][u](N-1'')[above of=N-1]{};
         \node[label=above:{$N-1$}][u](N-1''')[above of=N-1']{};
         \node[label=below:{1}][u](u11)[below of=u1]{};
         \node (dots)[below left of=N-1]{$\udots$};
         \node[label=below:{1}][u](1)[below left of=dots]{};
         \node (dots')[below right of=N-1']{$\ddots$};
         \node[label=below:{1}][u](1')[below right of=dots']{};
         \draw(1)--(dots);
         \draw(dots)--(N-1);
         \draw(N-1''')--(N-1');
         \draw(N-1)--(N-1'');
         \draw(N-1')--(u1);
         \draw(N-1)--(u1);
         \draw(N-1'')--(N-1''');
         \draw(N-1')--(dots');
         \draw(dots')--(1');
         \draw[double distance=2pt](u1)--(u11);
         \end{tikzpicture}
         \end{scriptsize}
    \end{array}=\begin{array}{c}
         \begin{scriptsize}
         \begin{tikzpicture}
         \node[label=above:{1}][u](u1){};
         \node[label=left:{$N-1$}][u](N-1)[above left of=u1]{};
         \node[label=right:{$N-1$}][u](N-1')[above right of=u1]{};
         \node[label=above:{$N-1$}][u](N-1'')[above of=N-1]{};
         \node[label=above:{$N-1$}][u](N-1''')[above of=N-1']{};
         \node (dots)[below left of=N-1]{$\udots$};
         \node[label=below:{1}][u](1)[below left of=dots]{};
         \node (dots')[below right of=N-1']{$\ddots$};
         \node[label=below:{1}][u](1')[below right of=dots']{};
         \draw(1)--(dots);
         \draw(dots)--(N-1);
         \draw(N-1''')--(N-1');
         \draw(N-1)--(N-1'');
         \draw(N-1')--(u1);
         \draw(N-1)--(u1);
         \draw(N-1'')--(N-1''');
         \draw(N-1')--(dots');
         \draw(dots')--(1');
         \end{tikzpicture}
         \end{scriptsize}
    \end{array}\times \begin{array}{c}
         \begin{scriptsize}
         \begin{tikzpicture}
         \node[label=below:{1}][u](1){};
         \node[label=above:{1}][u](11)[above of=1]{};
         \draw[double distance=2pt](1)--(11);
         \end{tikzpicture}
         \end{scriptsize}
    \end{array}\;,
\end{equation}
where the right hand side of the above is obtained after ungauging the overall decoupled U$(1)$ in the original quiver. The first quiver in the right hand side of the above is a height 2 nilpotent orbit, whose HWG is presented in \cite{Bourget:2019rtl}, while the second quiver is just $\mathcal{N}=4$ QED with 2 electrons. Turning to the second and third row, we again see that for $N=1$ the correspondence between the unitary and OSp quivers is obvious, which one may view as a further robustness of our proposal.

Now we have all the necessary ingredients to write down HWGs for the OSp quivers
\begin{equation}
\begin{split}
    \HWG_{3',3'}^{(\text{I})}&=\PE\left[\sum_{k=1}^N\left(\mu_k\mu_{2N+1-k}+\nu_k\nu_{2N+1-k}\right)t^{2k}\right]\;,\\
    \HWG_{3',3'}^{(\text{II})}&=\PE\left[\sum_{k=1}^{N-1}\left(\mu_k\mu_{2N+1-k}+\nu_k\nu_{2N+1-k}\right)t^{2k}+\left(\rho^2+\lambda^2\right)t^2\right]\;,\\
    \HWG_{3',3'}^{(\text{III})}&=\PE\left[\sum_{k=1}^N\mu_k\mu_{2N+1-k}t^{2k}+\sum_{j=1}^{N-1}\nu_j\nu_{2N+1-j}t^{2j}+\rho^2t^2\right]\;.
    \end{split}
\end{equation}
This proposal can be checked by a direct computation of the unrefined Hilbert series of the OSp quiver. For low values of $N$, results are given in table \ref{TableHSOSp}.

\subsection{The \texorpdfstring{$E_3'\times E_4$}{TEXT} sequence}
The $E_3'\times E_4$ sequence is the magnetic quiver for the fixed point limit of the 5d IR electric quiver
   \begin{equation}\label{E3xE4 electric OSp}
   \EQ_{3',4}=
    \begin{array}{c}
\begin{tikzpicture}
    \node{${\underset{\underset{\text{\large$\left[1\textbf{s}\right]$}}{\textstyle\vert}}{\text{SO}(4)}}-\text{USp(0)}-\text{SO(4)}-\cdots-\text{USp(0)}-{\underset{\underset{\text{\large$\left[2\textbf{s}+2\textbf{c}\right]$}}{\textstyle\vert}}{\text{SO}(4)}}$};
    
    \draw [thick,decorate,decoration={brace,amplitude=6pt},xshift=0pt,yshift=10pt]
(-3.75,0.5) -- (3.75,0.5)node [black,midway,xshift=0pt,yshift=15pt] {
$2N-1$};
    \end{tikzpicture}
    \end{array}\;.
\end{equation}
It can be engineered by the orientifold web diagram presented in figure \ref{fig:EQ34 orientifold web}.
\begin{figure}[!htb]
    \centering
           \begin{scriptsize}
         \begin{tikzpicture}
         
    \draw[thick](-3,0)--(2,0);
    \draw[thick, dashed](-4,0)--(-3,0);
    \draw[thick, dashed](2,0)--(4,0);
    \draw[thick](2,4)--(2,6);
    \node[label=below:{O5$^-$}] at (-3.5,0){};
    \node[label=below:{O5$^-$}] at (3.5,0){};
    \node[label=right:{$1$}] at (2,5.5){};
    \node[label=right:{$2$}] at (2,4.5){};
    \node[label=right:{$2N$}] at (2,2.5){};
    \node[label=right:{$2N+1$}] at (2,1.5){};
    \draw[thick](2,0)--(4,1);
    \draw[thick](2,0)--(2,3);
    \node[label=above right:{(2,1)}][7brane]at(4,1){};
    \node[7brane]at(2,2){};
    \node[7brane]at(2,3){};
    \node at (2,3.5) {$\vdots$};
    \node[7brane]at(2,4){};
    \node[7brane]at(2,5){};
    \node[7brane]at(2,6){};
    \node[label=below:{$\frac{1}{2}$}]at(-2.5,0){};
    \node[label=below:{$1$}]at(-1.5,0){};
    \node[label=below:{$\frac{3}{2}$}]at(-.5,0){};
    \node[label=below:{$2$}]at(.5,0){};
    \node[7brane]at(-1,0){};
    \node[7brane]at(-2,0){};
    \node[7brane]at(-3,0){};
    \node[7brane]at(0,0){};
    
    \end{tikzpicture}
    \end{scriptsize}
    \caption{Orientifold web for EQ$_{3',4}$.}
    \label{fig:EQ34 orientifold web}
\end{figure}
Alternatively we may reformulate the electric theory EQ$_{3,4}$ as a product of two unitary electric quivers
\begin{equation}
\EQ_{3',4}=\EQ_{3'}\times\EQ_4=\begin{array}{c}
\begin{tikzpicture}
    \node{$\text{SU}(2)-\text{SU(2)}-\cdots-{\underset{\underset{\text{\large$\left[2\textbf{F}\right]$}}{\textstyle\vert}}{\text{SU}(2)}}$};
    
    \draw [thick,decorate,decoration={brace,amplitude=6pt},xshift=0pt,yshift=10pt]
(-2.25,.5) -- (2.25,.5)node [black,midway,xshift=0pt,yshift=15pt] {
$N$};
    \end{tikzpicture}
    \end{array}\times \begin{array}{c}
\begin{tikzpicture}
    \node{${\underset{\underset{\text{\large$\left[1\textbf{F}\right]$}}{\textstyle\vert}}{\text{SU}(2)}}-\text{SU(2)}-\cdots-{\underset{\underset{\text{\large$\left[2\textbf{F}\right]$}}{\textstyle\vert}}{\text{SU}(2)}}$};
    
    \draw [thick,decorate,decoration={brace,amplitude=6pt},xshift=0pt,yshift=10pt]
(-2.25,.5) -- (2.25,.5)node [black,midway,xshift=0pt,yshift=15pt] {
$N$};
    \end{tikzpicture}
    \end{array}\;,
\end{equation}
where $\EQ_{3'}$ is engineered by the unitary web in figure \ref{fig:E3'xE3' brane webs}, while the unitary web engineering $\EQ_4$ is the one in figure \ref{fig:E4xE4 brane webs}. Given these webs, the magnetic quivers can be extracted using the rules in \cite{Akhond:2020vhc}. The Higgs branch of $\EQ_{3',4}$ at the fixed point is the union of two cones, whose magnetic quivers are given in table \ref{tab:MQ34}. The HWG that we propose for the OSp quivers are
\begin{equation}\begin{split}
    \HWG_{3',4}^{(\text{I})}&=\PE\left[\sum_{i=1}^N\mu_i\mu_{2N+1-i}t^{2i}+(\nu^2+1)t^2+\nu(\mu_Nq+\mu_{N+1}q^{-1})t^{N+1}-\nu^2\mu_{N}\mu_{N+1}t^{2N+2}\right]\\
    &\times \PE\left[\sum_{k=1}^N\rho_k\rho_{2N+1-k}t^{2k}\right]\\
    \HWG_{3',4}^{(\text{II})}&=\PE\left[\sum_{i=1}^N\mu_i\mu_{2N+1-i}t^{2i}+(\nu^2+1)t^2+\nu(\mu_Nq+\mu_{N+1}q^{-1})t^{N+1}-\nu^2\mu_{N}\mu_{N+1}t^{2N+2}\right]\\
    &\times\PE\left[\sum_{k=1}^N\lambda_k\lambda_{2N+1-k}t^{2k}\right]\PE\left[\eta^2t^2\right]
\end{split}\;
\end{equation}
As a check, we have also computed the unrefined Hilbert series for low values of $N$ which are given in table \ref{TableHSOSp}.

\begin{table}\centering
    \begin{tabular}{|c|c|c|}\hline MS&OSp&Unitary\\\hline
        $\MQ_{3',4}^{(\text{I})}$ & $\begin{array}{c}
         \begin{scriptsize}
         \begin{tikzpicture}
             \node[label=right:{1}][u](1){};
             \node (dots)[below of=1]{$\vdots$};
             \node[label=right:{$2N$}][u](2N-2)[below of=dots]{};
             \node[label=below:{$2N$}][sp](sp2N-2)[below of=2N-2]{};
             \node[label=below:{1}][u](u1)[left of=sp2N-2]{};
             \node[label=below:{2}][so](so2)[left of=u1]{};
             \draw(1)--(dots);
             \draw(dots)--(2N-2);
             \draw(2N-2)--(sp2N-2);
             \draw(sp2N-2)--(u1);
             \draw(u1)--(so2);
             \draw(u1)--(2N-2);
         \end{tikzpicture}
         \end{scriptsize}
    \end{array}$ &$\begin{array}{c}
        \begin{scriptsize}
             \begin{tikzpicture}
                 \node[label=above:{1}][u](1){};
                 \node[label=left:{1}][u](2)[below left of=1]{};
                 \node[label=right:{1}][u](5)[below right of=1]{};
                 \node[label=left:{$N$}][u](3)[below of=2]{};
                 \node[label=right:{$N$}][u](4)[below of=5]{};
                  \node (dots)[below of=4]{$\vdots$};
                   \node (dots')[below of=3]{$\vdots$};
                  \node[label=below:{1}][u](u1)[below of=dots']{};
                  \node[label=below:{1}][u](u11)[below of=dots]{};
                  \draw(3)--(dots');
                  \draw(4)--(dots);
                  \draw(dots)--(u11);
                  \draw(dots')--(u1);
                 \draw(1)--(2);
                 \draw(2)--(3);
                 \draw(3)--(4);
                 \draw(4)--(5);
                 \draw(5)--(1);
             \end{tikzpicture}
    \end{scriptsize}    \end{array}$
    $\begin{array}{c}
        \begin{scriptsize}
             \begin{tikzpicture}
                 \node[label=above:{1}][u](1){};
                 \node[label=below:{$N$}][u](3)[below left of=1]{};
                 \node[label=below:{$N$}][u](4)[below right of=1]{};
                  \node (dots)[below right of=4]{$\ddots$};
                   \node (dots')[below left of=3]{$\udots$};
                  \node[label=below:{1}][u](u1)[below left of=dots']{};
                  \node[label=below:{1}][u](u11)[below right of=dots]{};
                  \draw(3)--(dots');
                  \draw(4)--(dots);
                  \draw(dots)--(u11);
                  \draw(dots')--(u1);
                 \draw(1)--(4);
                 \draw(1)--(3);
                 \draw(3)--(4);
             \end{tikzpicture}
    \end{scriptsize}    \end{array}$\\\hline$\MQ_{3',4}^{(\text{II})}$
         &$\begin{array}{c}
         \begin{scriptsize}
         \begin{tikzpicture}
             \node[label=right:{1}][u](1){};
             \node (dots)[below of=1]{$\vdots$};
             \node[label=right:{$2N-1$}][u](2N-1)[below of=dots]{};
             \node[label=right:{$2N-1$}][u](2N-1')[below of=2N-1]{};
             \node[label=right:{$2N-2$}][sp](sp2N-2)[below of=2N-1']{};
             \node[label=below:{$1$}][u](1')[left of=2N-1]{};
             \node[label=below:{$2$}][sp](sp2)[left of=1']{};
             \node[label=below:{$2$}][so](so2)[left of=sp2]{};
                          \draw(1)--(dots);
             \draw(dots)--(2N-1);
             \draw(2N-1)--(2N-1');
             \draw(2N-1')--(sp2N-2);
             \draw[dashed](2N-1')--(1');
             \draw (1')--(2N-1);
             \draw[double distance=2pt](1')--(sp2);
             \draw(sp2)--(so2);
         \end{tikzpicture}
         \end{scriptsize}
    \end{array} $&$\begin{array}{c}
        \begin{scriptsize}
             \begin{tikzpicture}
                 \node[label=above:{1}][u](1){};
                 \node[label=left:{1}][u](2)[below left of=1]{};
                 \node[label=right:{1}][u](5)[below right of=1]{};
                 \node[label=left:{$N$}][u](3)[below of=2]{};
                 \node[label=right:{$N$}][u](4)[below of=5]{};
                  \node (dots)[below of=4]{$\vdots$};
                   \node (dots')[below of=3]{$\vdots$};
                  \node[label=below:{1}][u](u1)[below of=dots']{};
                  \node[label=below:{1}][u](u11)[below of=dots]{};
                  \draw(3)--(dots');
                  \draw(4)--(dots);
                  \draw(dots)--(u11);
                  \draw(dots')--(u1);
                 \draw(1)--(2);
                 \draw(2)--(3);
                 \draw(3)--(4);
                 \draw(4)--(5);
                 \draw(5)--(1);
             \end{tikzpicture}
    \end{scriptsize}    \end{array}
    \begin{array}{c}
         \begin{scriptsize}
         \begin{tikzpicture}
         \node[label=above:{1}][u](u1){};
         \node[label=left:{$N-1$}][u](N-1)[above left of=u1]{};
         \node[label=right:{$N-1$}][u](N-1')[above right of=u1]{};
         \node[label=above:{$N-1$}][u](N-1'')[above of=N-1]{};
         \node[label=above:{$N-1$}][u](N-1''')[above of=N-1']{};
         \node[label=below:{1}][u](u11)[below of=u1]{};
         \node (dots)[below left of=N-1]{$\udots$};
         \node[label=below:{1}][u](1)[below left of=dots]{};
         \node (dots')[below right of=N-1']{$\ddots$};
         \node[label=below:{1}][u](1')[below right of=dots']{};
         \draw(1)--(dots);
         \draw(dots)--(N-1);
         \draw(N-1''')--(N-1');
         \draw(N-1)--(N-1'');
         \draw(N-1')--(u1);
         \draw(N-1)--(u1);
         \draw(N-1'')--(N-1''');
         \draw(N-1')--(dots');
         \draw(dots')--(1');
         \draw[double distance=2pt](u1)--(u11);
         \end{tikzpicture}
         \end{scriptsize}
    \end{array}$\\\hline
    \end{tabular}
        \caption{Magnetic quivers for the two cones of the Higgs branch of $\EQ_{3',4}$.}
    \label{tab:MQ34}
\end{table}

\subsection{The \texorpdfstring{$E_4\times E_4$}{TEXT} sequence}
\begin{figure}[!htb]
    \centering
   \begin{scriptsize}
         \begin{tikzpicture}[scale=.75]
    \draw[thick](0,0)--(6,0);
    \draw[thick, dashed](-1,0)--(0,0);
    \draw[thick, dashed](7,0)--(6,0);
    \draw[thick](2,4)--(2,6);
    \node[label=below:{O5$^-$}] at (-.5,0){};
    \node[label=below:{O5$^-$}] at (6.5,0){};
    \node[label=right:{$1$}] at (2,5.5){};
    \node[label=right:{$2$}] at (2,4.5){};
    \node[label=right:{$2N$}] at (2,2.5){};
    \node[label=right:{$2N+1$}] at (2,1.5){};
    \draw[thick](2,0)--(1,1);
    \draw[thick](2,0)--(2,3);
    \node[label=above left:{(1,-1)}][7brane]at(1,1){};
    \node[7brane]at(2,2){};
    \node[7brane]at(2,3){};
    \node at (2,3.5) {$\vdots$};
    \node[7brane]at(2,4){};
    \node[7brane]at(2,5){};
    \node[7brane]at(2,6){};
    \node[label=below:{$\frac{1}{2}$}]at(5.5,0){};
    \node[label=below:{$1$}]at(4.5,0){};
    \node[label=below:{$\frac{3}{2}$}]at(3.5,0){};
    \node[label=below:{$2$}]at(2.5,0){};
    \node[label=below:{$\frac{1}{2}$}]at(.5,0){};
    \node[label=below:{$1$}]at(1.5,0){};
    \node[7brane]at(1,0){};
    \node[7brane]at(0,0){};
    \node[7brane]at(5,0){};
    \node[7brane]at(4,0){};
    \node[7brane]at(6,0){};
    \node[7brane]at(3,0){};
    
    \end{tikzpicture}
    \end{scriptsize}
    \begin{scriptsize}
    \begin{tikzpicture}[scale=.55]
    \draw[thick](-2.5,-2.5)--(2.5,2.5);
    \draw[thick](-3.5,-3.5)--(-5.5,-5.5);
    \draw[thick](3.5,3.5)--(5.5,5.5);
    \node[label=above right:{$(1,1)$}][7brane]at(5.5,5.5){};
    \node[7brane]at(4.5,4.5){};
    \node[7brane]at(3.5,3.5){};
    \node[7brane]at(2.5,2.5){};
    \node[7brane]at(1.5,1.5){};
        \node[label=below left:{$(1,1)$}][7brane]at(-5.5,-5.5){};
    \node[7brane]at(-4.5,-4.5){};
    \node[7brane]at(-3.5,-3.5){};
    \node[7brane]at(-2.5,-2.5){};
    \node[7brane]at(-1.5,-1.5){};
    \draw[thick](-4,0)--(3,0);
    \draw[thick](0,-2)--(0,0);
    \node[7brane]at(-2,0){};
    \node[7brane]at(-4,0){};
    \node[7brane]at(3,0){};
    \node[7brane]at(0,-2){};
    \node at (3,3){$\udots$};
    \node at (-3,-3){$\udots$};
    \node[label=above:{1}] at(4.75,4.75){};
    \node[label=above:{2}] at(3.75,3.75){};
    \node[label=above:{$N$}] at(1.75,1.75){};
    \node[label=above:{$N+1$}] at(.3,.7){};
    \node[label=above left:{1}] at(-4.75,-5.25){};
    \node[label=above left:{2}] at(-3.75,-4.25){};
    \node[label=above left:{$N-1$}] at(-1.75,-2.25){};
    \node[label=above left:{$N$}] at(-.75,-1.25){};
    \node[label=above:{1}]at(-3,0){};
    \node[label=above:{2}]at(-1,0){};
    \node[label=below:{1}]at(2,0){};
    \node[label=right:{1}]at(0,-1){};
    \end{tikzpicture}
         \end{scriptsize}
    \caption{Brane webs for EQ$_{4,4}$ (left) and EQ$_4$ (right).}
    \label{fig:E4xE4 brane webs}
\end{figure}
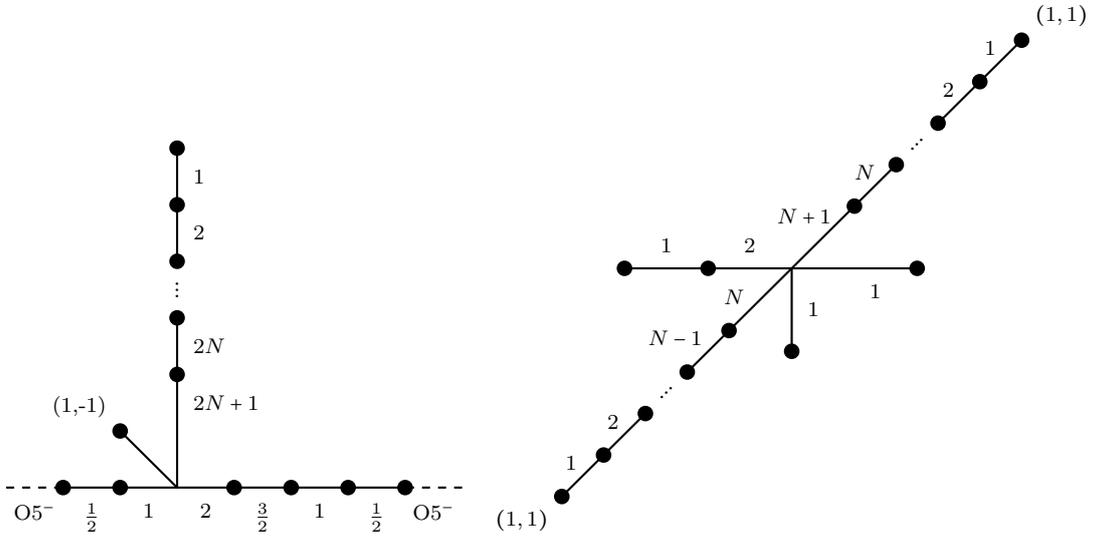
The $E_4\times E_4$ sequence are the magnetic quivers for the fixed point limit of the 5d electric quiver
    \begin{equation}\EQ_{4,4}=
\begin{array}{c}
\begin{tikzpicture}
    \node{$[1\text{s}+1\text{c}]-\text{SO(4)}-\text{USp(0)}-\text{SO(4)}-\cdots-\text{USp(0)}-\text{SO(4)}-[2\text{s}+2\text{c}]$};
    
    \draw [thick,decorate,decoration={brace,amplitude=6pt},xshift=0pt,yshift=10pt]
(-3.75,0) -- (3.75,0)node [black,midway,xshift=0pt,yshift=15pt] {
$2N-1$};
    \end{tikzpicture}
    \end{array}\;.
\end{equation}
We present the orientifold web that engineers this theory in figure \ref{fig:E4xE4 brane webs}. Alternatively we can write EQ$_{4,4}$ as the product of two unitary electric quivers
\begin{equation}
\EQ_{4,4}=\EQ_4^2=\left(\begin{array}{c}
\begin{tikzpicture}
    \node{${\underset{\underset{\text{\large$\left[1\textbf{F}\right]$}}{\textstyle\vert}}{\text{SU}(2)}}-\text{SU(2)}-\cdots-{\underset{\underset{\text{\large$\left[2\textbf{F}\right]$}}{\textstyle\vert}}{\text{SU}(2)}}$};
    
    \draw [thick,decorate,decoration={brace,amplitude=6pt},xshift=0pt,yshift=10pt]
(-2.25,.5) -- (2.25,.5)node [black,midway,xshift=0pt,yshift=15pt] {
$N$};
    \end{tikzpicture}
    \end{array}\right)^2\;,
\end{equation}
where each copy of EQ$_4$ can be engineered by the unitary web depicted in figure \ref{fig:E4xE4 brane webs}. Given the brane webs in figure \ref{fig:E4xE4 brane webs}, one can derive an OSp and a pair of unitary magnetic quivers whose Coulomb branches are expected to describe the same unique 5d Higgs branch, leading us to conjecture that
\begin{equation}
        \MQ_{4,4}=\begin{array}{c}
             \begin{scriptsize}
             \begin{tikzpicture}
             \node[label=below:{2}][so](so2){};
             \node[label=below:{2}][sp](sp2)[right of=so2]{};
             \node[label=below:{2}][so](so2')[below right of=sp2]{};
             \node[label=above:{1}][u](u11)[above right of=sp2]{};
             \node[label=below:{$2N$}][sp](sp2'')[right of=so2']{};
             \node[label=above:{$2N$}][u](u2)[right of=u11]{};
             \node (dots)[right of=u2]{$\cdots$};
             \node[label=above:{1}][u](u1)[right of=dots]{};
             \draw(so2)--(sp2);
             \draw(sp2)--(so2');
             \draw(so2')--(sp2'');
             \draw(sp2)--(u11);
             \draw(u11)--(u2);
             \draw(dots)--(u2);
             \draw(dots)--(u1);
             \draw(u2)--(sp2'');
             \end{tikzpicture}
             \end{scriptsize}
        \end{array}=\left(\begin{array}{c}
        \begin{scriptsize}
             \begin{tikzpicture}
                 \node[label=above:{1}][u](1){};
                 \node[label=left:{1}][u](2)[below left of=1]{};
                 \node[label=right:{1}][u](5)[below right of=1]{};
                 \node[label=below:{$N$}][u](3)[below of=2]{};
                 \node[label=below:{$N$}][u](4)[below of=5]{};
                  \node (dots)[right of=4]{$\cdots$};
                   \node (dots')[left of=3]{$\cdots$};
                  \node[label=below:{1}][u](u1)[left of=dots']{};
                  \node[label=below:{1}][u](u11)[right of=dots]{};
                  \draw(3)--(dots');
                  \draw(4)--(dots);
                  \draw(dots)--(u11);
                  \draw(dots')--(u1);
                 \draw(1)--(2);
                 \draw(2)--(3);
                 \draw(3)--(4);
                 \draw(4)--(5);
                 \draw(5)--(1);
             \end{tikzpicture}
    \end{scriptsize}    \end{array}\right)^2
    \end{equation}
The HWG for the unitary quiver appearing here was evaluated in \cite{Ferlito:2017xdq}. We can use this result to obtain an exact HWG for the OSp quiver by simply taking its square. Our claim is     
\begin{align}
\HWG_{4,4}=\PE\left[\sum_{i=1}^N\mu_i\mu_{2N+1-i}t^{2i}+(\nu^2+1)t^2+\nu(\mu_Nq+\mu_{N+1}q^{-1})t^{N+1}-\nu^2\mu_{N}\mu_{N+1}t^{2N+2}\right] \nonumber \\
\times \PE\left[\sum_{i=1}^N\eta_i\eta_{2N+1-i}t^{2i}+(\lambda^2+1)t^2+\lambda(\eta_Nr+\eta_{N+1}r^{-1})t^{N+1}-\lambda^2\eta_{N}\eta_{N+1}t^{2N+2}\right] ~.
\end{align}
The explicit unrefined Hilbert series computation for $N=1$ and $N=2$ for the OSp quiver is given in table \ref{TableHSOSp}. 
\subsection{The \texorpdfstring{$E_5\times E_5$}{TEXT} sequence}
The $E_5\times E_5$ sequence is obtained by taking the fixed point limit of the electric quiver given by
\begin{equation}\EQ_{5,5}=
\begin{array}{c}
\begin{tikzpicture}
    \node{$[2\textbf{s}+2\textbf{c}]-\text{SO(4)}-\text{USp(0)}-\text{SO(4)}-\cdots-\text{USp(0)}-\text{SO(4)}-[2\textbf{s}+2\textbf{c}]$};
    
    \draw [thick,decorate,decoration={brace,amplitude=6pt},xshift=0pt,yshift=10pt]
(-3.75,0) -- (3.75,0)node [black,midway,xshift=0pt,yshift=15pt] {
$2N-1$};
    \end{tikzpicture}
    \end{array}\;,
\end{equation}
which can be engineered using the following orientifold web:
\begin{equation}\label{EQ55 orientifold web}
    \begin{array}{c}
         \begin{scriptsize}
         \begin{tikzpicture}[scale=.75]
             \draw[dashed](0,0)--(1,0);
             \draw[thick](1,0)--(9,0);
             \node[7brane]at(1,0){};
             \node[7brane]at(2,0){};
             \node[7brane]at(3,0){};
             \node[7brane]at(4,0){};
             \node[7brane]at(7,0){};
             \node[7brane]at(8,0){};
             \draw[thick,dashed](9,0)--(10,0);
             \node[7brane]at(9,0){};
             \node[7brane]at(6,0){};
             \draw[thick](5,0)--(5,2.5);
             \node[7brane]at(5,1.5){};
             \node[7brane]at(5,2.5){};
             \node at(5,3){$\vdots$};
             \draw[thick](5,3.5)--(5,5.5);
             \node[7brane]at(5,3.5){};
             \node[7brane]at(5,4.5){};
             \node[7brane]at(5,5.5){};
             \node[label=below:{O5$^-$}] at(.5,0){};
             \node[label=below:{O5$^-$}] at(9.5,0){};
             \node[label=below:{$\frac{1}{2}$}] at(1.5,0){};
             \node[label=below:{1}] at(2.5,0){};
             \node[label=below:{$\frac{3}{2}$}] at(3.5,0){};
             \node[label=below:{2}] at(4.5,0){};
             \node[label=below:{$\frac{1}{2}$}] at(8.5,0){};
             \node[label=below:{1}] at(7.5,0){};
             \node[label=below:{$\frac{3}{2}$}] at(6.5,0){};
             \node[label=below:{2}] at(5.5,0){};
             \node[label=right:{1}]at(5,5){};
             \node[label=right:{2}]at(5,4){};
             \node[label=right:{$2N+2$}]at(5,1){};
             \node[label=right:{$2N+1$}]at(5,2){};
         \end{tikzpicture}
         \end{scriptsize}
    \end{array}\;.
\end{equation}
Alternatively, we can rewrite EQ$_{5,5}$ as the product of two unitary electric quivers
\begin{equation}
\EQ_{5,5}=\EQ_5^2=\left(\begin{array}{c}
\begin{tikzpicture}
    \node{${\underset{\underset{\text{\large$\left[2\textbf{F}\right]$}}{\textstyle\vert}}{\text{SU}(2)}}-\text{SU(2)}-\cdots-{\underset{\underset{\text{\large$\left[2\textbf{F}\right]$}}{\textstyle\vert}}{\text{SU}(2)}}$};
    
    \draw [thick,decorate,decoration={brace,amplitude=6pt},xshift=0pt,yshift=10pt]
(-2.25,.5) -- (2.25,.5)node [black,midway,xshift=0pt,yshift=15pt] {
$N$};
    \end{tikzpicture}
    \end{array}\right)^2\;,
\end{equation}
each of which is engineered by taking one copy of the following unitary brane web
\begin{equation}\label{EQ5 unitary web}
\begin{array}{c}
        \begin{scriptsize}
    \begin{tikzpicture}[scale=.55]
    \draw[thick](-2.5,-2.5)--(2.5,2.5);
    \draw[thick](-3.5,-3.5)--(-5.5,-5.5);
    \draw[thick](3.5,3.5)--(5.5,5.5);
    \node[label=right:{$(1,1)$}][7brane]at(5.5,5.5){};
    \node[7brane]at(4.5,4.5){};
    \node[7brane]at(3.5,3.5){};
    \node[7brane]at(2.5,2.5){};
    \node[7brane]at(1.5,1.5){};
        \node[label=left:{$(1,1)$}][7brane]at(-5.5,-5.5){};
    \node[7brane]at(-4.5,-4.5){};
    \node[7brane]at(-3.5,-3.5){};
    \node[7brane]at(-2.5,-2.5){};
    \node[7brane]at(-1.5,-1.5){};
    \draw[thick](-4,0)--(4,0);
    \node[7brane]at(-2,0){};
    \node[7brane]at(-4,0){};
    \node[7brane]at(2,0){};
    \node[7brane]at(4,0){};
    \node at (3,3){$\udots$};
    \node at (-3,-3){$\udots$};
    \node[label=above:{1}] at(4.75,4.75){};
    \node[label=above:{2}] at(3.75,3.75){};
    \node[label=above:{$N$}] at(1.75,1.75){};
    \node[label=above:{$N+1$}] at(.4,.7){};
    \node[label=above left:{1}] at(-4.75,-5.25){};
    \node[label=above left:{2}] at(-3.75,-4.25){};
    \node[label=above left:{$N$}] at(-1.75,-2.25){};
    \node at(-1.55,-.7){$N+1$};
    \node[label=above:{1}]at(-3,0){};
    \node[label=above:{2}]at(-1,0){};
    \node[label=below:{2}]at(1,0){};
    \node[label=below:{1}]at(3,0){};
    \end{tikzpicture}
         \end{scriptsize}\end{array}\;.
\end{equation}
From the brane webs in \eqref{EQ55 orientifold web} and \eqref{EQ5 unitary web}, we obtain the two corresponding magnetic quivers which then imply 
 \begin{equation}\MQ_{5,5}=\MQ_5^2=
        \begin{array}{c}
             \begin{scriptsize}
             \begin{tikzpicture}
                 \node[label=right:{1}][u](1){};
                 \node (dots)[below of=1]{$\vdots$};
                 \node[label=right:{$2N+1$}][u](2N-1)[below of=dots]{};
                 \node[label=right:{$2N+2$}][sp](sp2N)[below of=2N-1]{};
                 \node[label=below:{4}][so](so4)[below of=sp2N]{};
                 \node[label=below:{2}][sp](sp2)[right of=so4]{};
                 \node[label=below:{2}][sp](sp2')[left of=so4]{};
                 \node[label=below:{2}][so](so2)[right of=sp2]{};
                 \node[label=below:{2}][so](so2')[left of=sp2']{};
                 \draw(1)--(dots);
                 \draw(dots)--(2N-1);
                 \draw(2N-1)--(sp2N);
                 \draw(sp2N)--(so4);
                 \draw(so4)--(sp2);
                 \draw(so4)--(sp2');
                 \draw(sp2)--(so2);
                 \draw(so2')--(sp2');
             \end{tikzpicture}
             \end{scriptsize}
        \end{array}=\left(\begin{array}{c}
             \begin{scriptsize}
             \begin{tikzpicture}
                 \node[label=below:{1}][u](1){};
                 \node (dots)[right of=1]{$\cdots$};
                 \node[label=below:{$N+1$}][u](N)[right of=dots]{};
                 \node (dots')[right of=N]{$\cdots$};
                 \node[label=below:{1}][u](1')[right of=dots']{};
                 \node[label=above:{2}][u](2)[above of=N]{};
                 \node[label=above:{1}][u](u1)[right of=2]{};
                 \node[label=above:{1}][u](u1')[left of=2]{};
                 \draw(1)--(dots);
                 \draw(dots)--(N);
                 \draw(N)--(dots');
                 \draw(dots')--(1');
                 \draw(N)--(2);
                 \draw(2)--(u1);
                 \draw(2)--(u1');
                 
             \end{tikzpicture}
             \end{scriptsize}
        \end{array}\right)^2\;.
    \end{equation}
The unitary quiver appearing here has been studied previously, and its HWG was given in \cite{Ferlito:2017xdq}. We can now obtain the HWG for the OSp quiver by simply squaring that expression to obtain
    \begin{align}
        \HWG_{5,5}&=\PE\left[\sum_{i=1}^{N+1}\mu_i\mu_{2N+2-i}t^{2i}+(\nu_1^2+\nu_2^2)t^2+t^4+\nu_1\nu_2\mu_{N+1}(t^{N+1}+t^{N+3})-\nu_1^2\nu_2^2\mu_{N+1}^2t^{2N+6}\right]\nonumber \\
        &\times \PE\left[\sum_{i=1}^{N+1}\lambda_i\lambda_{2N+2-i}t^{2i}+(\rho_1^2+\rho_2^2)t^2+t^4+\rho_1\rho_2\lambda_{N+1}(t^{N+1}+t^{N+3})-\rho_1^2\rho_2^2\lambda_{N+1}^2t^{2N+6}\right] ~.
    \end{align}

\subsection{The \texorpdfstring{$E_5'\times E_5'$}{TEXT} sequence}
The $E_{5'}\times E_{5'}$ sequence is obtained by considering the fixed point limit of the following 5d electric quiver:
  \begin{equation}\EQ_{5',5'}=
\begin{array}{c}
\begin{tikzpicture}
    \node{$\text{SO(4)}-\text{USp(0)}-\text{SO(4)}-\cdots-\text{USp(0)}-\text{SO(4)}-[4\textbf{s}+4\textbf{c}]$};
    
    \draw [thick,decorate,decoration={brace,amplitude=6pt},xshift=0pt,yshift=10pt]
(-4.5,0) -- (3,0)node [black,midway,xshift=0pt,yshift=15pt] {
$2N-1$};
    \end{tikzpicture}
    \end{array}\;,
\end{equation}
which can be engineered using the orientifold web:
\begin{equation}
\begin{array}{c}
\begin{scriptsize}
\begin{tikzpicture}[scale=1.075]
\draw[thick,dashed](-2,0)--(-3,0);
\draw[thick,dashed](6,0)--(7,0);
    \draw[thick](-2,0)--(3,0);
    \draw[thick](4,0)--(6,0);
    \draw[thick](0,0)--(0,1.5);
    
    \node at(3.5,0){$\cdots$};
    \node[label=below:{$2N$}] at(-.5,0){};
    \node[label=below:{$2N+2$}] at(.5,0){};
    \node at(1.5,-.3){$\frac{4N+3}{2}$};
    \node[label=below:{$2N+1$}] at(2.5,0){};
    \node at(4.5,-.3){1};
    \node at(5.5,-.3){$\frac{1}{2}$};
    \node at(-1.5,-.3){$N$};
    \node[label=right:{$2N+1$}] at(0,1){};
    \node[label=below:{O5$^-$}] at(6.5,0){};
    \node[label=below:{O5$^+$}] at(-2.5,0){};
    \node[7brane]at(0,1.5){};
    \node[7brane]at(-1,0){};
    \node[7brane]at(-2,0){};
    \node[7brane]at(1,0){};
    \node[7brane]at(2,0){};
    \node[7brane]at(3,0){};
    \node[7brane]at(4,0){};
    \node[7brane]at(5,0){};
    \node[7brane]at(6,0){};

    \end{tikzpicture}
    \end{scriptsize}
    \end{array}\;.
\end{equation}
Alternatively we may rewrite $\EQ_{5',5'}$ as two copies of a single electric quiver with SU(2) gauge nodes, namely
\begin{equation}
\EQ_{5'}^2=
    \left(\begin{array}{c}
\begin{tikzpicture}
    \node{$\text{SU}(2)-\text{SU(2)}-\cdots-{\underset{\underset{\text{\large$\left[4\textbf{F}\right]$}}{\textstyle\vert}}{\text{SU}(2)}}$};
    
    \draw [thick,decorate,decoration={brace,amplitude=6pt},xshift=0pt,yshift=10pt]
(-2.25,.5) -- (2.25,.5)node [black,midway,xshift=0pt,yshift=15pt] {
$N$};
    \end{tikzpicture}
    \end{array}\right)^2\;,
\end{equation}
each copy of which can now be engineered using an ordinary brane web:
\begin{equation}\label{EQ5' unitary web}
    \begin{array}{c}
         \begin{scriptsize}
         \begin{tikzpicture}[scale=.8]
             \draw[thick](0.5,0.5)--(2.5,2.5);
             \draw[thick](3,3)--(5,5);
             \draw[thick](5,5)--(8,5);
             \draw[thick](4,6)--(5,5);
             \draw[thick](5,7.5)--(5,5);
             \node[7brane]at(0.5,0.5){};
             \node[7brane]at(1.5,1.5){};
             \node[7brane]at(2.5,2.5){};
             \node[7brane]at(3,3){};
             \node[7brane]at(4,4){};
             \node[7brane]at(5,6.5){};
             \node[7brane]at(5,7.5){};
             \node[7brane]at(6.5,5){};
             \node[7brane]at(8,5){};
             \node[7brane]at(4,6){};
             \node at (2.75,2.75){$\udots$};
             \node[label=below:{1}] at(0.7,1.5){};
             \node[label=below:{2}] at(1.7,2.5){};
             \node at(3.0,3.5){$2N$};
             \node at(3.9,4.5){$2N+1$};
             \node[label=below:{$N+1$}] at(7.25,5){};
             \node[label=below:{$2N+2$}] at(5.75,5){};
             \node[label=above:{$1$}] at(4,5.25){};
             \node[label=right:{$N$}] at(5,7){};
             \node[label=right:{$2N$}] at(5,5.75){};
         \end{tikzpicture}
         \end{scriptsize}
    \end{array}\;.
\end{equation}
This leads us to conjecture the equivalence of the following 3d magnetic quivers
\begin{equation}\begin{split}
\MQ_{5',5'}&=
    \begin{array}{c}
         \begin{scriptsize}
         \begin{tikzpicture}
         \node[label=below:{1}][so](o1){};
         \node[label=below:{$2N$}][sp](sp4)[right of=o1]{};
         \node[label=below:{$4N+1$}][so](so5)[right of=sp4]{};
         \node[label=below:{$4N$}][sp](sp4')[right of=so5]{};
         \node[label=below:{$4N+1$}][so](so5')[right of=sp4']{};
         \node[label=below:{$4N$}][sp](sp4'')[right of=so5']{};
         \node[label=below:{$4N$}][so](so4)[right of=sp4'']{};
         \node (dots)[right of=so4]{$\cdots$};
         \node[label=below:{2}][sp](sp2)[right of=dots]{};
         \node[label=below:{2}][so](so2)[right of=sp2]{};
         \node[label=above:{1}][sof](so1)[above of=sp4'']{};
         \node[label=right:{$2N$}][sp](sp2')[above of=so5]{};
         \node[label=above:{1}][sof](so3f)[above of=sp2']{};
         \draw(o1)--(sp4);
         \draw(sp4)--(so5);
         \draw(so5)--(sp2');
         \draw(sp2')--(so3f);
         \draw(so5)--(sp4');
         \draw(sp4')--(so5');
         \draw(so5')--(sp4'');
         \draw(sp4'')--(so1);
         \draw(sp4'')--(so4);
         \draw(so4)--(dots);
         \draw(dots)--(sp2);
         \draw(sp2)--(so2);
         \end{tikzpicture}
         \end{scriptsize}
    \end{array}\\=\MQ_{5'}^2&=\left(\begin{array}{c}
        \begin{scriptsize}
        \begin{tikzpicture}
         \node[label=below:{1}][u](1){};   
         \node (dots)[left of=1]{$\cdots$};
         \node[label=below:{$2N$}][u](2N)[left of=dots]{};
         \node[label=below:{$2N$}][u](2N')[left of=2N]{};
         \node[label=below:{$N$}][u](N)[left of=2N']{};
         \node[label=above:{$N$}][u](N+1)[above of=2N']{};
         \node[label=above:{$1$}][u](1')[above of=2N]{};
         \draw(1)--(dots);
         \draw(dots)--(2N);
         \draw(2N)--(2N');
         \draw(2N')--(N+1);
         \draw(2N')--(N);
         \draw(2N)--(1');
        \end{tikzpicture}
        \end{scriptsize}
    \end{array}\right)^2\;.
    \end{split}
\end{equation}
We now want to write down the HWG for the Coulomb branch Hilbert series of MQ$_{5',5'}$, via the conjectured relation to the unitary quiver. 
The HWG for the unitary quiver $\MQ_{5'}$ can be found in \cite{Ferlito:2016grh} (also see table 18 of \cite{Hanany:2016gbz}), which we report in table \ref{TableHWGUnitaryQuivers}. Consequently, the HWG for MQ$_{5',5'}$ is obtained by squaring this result, namely
\begin{equation}
    \HWG_{5',5'}=\PE\left[\sum_{k=1}^N\left(\mu_{2k}+\nu_{2k}\right)t^{2k}\right]\;.
\end{equation}
\subsection{The \texorpdfstring{$E_{5'}\times E_6$}{TEXT} sequence}
The $E_{5'}\times E_6$ sequence is obtained by taking the fixed point limit of the 5d electric quiver given by
    \begin{equation}\EQ_{5',6}=
\begin{array}{c}
\begin{tikzpicture}
    \node{$[1\textbf{s}]-\text{SO(4)}-\text{USp(0)}-\text{SO(4)}-\cdots-\text{USp(0)}-\text{SO(4)}-[4\textbf{s}+4\textbf{c}]$};
    
    \draw [thick,decorate,decoration={brace,amplitude=6pt},xshift=0pt,yshift=10pt]
(-3.75,0) -- (3.75,0)node [black,midway,xshift=0pt,yshift=15pt] {
$2N-1$};
    \end{tikzpicture}
    \end{array}\;,
\end{equation}
which can be engineered by the following orientifold web diagram
\begin{equation}\label{E5'xE6 orientifold web}
    \begin{array}{c}
         \begin{scriptsize}
         \begin{tikzpicture}[scale=.75]
    \draw[thick](-1,0)--(5,0);
    \draw[thick](8,0)--(6,0);
    \draw[thick, dashed](-1,0)--(-2,0);
    \draw[thick, dashed](9,0)--(8,0);
    \draw[thick](2,4)--(2,6);
    \node[label=below:{O5$^-$}] at (-1.5,0){};
    \node[label=below:{O5$^-$}] at (8.5,0){};
    \node[label=right:{$1$}] at (2,5.5){};
    \node[label=right:{$2$}] at (2,4.5){};
    \node[label=right:{$2N$}] at (2,2.5){};
    \node[label=right:{$2N+1$}] at (2,1.5){};
    \draw[thick](2,0)--(1,1);
    \draw[thick](2,0)--(2,3);
    \node[label=above left:{(1,-1)}][7brane]at(1,1){};
    \node[7brane]at(2,2){};
    \node[7brane]at(2,3){};
    \node at (2,3.5) {$\vdots$};
    \node[7brane]at(2,4){};
    \node[7brane]at(2,5){};
    \node[7brane]at(2,6){};
    \node[label=below:{$\frac{1}{2}$}]at(7.5,0){};
    \node[label=below:{$1$}]at(6.5,0){};
    \node[label=below:{$\frac{4N+3}{2}$}]at(4.5,0){};
    \node[label=below:{$2N+2$}]at(3,0){};
    \node[label=below:{$\frac{2N+1}{2}$}]at(-.5,0){};
    \node[label=below:{$2N+1$}]at(1,0){};
    \node[7brane]at(-1,0){};
    \node[7brane]at(0,0){};
    \node[7brane]at(5,0){};
    \node[7brane]at(4,0){};
    \node[7brane]at(6,0){};
    \node[7brane]at(7,0){};
    \node[7brane]at(8,0){};
    
    \end{tikzpicture}
    \end{scriptsize}
    \end{array}\;.
\end{equation}
Alternatively we may reformulate EQ$_{5',6}$ as the product of two unitary electric quivers
\begin{equation}
\EQ_{5',6}=\EQ_{5'}\times\EQ_6=\begin{array}{c}
\begin{tikzpicture}
    \node{$\text{SU}(2)-\text{SU(2)}-\cdots-{\underset{\underset{\text{\large$\left[4\textbf{F}\right]$}}{\textstyle\vert}}{\text{SU}(2)}}$};
    
    \draw [thick,decorate,decoration={brace,amplitude=6pt},xshift=0pt,yshift=10pt]
(-2.25,.5) -- (2.25,.5)node [black,midway,xshift=0pt,yshift=15pt] {
$N$};
    \end{tikzpicture}
    \end{array}\times \begin{array}{c}
\begin{tikzpicture}
    \node{${\underset{\underset{\text{\large$\left[1\textbf{F}\right]$}}{\textstyle\vert}}{\text{SU}(2)}}-\text{SU(2)}-\cdots-{\underset{\underset{\text{\large$\left[4\textbf{F}\right]$}}{\textstyle\vert}}{\text{SU}(2)}}$};
    
    \draw [thick,decorate,decoration={brace,amplitude=6pt},xshift=0pt,yshift=10pt]
(-2.25,.5) -- (2.25,.5)node [black,midway,xshift=0pt,yshift=15pt] {
$N$};
    \end{tikzpicture}
    \end{array}\;,
\end{equation}
where EQ$_{5'}$ is engineered by the unitary web in \eqref{EQ5' unitary web}, and EQ$_6$ is the IR quiver description of the web diagram given by
\begin{equation}\label{EQ6 unitary web}
    \begin{array}{c}
         \begin{scriptsize}
         \begin{tikzpicture}[scale=.8]
             \draw[thick](0.5,0.5)--(2.5,2.5);
             \draw[thick](3,3)--(5,5);
             \draw[thick](5,5)--(8,5);
             \draw[thick](2,5)--(5,5);
             \draw[thick](5,8.5)--(5,5);
             \node[7brane]at(0.5,0.5){};
             \node[7brane]at(1.5,1.5){};
             \node[7brane]at(2.5,2.5){};
             \node[7brane]at(3,3){};
             \node[7brane]at(4,4){};
             \node[7brane]at(5,6.5){};
             \node[7brane]at(5,7.5){};
             \node[7brane]at(5,8.5){};
             \node[7brane]at(6.5,5){};
             \node[7brane]at(8,5){};
             \node[7brane]at(2,5){};
             \node at (2.75,2.75){$\udots$};
             \node[label=below:{1}] at(1,1){};
             \node[label=below:{2}] at(2,2){};
             \node at(3.5,3.5){$2N$};
             \node at(4.5,4.5){$2N+1$};
             \node[label=below:{$N+1$}] at(7.25,5){};
             \node[label=below:{$2N+2$}] at(5.75,5){};
             \node[label=above:{$1$}] at(2.75,5){};
             \node[label=right:{$N+1$}] at(5,7){};
             \node[label=right:{$1$}] at(5,8){};
             \node[label=right:{$2N+1$}] at(5,5.75){};
         \end{tikzpicture}
         \end{scriptsize}
    \end{array}\;.
\end{equation}
Reading off the OSp magnetic quiver from \eqref{E5'xE6 orientifold web} and the unitary quivers from \eqref{EQ5' unitary web}, \eqref{EQ6 unitary web}, we arrive at the conjecture 
\begin{equation}\begin{split}\label{MQ5'6}\MQ_{5',6}&=
    \begin{array}{c}
         \begin{scriptsize}
         \begin{tikzpicture}
             \node[label=below:{2}][so](so2){};
             \node[label=below:{2}][sp](sp2)[left of=so2]{};
             \node (dots)[left of=sp2]{$\cdots$};
             \node[label=below:{$4N+2$}][so](so4N+2)[left of=dots]{};
             \node[label=below:{$4N+2$}][sp](sp4N+2)[left of=so4N+2]{};
             \node[label=below:{$4N+2$}][so](so4N+2')[left of=sp4N+2]{};
             \node[label=above:{$2N$}][sp](sp2N)[above of=so4N+2']{};
             \node[label=below:{$2N$}][sp](sp2N')[left of=so4N+2']{};
             \node[label=above:{$1$}][u](u1)[above of=sp4N+2]{};
             \draw(so2)--(sp2);
             \draw(sp2)--(dots);
             \draw(dots)--(so4N+2);
             \draw(so4N+2)--(sp4N+2);
             \draw(sp4N+2)--(so4N+2');
             \draw(so4N+2')--(sp2N);
             \draw(so4N+2')--(sp2N');
             \draw(sp4N+2)--(u1);
         \end{tikzpicture}
         \end{scriptsize}
    \end{array}=\\
  =\MQ_{5'}\times\MQ_6 &=\begin{array}{c}
        \begin{scriptsize}
        \begin{tikzpicture}
         \node[label=below:{1}][u](1){};   
         \node (dots)[left of=1]{$\cdots$};
         \node[label=below:{$2N$}][u](2N)[left of=dots]{};
         \node[label=below:{$2N$}][u](2N')[left of=2N]{};
         \node[label=below:{$N$}][u](N)[left of=2N']{};
         \node[label=above:{$N$}][u](N+1)[above of=2N']{};
         \node[label=above:{$1$}][u](1')[above of=2N]{};
         \draw(1)--(dots);
         \draw(dots)--(2N);
         \draw(2N)--(2N');
         \draw(2N')--(N+1);
         \draw(2N')--(N);
         \draw(2N)--(1');
        \end{tikzpicture}
        \end{scriptsize}
    \end{array}\times \begin{array}{c}
         \begin{scriptsize}
         \begin{tikzpicture}
             \node[label=below:{1}][u](1){};
             \node (dots)[left of=1]{$\cdots$};
             \node[label=below:{$2N+1$}][u](2N-1)[left of=dots]{};
             \node[label=below:{$N+1$}][u](N)[left of=2N-1]{};
             \node[label=right:{$N+1$}][u](N')[above of=2N-1]{};
             \node[label=below:{$1$}][u](u1)[left of=N]{};
             \node[label=right:{$1$}][u](u1')[above of=N']{};
             \draw(1)--(dots);
             \draw(dots)--(2N-1);
             \draw(2N-1)--(N);
             \draw(N)--(u1);
             \draw(2N-1)--(N');
             \draw(N')--(u1');
         \end{tikzpicture}
         \end{scriptsize}
    \end{array}
    \end{split}
\end{equation}
Obtaining the HWG for the Coulomb branch of MQ$_{5',6}$ is now straightforward, given the above relation. The HWG for MQ$_5'$ was worked out in \cite{Ferlito:2016grh, Hanany:2016gbz} 
and the HWG for MQ$_6$ appears in \cite{Ferlito:2017xdq}. For the OSp quiver MQ$_{5',6}$ in \eqref{MQ5'6} we simply take the product of these two results
\begin{equation}
    \HWG_{5',6}=\PE\left[\sum_{i=1}^{N}\left(\mu_{2i}+\nu_{2i}\right)t^{2i}+t^2+\left(\mu_{2N+2}q+\mu_{2N+3}q^{-1}\right)t^{N+1}\right] ~.
\end{equation}
This can be verified by computing the unrefined Coulomb branch Hilbert series, the results of which are given in table \ref{TableHSOSp} for low values of $N$.
\subsection{The \texorpdfstring{$E_6\times E_6$}{TEXT} sequence}
\label{E6E6section}

The electric quiver for the IR limit of the $E_6\times E_6$ sequence reads
    \begin{equation}\EQ_{6,6}=
\begin{array}{c}
\begin{tikzpicture}
    \node{$[1\textbf{s}+1\textbf{c}]-\text{SO(4)}-\text{USp(0)}-\text{SO(4)}-\cdots-\text{USp(0)}-\text{SO(4)}-[4\textbf{s}+4\textbf{c}]$};
    
    \draw [thick,decorate,decoration={brace,amplitude=6pt},xshift=0pt,yshift=10pt]
(-3.75,0) -- (3.75,0)node [black,midway,xshift=0pt,yshift=15pt] {
$2N-1$};
    \end{tikzpicture}
    \end{array}\;.
\end{equation}
It can be engineered using the following orientifold web
\begin{equation}\label{EQ66 orientifold web}
\begin{array}{c}
\begin{scriptsize}
\begin{tikzpicture}[scale=1.075]
\draw[thick,dashed](-4,0)--(-5,0);
\draw[thick,dashed](6,0)--(7,0);
    \draw[thick](-4,0)--(3,0);
    \draw[thick](4,0)--(6,0);
    \draw[thick](0,0)--(0,2.5);
    
    \node at(3.5,0){$\cdots$};
    \node[label=below:{$2N+2$}] at(-.5,0){};
    \node[label=below:{$2N+2$}] at(.5,0){};
    \node at(1.5,-.3){$\frac{4N+3}{2}$};
    \node[label=below:{$2N+1$}] at(2.5,0){};
    \node at(4.5,-.3){1};
    \node at(5.5,-.3){$\frac{1}{2}$};
    \node at(-1.5,-.3){$\frac{2N+3}{2}$};
    \node at(-2.5,-.3){1};
    \node at(-3.5,-.3){$\frac{1}{2}$};
    \node[label=right:{$2N+2$}] at(0,1){};
    \node[label=right:{$1$}] at(0,2){};
    \node[label=below:{O5$^-$}] at(6.5,0){};
    \node[label=below:{O5$^-$}] at(-4.5,0){};
    \node[7brane]at(0,1.5){};
    \node[7brane]at(0,2.5){};
    \node[7brane]at(-1,0){};
    \node[7brane]at(-2,0){};
    \node[7brane]at(-3,0){};
    \node[7brane]at(-4,0){};
    \node[7brane]at(1,0){};
    \node[7brane]at(2,0){};
    \node[7brane]at(3,0){};
    \node[7brane]at(4,0){};
    \node[7brane]at(5,0){};
    \node[7brane]at(6,0){};

    \end{tikzpicture}
    \end{scriptsize}
    \end{array}\;.
\end{equation}
Alternatively we may present the electric quiver as a product of two unitary quivers:
\begin{equation}
\EQ_{6,6}=\EQ_6^2=\left(\begin{array}{c}
\begin{tikzpicture}
    \node{${\underset{\underset{\text{\large$\left[1\textbf{F}\right]$}}{\textstyle\vert}}{\text{SU}(2)}}-\text{SU(2)}-\cdots-{\underset{\underset{\text{\large$\left[4\textbf{F}\right]$}}{\textstyle\vert}}{\text{SU}(2)}}$};
    
    \draw [thick,decorate,decoration={brace,amplitude=6pt},xshift=0pt,yshift=10pt]
(-2.25,.5) -- (2.25,.5)node [black,midway,xshift=0pt,yshift=15pt] {
$N$};
    \end{tikzpicture}
    \end{array}\right)^2\;,
\end{equation}
where each copy is engineered by the web diagram in \eqref{EQ6 unitary web}. We can then obtain the magnetic quivers from the brane webs in \eqref{EQ66 orientifold web} and \eqref{EQ6 unitary web}, that lead us to the conjecture that
\begin{equation}
\label{MQ E6E6}
\MQ_{6,6}=\MQ_6^2=
      \begin{array}{c}
         \begin{scriptsize}
         \begin{tikzpicture}
         \node[label=below:{2}][so](so2){};
         \node[label=below:{$2N+2$}][sp](sp4)[right of=so2]{};
         \node[label=below:{$4N+4$}][so](so8)[right of=sp4]{};
         \node[label=below:{$4N+2$}][sp](sp6)[right of=so8]{};
         \node[label=below:{$4N+2$}][so](so6)[right of=sp6]{};
         \node(dots)[right of=so6]{$\cdots$};
         \node[label=below:{2}][sp](sp2)[right of=dots]{};
         \node[label=below:{2}][so](so2')[right of=sp2]{};
         \node[label=right:{$2N+2$}][sp](sp4'')[above of=so8]{};
         \node[label=right:{1}][u](u1)[above of=sp4'']{};
         \draw(so2)--(sp4);
         \draw(sp4)--(so8);
         \draw(so8)--(sp6);
         \draw(sp6)--(so6);
         \draw(sp2)--(dots);
         \draw(so6)--(dots);
         \draw(sp2)--(so2');
         \draw(so8)--(sp4'');
         \draw(sp4'')--(u1);
         \end{tikzpicture}
         \end{scriptsize}
    \end{array}=\left(\begin{array}{c}
         \begin{scriptsize}
         \begin{tikzpicture}
             \node[label=below:{1}][u](1){};
             \node (dots)[left of=1]{$\cdots$};
             \node[label=below:{$2N+1$}][u](2N-1)[left of=dots]{};
             \node[label=below:{$N+1$}][u](N)[left of=2N-1]{};
             \node[label=right:{$N+1$}][u](N')[above of=2N-1]{};
             \node[label=below:{$1$}][u](u1)[left of=N]{};
             \node[label=right:{$1$}][u](u1')[above of=N']{};
             \draw(1)--(dots);
             \draw(dots)--(2N-1);
             \draw(2N-1)--(N);
             \draw(N)--(u1);
             \draw(2N-1)--(N');
             \draw(N')--(u1');
         \end{tikzpicture}
         \end{scriptsize}
    \end{array}\right)^2\;.
\end{equation}
This has an immediate corollary which allows us to extract the HWG for the OSp quiver appearing above by squaring the known result \cite{Ferlito:2017xdq} for the unitary quiver:
\begin{align}
    \HWG_{6,6}&=\PE\left[\sum_{i=1}^{N}\mu_{2i}t^{2i}+t^2+\left(\mu_{2N+2}q_1+\mu_{2N+3}q_1^{-1}\right)t^{N+1}\right] \nonumber \\
    &\times\PE\left[\sum_{i=1}^{N}\nu_{2i}t^{2i}+t^2+\left(\nu_{2N+2}q_2+\nu_{2N+3}q_2^{-1}\right)t^{N+1}\right]
    ~.
\end{align}
Indeed the unrefined Hilbert series for the OSp quiver for $N=1$ was computed in \cite{Bourget:2020xdz}, and is in agreement with our claim. For higher values of $N$ we were not able to perform an explicit computation due to the high rank of the OSp quiver. This is one instance in which our conjecture proves powerful, as it gives an exact expression for the Hilbert series of a quiver which would otherwise be very challenging to compute. 

We further point out an interesting fact about this theory, namely the existence of a 4d $\mathcal{N}=2$ theory with very similar properties. As noticed in \cite{Chacaltana:2011ze}, there is one class-S theory of $D_4$ type in which a single three-punctured sphere realizes a product SCFT, where both factors are the $E_6$ Minahan-Nemeschansky (MN) theory \cite{Minahan:1996fg}. We recall that the $E_6$ Minahan-Nemeschansky theory is a 4d $\mathcal{N}=2$ SCFT of rank $1$, with flavor symmetry group $E_6$, and central charges
\begin{equation}
a_{E_6}=\dfrac{41}{24}, \qquad c_{E_6}=\dfrac{13}{6},
\end{equation}

We report the partitions labeling the punctures in table \ref{tab::E6E6}, together with their contribution to the effective number of hypermutiplets and vector multiplets.

\begin{table}[h]
	\begin{center}
		\begin{tabular}{|c|c|c|}
			\hline
			Nahm partition	&  $(\delta n_h, \delta n_v)$ \\
			\hline
			\hline
			$[1^8]$  & $(112,100)$  \\
			\hline
			$[3^2,1^2]$	 & $(72,69)$  \\
			\hline
			$[3^2,1^2]$	 & $(72,69)$  \\
			\hline
		\end{tabular}
	\end{center}
\caption{Table containing the data defining the punctures for the 4d $E_6\times E_6$ theory.}
	\label{tab::E6E6}
\end{table}

From this data it is easy to compute the central charges $a$ and $c$ of this theory\footnote{For a small review see appendix \ref{appendixC1}}, finding 
\begin{equation}
a_{E_6\times E_6}=\dfrac{41}{12}, \qquad c_{E_6\times E_6}=\dfrac{13}{3}
\end{equation}
as it should be for two copies of the $E_6$ MN theory. By applying the procedure to write the 3d mirror for this theory\footnote{For a small review see appendix \ref{appendixC2}} we find that the full puncture $[1^8]$ is associated to the quiver tail \eqref{leg_full8} while the puncture $[3^2,1^2]$ is associated to the quiver tail \eqref{leg_3311}, 

\begin{equation}
\label{leg_full8}
\begin{array}{c}
\begin{scriptsize}
\begin{tikzpicture}
\node[label=below:{2}][so](so2){};
\node[label=below:{$2$}][sp](sp2)[right of=so2]{};
\node[label=below:{$4$}][so](so4)[right of=sp2]{};
\node[label=below:{$4$}][sp](sp4)[right of=so4]{};
\node[label=below:{$6$}][so](so6)[right of=sp4]{};
\node[label=below:{$6$}][sp](sp6)[right of=so6]{};
\node[label=below:{$8$}][sof](so8)[right of=sp6]{};
\draw(so2)--(sp2);
\draw(sp2)--(so4);
\draw(so4)--(sp4);
\draw(sp4)--(so6);
\draw(so6)--(sp6);
\draw(sp6)--(so8);
\end{tikzpicture}
\end{scriptsize}
\end{array}
\end{equation}

\begin{equation}
\label{leg_3311}
\begin{array}{c}
\begin{scriptsize}
\begin{tikzpicture}
\node[label=below:{2}][so](so2){};
\node[label=below:{$4$}][sp](sp4)[right of=so2]{};
\node[label=below:{$8$}][sof](so8)[right of=sp4]{};
\draw(so2)--(sp4);
\draw(sp4)--(so8);
\end{tikzpicture}
\end{scriptsize}
\end{array} .
\end{equation}

Gluing the three tails together results in the magnetic quiver for the $5$d $E_6\times E_6$ depicted in \eqref{MQ E6E6}. Therefore the magnetic quiver of the 5d $E_6\times E_6$ theory is the 3d mirror theory of the 4d $E_6\times E_6$ theory above described. It is then tempting to conjecture that the 5d $E_6\times E_6$ theory reduces to 4d to this $D_4$ type class-S theory, giving two copies of $E_6$ Minahan-Nemeschansky.

Having derived the magnetic quiver for the $E_6\times E_6$ sequence from the brane web, for any $N\in \mathbb{N}$, we can use the same argument as the paragraphs above to conjecture that all class-S theories of $D_{2N+2}$ type given by a three punctured sphere with regular punctures given by $[1^{4N+4}], [2N+1,2N+1,1,1], [2N+1,2N+1,1,1]$ will be a factorized SCFT.
We conjecture that it will decompose into two copies of three punctured $A_{2N}$ spheres, with regular punctures given by $[1^{2N+1}],[N^2,1],[N^2,1]$.
It will be interesting to further check this proposal.

\subsection{The \texorpdfstring{$E_7\times E_7$}{TEXT} sequence}
\label{E7E7section}

The $E_7\times E_7$ sequence corresponds to the fixed point limit of the following IR quiver 
    \begin{equation}\EQ_{7,7}=
\begin{array}{c}
\begin{tikzpicture}
    \node{$[2\textbf{s}+2\textbf{c}]-\text{SO(4)}-\text{USp(0)}-\text{SO(4)}-\cdots-\text{USp(0)}-\text{SO(4)}-[4\textbf{s}+4\textbf{c}]$};
    
    \draw [thick,decorate,decoration={brace,amplitude=6pt},xshift=0pt,yshift=10pt]
(-3.75,0) -- (3.75,0)node [black,midway,xshift=0pt,yshift=15pt] {
$2N-1$};
    \end{tikzpicture}
    \end{array}\;.
\end{equation}
It can be engineered using the orientifold web given by
\begin{equation}
\begin{array}{c}
\begin{scriptsize}
\begin{tikzpicture}[scale=1.075]
\draw[thick,dashed](-6,0)--(-7,0);
\draw[thick,dashed](6,0)--(7,0);
    \draw[thick](-6,0)--(3,0);
    \draw[thick](4,0)--(6,0);
    \draw[thick](0,0)--(0,2);
    
    \node at(3.5,0){$\cdots$};
    \node[label=below:{$2N+3$}] at(-.5,0){};
    \node[label=below:{$2N+3$}] at(.5,0){};
    \node at(1.5,-.3){$\frac{4N+5}{2}$};
    \node[label=below:{$2N+4$}] at(2.5,0){};
    \node at(4.5,-.3){1};
    \node at(5.5,-.3){$\frac{1}{2}$};
    \node at(-1.5,-.3){$\frac{2N+5}{2}$};
    \node at(-2.5,-.3){2};
    \node at(-3.5,-.3){$\frac{3}{2}$};
    \node at(-4.5,-.3){1};
    \node at(-5.5,-.3){$\frac{1}{2}$};
    \node[label=right:{$2N+2$}] at(0,1){};
    \node[label=below:{O5$^-$}] at(6.5,0){};
    \node[label=below:{O5$^-$}] at(-6.5,0){};
    \node[7brane]at(0,2){};
    \node[7brane]at(-1,0){};
    \node[7brane]at(-2,0){};
    \node[7brane]at(-3,0){};
    \node[7brane]at(-4,0){};
    \node[7brane]at(-5,0){};
    \node[7brane]at(-6,0){};
    \node[7brane]at(1,0){};
    \node[7brane]at(2,0){};
    \node[7brane]at(3,0){};
    \node[7brane]at(4,0){};
    \node[7brane]at(5,0){};
    \node[7brane]at(6,0){};

    \end{tikzpicture}
    \end{scriptsize}
    \end{array}
\end{equation}
This web can be converted to a magnetic quiver following \cite{Akhond:2020vhc}, which results in
\begin{equation}\label{MQ77 OSp}
   \MQ_{7,7}= \begin{array}{c}
         \begin{scriptsize}
         \begin{tikzpicture}
         \node[label=below:{2}][so](so2){};
         \node[label=below:{2}][sp](sp2)[right of=so2]{};
         \node[label=below:{4}][so](so4)[right of=sp2]{};
         \node[label=below:{$2N+4$}][sp](sp6)[right of=so4]{};
         \node[label=below:{$4N+6$}][so](so10)[right of=sp6]{};
         \node[label=below:{$4N+4$}][sp](sp8)[right of=so10]{};
         \node[label=below:{$4N+4$}][so](so8)[right of=sp8]{};
         \node (dots)[right of=so8]{$\cdots$};
         \node[label=below:{2}][sp](sp2')[right of=dots]{};
         \node[label=below:{2}][so](so2')[right of=sp2']{};
         \node[label=above:{$2N+2$}][sp](sp4)[above of=so10]{};
         \draw(so2)--(sp2);
         \draw(sp2)--(so4);
         \draw(so4)--(sp6);
         \draw(sp6)--(so10);
         \draw(so10)--(sp4);
         \draw(so10)--(sp8);
         \draw(sp8)--(so8);
         \draw(so8)--(dots);
         \draw(dots)--(sp2');
         \draw(sp2')--(so2');
         \end{tikzpicture}
         \end{scriptsize}
    \end{array}\;.
\end{equation}
Now we use the alternative description of the EQ$_{7,7}$ as a product of a pair of linear quivers with SU(2) nodes, namely
\begin{equation}
\EQ_{7,7}=\EQ_7^2=\left(\begin{array}{c}
\begin{tikzpicture}
    \node{${\underset{\underset{\text{\large$\left[2\textbf{F}\right]$}}{\textstyle\vert}}{\text{SU}(2)}}-\text{SU(2)}-\cdots-{\underset{\underset{\text{\large$\left[4\textbf{F}\right]$}}{\textstyle\vert}}{\text{SU}(2)}}$};
    
    \draw [thick,decorate,decoration={brace,amplitude=6pt},xshift=0pt,yshift=10pt]
(-2.25,.5) -- (2.25,.5)node [black,midway,xshift=0pt,yshift=15pt] {
$N$};
    \end{tikzpicture}
    \end{array}\right)^2\;.
\end{equation}
Each individual factor can be engineered using the following unitary 5-brane web
\begin{equation}
    \begin{array}{c}
         \begin{scriptsize}
         \begin{tikzpicture}[scale=.8]
             \draw[thick](0.5,0.5)--(2.5,2.5);
             \draw[thick](3,3)--(5,5);
             \draw[thick](5,5)--(8,5);
             \draw[thick](2,5)--(5,5);
             \draw[thick](5,8)--(5,5);
             \node[7brane]at(0.5,0.5){};
             \node[7brane]at(1.5,1.5){};
             \node[7brane]at(2.5,2.5){};
             \node[7brane]at(3,3){};
             \node[7brane]at(4,4){};
             \node[7brane]at(5,6.5){};
             \node[7brane]at(5,8){};
             \node[7brane]at(6.5,5){};
             \node[7brane]at(8,5){};
             \node[7brane]at(2,5){};
             \node[7brane]at(3.5,5){};
             \node at (2.75,2.75){$\udots$};
             \node[label=below:{1}] at(1.3,1.3){};
             \node[label=below:{2}] at(2.3,2.3){};
             \node at(4.2,3.5){$2N+1$};
             \node at(5.2,4.5){$2N+2$};
             \node[label=below:{$N+2$}] at(7.25,5.7){};
             \node[label=below:{$2N+4$}] at(5.75,5.7){};
             \node[label=above:{$2$}] at(4.25,4.9){};
             \node[label=above:{$1$}] at(2.75,4.9){};
             \node[label=right:{$N+1$}] at(5,7.55){};
             \node[label=right:{$2N+2$}] at(5,6.05){};
         \end{tikzpicture}
         \end{scriptsize}
    \end{array}\;.
\end{equation}
The magnetic quiver one obtains from this unitary web leads us to the following conjecture
\begin{equation}\MQ_{7,7}=\MQ_7^2=
    \left(\begin{array}{c}
         \begin{scriptsize}
         \begin{tikzpicture}
             \node[label=below:{1}][u](1){};
             \node[label=below:{2}][u](2)[right of=1]{};
             \node[label=below:{$N+2$}][u](N+1)[right of=2]{};
             \node[label=below:{$2N+2$}][u](2N)[right of=N+1]{};
             \node[label=above:{$N+1$}][u](N)[above of=2N]{};
             \node[label=below:{$2N+1$}][u](2N-1)[right of=2N]{};
             \node (dots)[right of=2N-1]{$\cdots$};
             \node[label=below:{$1$}][u](1')[right of=dots]{};
             \draw(1)--(2);
             \draw(2)--(N+1);
             \draw(N+1)--(2N);
             \draw(2N)--(N);
             \draw(2N)--(2N-1);
             \draw(2N-1)--(dots);
             \draw(dots)--(1');
         \end{tikzpicture}
         \end{scriptsize}
    \end{array}\right)^2\;.
\end{equation}
The HWG for the unitary quiver appearing above was conjectured in \cite{Ferlito:2017xdq}. We will use this result and square it to obtain the HWG for the OSp quiver MQ$_{7,7}$ \eqref{MQ77 OSp}: 
\begin{align}
    \HWG_{7,7}&=\PE\left[\sum_{i=1}^{N+1}\mu_{2i}t^{2i}+t^4+\nu^2t^2+\nu\mu_{2N+4}(t^{N+1}+t^{N+3})+\mu_{2N+4}^2t^{2N+4}-\nu^2\mu_{2N+4}t^{2N+6}\right]\nonumber \\
    &\times\PE\left[\sum_{i=1}^{N+1}\lambda_{2i}t^{2i}+t^4+\rho^2t^2+\rho\lambda_{2N+4}(t^{N+1}+t^{N+3})+\lambda_{2N+4}^2t^{2N+4}-\rho^2\lambda_{2N+4}t^{2N+6}\right] ~.
    \end{align}
We further point out an interesting fact about this theory, namely the existence of a 4d $\mathcal{N}=2$ theory with very similar properties. As noticed in \cite{Chacaltana:2011ze}, there is one class-S theory of $D_5$ type in which a single three-punctured sphere realizes a product SCFT, where both factors are the $E_7$ Minahan-Nemeschansky theory \cite{Minahan:1996cj}. We recall that the $E_7$ Minahan-Nemeschansky theory is a 4d $\mathcal{N}=2$ SCFT of rank $1$, with flavor symmetry group $E_7$, and central charges
\begin{equation}
a_{E_7}=\dfrac{59}{24}, \qquad c_{E_7}=\dfrac{19}{6}.
\end{equation}

We report the partitions labeling the punctures in table \ref{tab::E7E7}, together with their contribution to the effective number of hypermutiplets and vector multiplets.

\begin{table}[h]
	\begin{center}
		\begin{tabular}{|c|c|c|}
			\hline
			Nahm partition	&  $(\delta n_h, \delta n_v)$ \\
			\hline
			\hline
			$[1^{10}]$  & $(240,220)$  \\
			$[5^2]$  &  $(104,102)$ \\
			$[3^2,1^4]$  &  $(184,177)$ \\
			\hline
		\end{tabular}
	\end{center}
\caption{Table containing the data defining the punctures for the 4d $E_7\times E_7$ theory.}
	\label{tab::E7E7}
\end{table}

From this data it is easy to compute the central charges $a$ and $c$ of this theory, finding 
\begin{equation}
a_{E_7\times E_7}=\dfrac{59}{12}, \qquad c_{E_7\times E_7}=\dfrac{19}{3}
\end{equation}
as it should be for two copies of the $E_7$ MN theory.

By applying the procedure to write the 3d mirror for this theory we find that the full puncture $[1^{10}]$ is associated to the quiver tail \eqref{leg_full10}, the puncture $[3^2,1^4]$ is associated to the quiver tail \eqref{leg_331111}, and the puncture $[5^2]$ is associated to the quiver tail \eqref{leg_55}, 

\begin{equation}
\label{leg_full10}
\begin{array}{c}
\begin{scriptsize}
\begin{tikzpicture}
\node[label=below:{2}][so](so2){};
\node[label=below:{$2$}][sp](sp2)[right of=so2]{};
\node[label=below:{$4$}][so](so4)[right of=sp2]{};
\node[label=below:{$4$}][sp](sp4)[right of=so4]{};
\node[label=below:{$6$}][so](so6)[right of=sp4]{};
\node[label=below:{$6$}][sp](sp6)[right of=so6]{};
\node[label=below:{$8$}][so](so8)[right of=sp6]{};
\node[label=below:{$8$}][sp](sp8)[right of=so8]{};
\node[label=below:{$10$}][sof](so10)[right of=sp8]{};
\draw(so2)--(sp2);
\draw(sp2)--(so4);
\draw(so4)--(sp4);
\draw(sp4)--(so6);
\draw(so6)--(sp6);
\draw(sp6)--(so8);
\draw(so8)--(sp8);
\draw(sp8)--(so10);
\end{tikzpicture}
\end{scriptsize}
\end{array}
\end{equation}

\begin{equation}
\label{leg_331111}
\begin{array}{c}
\begin{scriptsize}
\begin{tikzpicture}
\node[label=below:{2}][so](so2){};
\node[label=below:{$2$}][sp](sp2)[right of=so2]{};
\node[label=below:{$4$}][so](so4)[right of=sp2]{};
\node[label=below:{$8$}][sp](sp6)[right of=so4]{};
\node[label=below:{$10$}][sof](so10)[right of=sp6]{};
\draw(so2)--(sp2);
\draw(sp2)--(so4);
\draw(so4)--(sp6);
\draw(sp6)--(so10);
\end{tikzpicture}
\end{scriptsize}
\end{array}
\end{equation}

\begin{equation}
\label{leg_55}
\begin{array}{c}
\begin{scriptsize}
\begin{tikzpicture}
\node[label=below:{$4$}][sp](sp4){};
\node[label=below:{$10$}][sof](so10)[right of=sp4]{};
\draw(sp4)--(so10);
\end{tikzpicture}
\end{scriptsize}
\end{array} .
\end{equation}

Gluing the three tails together results in the magnetic quiver for the $5$d $E_7\times E_7$ depicted in (\ref{MQ77 OSp}) for $N=1$. Therefore the magnetic quiver of the 5d $E_7\times E_7$ theory is the 3d mirror theory of the 4d $E_7\times E_7$ theory above described. It is then tempting to conjecture that the 5d $E_7\times E_7$ theory reduces to 4d to this $D_5$ type class-S theory, giving two copies of $E_7$ Minahan-Nemeschansky.

Having derived the magnetic quiver for the $E_7\times E_7$ sequence from the brane web, for any $N\in \mathbb{N}$, we can use the same argument as the paragraphs above to conjecture that all class-S theories of $D_{2N+3}$ type given by a three punctured sphere with regular punctures given by $[1^{4N+6}], [2N+1,2N+1,1^4], [2N+3,2N+3]$ will be a factorized SCFT.
We conjecture that it will decompose into two copies of three punctured $A_{2N+1}$ spheres, with regular punctures given by $[1^{2N+2}],[N+1,N+1],[N,N,1,1]$.
It will be interesting to further check this proposal.

\subsection{An outlier: the \texorpdfstring{$E_8\times E_8$}{TEXT} theory}
\label{E8E8section}

While not explicitly written\footnote{But surely noticed by the authors of such paper. See for example \cite{Distler:2017xba} and \cite{Distler:2018gbc} for discussions about product SCFTs in class-S.} in \cite{Chacaltana:2011ze}, it is easy to use the methods of such paper to find a choice of punctures in the $D_6$ theory, such that we realize the product of two copies of the $E_8$ Minahan-Nemeschansky theory \cite{Minahan:1996cj}. We recall that the $E_8$ Minahan-Nemeschansky theory is a 4d $\mathcal{N}=2$ SCFT of rank $1$, with flavor symmetry group $E_8$, and central charges
\begin{equation}
a_{E_8}=\dfrac{95}{24}, \qquad c_{E_8}=\dfrac{31}{6}.
\end{equation}

We report the partitions labeling the punctures which we believe engineer this product SCFT in table \ref{tab::E8E8}, together with their contribution to the effective number of hypermutiplets and vector multiplets.

\begin{table}[h]
	\begin{center}
		\begin{tabular}{|c|c|c|}
			\hline
			Nahm partition	&  $(\delta n_h, \delta n_v)$ \\
			\hline
			\hline
			$[1^{12}]$  & $(440,410)$  \\
			$[3,1^{9}]$  &  $(400,380)$ \\
			$[9,1^{3}]$  &  $(120,118)$ \\
			\hline
		\end{tabular}
	\end{center}
\caption{Table containing the data defining the punctures for the 4d $E_8\times E_8$ theory.}
	\label{tab::E8E8}
\end{table}

As a check that such 4d theory is really the product of two copies of the $E_8$ Minahan-Nemeschansky theory, we compute the central charges from the data defining the punctures. We get
\begin{equation}
a_{E_8\times E_8}=\dfrac{95}{12}, \qquad c_{E_8\times E_8}=\dfrac{31}{3}
\end{equation}
as it should be for two copies of the $E_8$ MN theory. We also check that there exist no other choice of three punctures, in the $D_6$ theory, that realizes these correct central charges.

By applying the procedure to write the 3d mirror for this theory we find that the full puncture $[1^{12}]$ is associated to the quiver tail \eqref{leg_full12}, the puncture $[3,1^9]$ is associated to the quiver tail \eqref{leg_31111111111}, and the puncture $[9,1^3]$ is associated to the quiver tail \eqref{leg_9111}, 
\begin{equation}
\label{leg_full12}
\begin{array}{c}
\begin{scriptsize}
\begin{tikzpicture}
\node[label=below:{2}][so](so2){};
\node[label=below:{$2$}][sp](sp2)[right of=so2]{};
\node[label=below:{$4$}][so](so4)[right of=sp2]{};
\node[label=below:{$4$}][sp](sp4)[right of=so4]{};
\node[label=below:{$6$}][so](so6)[right of=sp4]{};
\node[label=below:{$6$}][sp](sp6)[right of=so6]{};
\node[label=below:{$8$}][so](so8)[right of=sp6]{};
\node[label=below:{$8$}][sp](sp8)[right of=so8]{};
\node[label=below:{$10$}][so](so10)[right of=sp8]{};
\node[label=below:{$10$}][sp](sp10)[right of=so10]{};
\node[label=below:{$12$}][sof](so12)[right of=sp10]{};
\draw(so2)--(sp2);
\draw(sp2)--(so4);
\draw(so4)--(sp4);
\draw(sp4)--(so6);
\draw(so6)--(sp6);
\draw(sp6)--(so8);
\draw(so8)--(sp8);
\draw(sp8)--(so10);
\draw(so10)--(sp10);
\draw(sp10)--(so12);
\end{tikzpicture}
\end{scriptsize}
\end{array}
\end{equation}

\begin{equation}
\label{leg_31111111111}
\begin{array}{c}
\begin{scriptsize}
\begin{tikzpicture}
\node[label=below:{2}][so](so2){};
\node[label=below:{$2$}][sp](sp2)[right of=so2]{};
\node[label=below:{$4$}][so](so4)[right of=sp2]{};
\node[label=below:{$4$}][sp](sp4)[right of=so4]{};
\node[label=below:{$6$}][so](so6)[right of=sp4]{};
\node[label=below:{$6$}][sp](sp6)[right of=so6]{};
\node[label=below:{$8$}][so](so8)[right of=sp6]{};
\node[label=below:{$8$}][sp](sp8)[right of=so8]{};
\node[label=below:{$12$}][sof](so12)[right of=sp8]{};
\draw(so2)--(sp2);
\draw(sp2)--(so4);
\draw(so4)--(sp4);
\draw(sp4)--(so6);
\draw(so6)--(sp6);
\draw(sp6)--(so8);
\draw(so8)--(sp8);
\draw(sp8)--(so12);
\end{tikzpicture}
\end{scriptsize}
\end{array}
\end{equation}

\begin{equation}
\label{leg_9111}
\begin{array}{c}
\begin{scriptsize}
\begin{tikzpicture}
\node[label=below:{2}][so](so2){};
\node[label=below:{$2$}][sp](sp2)[right of=so2]{};
\node[label=below:{$12$}][sof](so12)[right of=sp2]{};
\draw(so2)--(sp2);
\draw(sp2)--(so12);
\end{tikzpicture}
\end{scriptsize}
\end{array} .
\end{equation}

Gluing the three tails together results in the quiver depicted in (\ref{E8E8quiver}). Given the similarity of this case to the previous cases of $E_6\times E_6$ and $E_7\times E_7$ theory, discussed respectively in sections \ref{E6E6section} and \ref{E7E7section}, it is natural to pose the question whether there exist a 5d $E_8\times E_8$ theory, whose magnetic quiver coincides with the one of (\ref{E8E8quiver}), which we derived here from 3d mirror symmetry applied to the class-S construction of the 4d $E_8\times E_8$ theory,

\begin{equation}
\label{E8E8quiver}
\begin{array}{c}
    \begin{scriptsize}
    \begin{tikzpicture}
    \node[label=below:{2}][so](so2){};
    \node[label=below:{2}][sp](sp2)[right of=so2]{};
    \node (dots)[right of=sp2]{$\cdots$};
    \node[label=below:{10}][so](so10)[right of=dots]{};
    \node[label=below:{10}][sp](sp10)[right of=so10]{};
    \node[label=below:{12}][so](so12)[right of=sp10]{};
    \node[label=below:{8}][sp](sp8)[right of=so12]{};
    \node[label=below:{8}][so](so8)[right of=sp8]{};
    \node (dots')[right of=so8]{$\cdots$};
    \node[label=below:{2}][sp](sp2')[right of=dots']{};
    \node[label=below:{2}][so](so2')[right of=sp2']{};
    \node[label=left:{2}][sp](sp2'')[above of=so12]{};
    \node[label=left:{2}][so](so2'')[above of=sp2'']{};
    \draw(so2)--(sp2);
    \draw(sp2)--(dots);
    \draw(dots)--(so10);
    \draw(so10)--(sp10);
    \draw(sp10)--(so12);
    \draw(so12)--(sp8);
    \draw(sp8)--(so8);
    \draw(so8)--(dots');
    \draw(dots')--(sp2');
    \draw(sp2')--(so2');
    \draw(so2'')--(sp2'');
    \draw(sp2'')--(so12);
    \end{tikzpicture}
    \end{scriptsize}
    \end{array} \; .
\end{equation}
We would like to mention that we were not able to compute the Coulomb branch Hilbert series of this quiver. It would be interesting to verify the matching of the Hilbert series with that of the unitary quiver as in the other cases.
\section{Discussion}\label{discussion}
In this paper we studied 3d $\mathcal{N}=4$ OSp quivers whose moduli space of vacua is comprised of two decoupled sectors. These OSp quivers were derived, using brane webs with O5-planes, as the magnetic quivers for the infinite gauge coupling limit of 5d $\mathcal{N}=1$ gauge theories whose gauge group is a product of SO(4) factors and contain matter hypermultiplets transforming as spinors of opposite chirality under the gauge group factors. We argued for the proposed factorisation, exploiting an accidental isomorphism between the Lie algebra of SO(4) and SU(2)$\times$ SU(2), by rewriting the gauge theory in terms of the group SU(2) and then taking the infinite coupling limit on both sides. The resulting theory in terms of SU(2) gauge groups is generically comprised of two decoupled sectors, each of which we also engineered using ordinary brane webs, without O5-planes. The ordinary brane webs were subsequently used to derive unitary magnetic quivers, which we then used to propose as the components to which the OSp quivers factorise. We further used this correspondence to extract highest weight generating functions for the Coulomb branch Hilbert series of the OSp quivers, relying on existing results for the unitary quivers. In some cases where the unitary quivers had not previously appeared in the literature, we also computed the highest weight generators. In order to test our proposal for the factorisation, and consequently the conjectured highest weight generators for the OSp quivers, we also computed the unrefined Coulomb branch Hilbert series of the OSp quivers directly in a perturbative manner and found an agreement with the proposed HWGs. We further illustrated the matching of the Higgs branch Hilbert series in two cases where we were able to perform the computation exactly on the OSp side. These too were in agreement with the results of the Higgs branch Hilbert series of the unitary side.

Although the higher dimensional intuition has led us to derive these results. It begs the question, whether a truly three-dimensional logic can be used to argue for or provide an explanation for the factorisation property of these OSp quivers. Moreover, all the quantitative checks performed in this paper probe the moduli space of the quivers studied. It would be interesting to ask whether the relationship between the OSp and unitary quivers in this paper are full-fledged dualities, or just a formal relation between their moduli spaces of vacua. One potential check that can be performed to illuminate this question would be to compute other observables, such as the superconformal index, or the three-sphere partition function for the theories in question.

\acknowledgments
We thank Antoine Bourget, Julius Eckhard, Sakura Schafer-Nameki and Zhenghao Zhong for stimulating questions, discussion and correspondence. SSK thanks APCTP, KIAS and POSTECH for his visit where part of this work is done. 
The work of HH is supported in part by JSPS KAKENHI Grant Number JP18K13543. 
FY is supported by the NSFC grant No. 11950410490, by Fundamental Research Funds for the Central Universities A0920502051904-48, by Start-up research grant A1920502051907-2-046, 
in part by NSFC grant No. 11501470 and No. 11671328, and by Recruiting Foreign Experts Program No. T2018050 granted by SAFEA. F.C. is supported by STFC consolidated grant ST/T000708/1. MA is supported by STFC grant ST/S505778/1. SD is supported by the NSFC grants No. 12050410249 and No. 11975158. 
\vspace{1cm}
\appendix
\section{Higgs branch of \texorpdfstring{$E_1\times E_1$}{TEXT} sequence}\label{appendixA}
In this appendix we give a derivation of the formula \eqref{MQ11 HB formula}, for the Higgs branch Hilbert series of $\MQ_{1,1}$ \eqref{MQ11}. The Higgs branch of the quiver \eqref{MQ11} can be computed using the gluing technique, following the discussion in appendix A of \cite{Hanany:2011db}. The first step is to break down the OSp quiver in \eqref{MQ11} into pieces as follows
\begin{equation}
\begin{array}{c}
     \begin{scriptsize}
     \begin{tikzpicture}
     \node[label=below:{1}][u](1){};
     \node (dots)[right of=1]{$\cdots$};
     \node[label=below:{$2N-2$}][u](2N-2)[right of=dots]{};
     \node[label=below:{$2N-1$}][uf](2N-1)[right of=2N-2]{};
     \node[label=below:{$2N-1$}][uf](2N-1')[right of=2N-1]{};
     \node[label=below:{$2N$}][spf](sp2N)[right of=2N-1']{};
     \node[label=below:{$2N$}][spf](sp2N')[right of=sp2N]{};
     \node[label=below:{$4$}][sof](so4)[right of=sp2N']{};
     \draw(1)--(dots);
     \draw(dots)--(2N-2);
     \draw(2N-2)--(2N-1);
     \draw(2N-1')--(sp2N);
     \draw(sp2N')--(so4);
     \end{tikzpicture}
     \end{scriptsize}
\end{array}\;.
\end{equation}
The Higgs Branch Hilbert series of the original quiver \eqref{MQ11} is then obtained by taking the product of the Hilbert series of the individual factors above, together with the gluing factors associated with the U$(2N-1)$ and USp$(2N)$ nodes which are to be gauged, all integrated with the appropriate Haar measure for the aforementioned gauge groups
\begin{equation}\label{HB MQ11 initial integral}
    \int d\mu_{\text{U}(2N-1)}\int d\mu_{C_N}H_{T\left[\text{SU}(2N-1)\right]}H_{\text{glue}}^{(2N-1)}H_{\left[2N-1\right]-\left[C_N\right]}H_{\text{glue}}^{(C_N)}H_{\left[C_N\right]-\left[D_2\right]}\;.
\end{equation}
The individual pieces in the above integrand are as follows
\begin{equation}
    \begin{split}
        H_{T\left[\text{SU}(2N-1)\right]}&=\prod_{q=2}^{2N-1}(1-t^{2q})\PE\left[\chi_{\left[1,0,\cdots,0,1\right]}^{\text{SU}(2N-1)}t^2\right]\;,\qquad H_{\left[C_N\right]-\left[D_2\right]}=\PE\left[4\chi_{\left[1,0,\cdots,0\right]}^{C_N}t\right]\;,\\
        H_{\left[2N-1\right]-\left[C_N\right]}&=\PE\left[\left(\chi_{\left[1,0,\cdots,0\right]}^{\text{SU}(2N-1)}q+\chi_{\left[0,\cdots,0,1\right]}^{\text{SU}(2N-1)}q^{-1}\right)\chi_{\left[1,0,\cdots,0\right]}^{C_N}t\right]\;,\\
        H_{\text{glue}}^{(2N-1)}&=\frac{(1-t^2)}{\PE\left[\chi_{\left[1,0,\cdots,0,1\right]}^{\text{SU}(2N-1)}t^2\right]}\;,\qquad H_{\text{glue}}^{(C_N)}=\frac{1}{\PE\left[\chi_{\left[2,0,\cdots,0\right]}^{C_N}t^2\right]}\;.
    \end{split}
\end{equation}
Plugging these into the integral \eqref{HB MQ11 initial integral} we arrive at
\begin{equation}\label{A4}
    \prod_{q=1}^{2N-1}(1-t^{2q})\int d\mu_{\text{U}(2N-1)}\int d\mu_{C_N} H_{\left[2N-1\right]-\left[C_N\right]}H_{\text{glue}}^{(C_N)}H_{\left[C_N]\right]-\left[D_2\right]}\;.
\end{equation}
Next we perform the integral over the U$(2N-1)$ group, by counting the gauge invariants of the free theory $\left[2N-1\right]-\left[C_N\right]$, which is the only part of the integrand that sees this integral (see appendix A of \cite{Hanany:2011db} for more details)
\begin{equation}
    \int d\mu_{\text{U}(2N-1)}H_{\left[2N-1\right]-\left[C_N\right]}=(1-t^{4N})\PE\left[\left(1+\chi_{\left[2,0,\cdots,0\right]}^{C_N}+\chi_{\left[0,1,\cdots,0\right]}^{C_N}\right)t^2\right].
\end{equation}
Substituting this back into \eqref{A4} one obtains the desired formula \eqref{MQ11 HB formula} for the Higgs branch Hilbert series of $\MQ_{1,1}$ \eqref{MQ11}.

\section{List of HWG of unitary quivers}\label{appendixB}
In this appendix we tabulate the highest weight generating functions of all the unitary quivers appearing in the previous sections.

\captionsetup{width=15cm}
\begin{longtable}{|c|C{2.5cm}|C{6.5cm}|}
        \caption{List of the HWG for the unitary quivers appearing in the earlier sections. The corresponding fugacities are denoted by subscripts in the symmetry groups. Note that for $N=1$, there is a possible enhancement in the symmetry which can be read from the set of balanced nodes in the quivers for $N=1$.}
			\label{TableHWGUnitaryQuivers} \\ \hline 
					Quiver&Symmetry&PL[HWG]\\\hline 
      $\begin{array}{c}
             \begin{scriptsize}
             \begin{tikzpicture}
             \node[label=below:{1}][u](2){};
             \node (dots)[right of=2]{$\cdots$};
             \node[label=below:{$N$}][u](N)[right of=dots]{};
             \node (dots')[right of=N]{$\cdots$};
             \node[label=below:{1}][u](2')[right of=dots']{};
             \node[label=above:{1}][u](11)[above of=N]{};
             \draw(2)--(dots);
             \draw(dots)--(N);
             \draw[double distance=2pt](N)--(11);
             \draw(N)--(dots');
             \draw(dots')--(2');
             \end{tikzpicture}
             \end{scriptsize}
        \end{array}$ & SU($2N$)$_{\mu}$ & $\sum\limits_{k=1}^N\mu_k\,\mu_{2N-k}\,t^{2k}$ \\\hline 
				$\begin{array}{c}
				    \begin{scriptsize} 
						\begin{tikzpicture}
             \node[label=below:{1}][u]{};
             \node (dots)[above right of=1]{$\udots$};
             \node[label=below:{$N$}][u](N)[above right of=dots]{};
             \node (dots')[below right of=N]{$\ddots$};
             \node[label=below:{1}][u](11)[below right of=dots']{};
             \node[label=above:{1}][u](u1)[above left of=N]{};
             \node[label=above:{1}][u](u11)[above right of=N]{};
             \draw(1)--(dots);
             \draw(dots)--(N);
             \draw(N)--(dots');
             \draw(dots')--(11);
             \draw(N)--(u1);
             \draw(N)--(u11);
             \draw(u1)--(u11);
         \end{tikzpicture}
         \end{scriptsize}
        \end{array}$ & SU($2N$)$_{\mu}$ $\times$ U(1)$_{q}$ & $t^2+\left(q+q^{-1}\right)\mu_N\,t^{N+1}
   +\sum\limits_{k=1}^{N}\mu_k\,\mu_{2N-k}\,t^{2k} -\mu_N^2\,t^{2N+2} $ \\\hline 
		$\begin{array}{c}
		\begin{scriptsize}
              \begin{tikzpicture}
                  \node[label=below:{1}][u](1){};
                  \node(dots)[above of=1]{$\vdots$};
                  \node[label=left:{$N-1$}][u](N-1)[above of=dots]{};
                  \node[label=below:{$N-1$}][u](N-1')[right of=N-1]{};
                  \node[label=right:{$N-1$}][u](N-1'')[right of=N-1']{};
                  \node (dots')[below of=N-1'']{$\vdots$};
                  \node[label=below:{1}][u](1')[below of=dots']{};
                  \node[label=above:{1}][u](u1)[above of=N-1]{};
                  \node[label=above:{1}][u](u11)[above of=N-1'']{};
                  \draw(1)--(dots);
                  \draw(dots)--(N-1);
                  \draw(N-1)--(N-1');
                  \draw(N-1')--(N-1'');
                  \draw(N-1'')--(dots');
                  \draw(dots')--(1');
                  \draw(N-1)--(u1);
                  \draw(N-1'')--(u11);
                  \draw[double distance=2pt](u1)--(u11);
              \end{tikzpicture}
              \end{scriptsize}
             \end{array}$ & SU($2N$)$_{\mu}$ $\times$ U(1)$_{q}$ & $t^2+\left(\mu_{N+1}q+\mu_{N-1}q^{-1}\right)\,t^{N+1}+\sum\limits_{k=1}^{N-1}\mu_k\,\mu_{2N-k}\,t^{2k} -\mu_{N+1}\,\mu_{N-1}\,t^{2N+2} $ \\\hline 
		$\begin{array}{c}
        \begin{scriptsize}
             \begin{tikzpicture}
                 \node[label=above:{1}][u](1){};
                 \node[label=below:{$N$}][u](3)[below left of=1]{};
                 \node[label=below:{$N$}][u](4)[below right of=1]{};
                  \node (dots)[below right of=4]{$\ddots$};
                   \node (dots')[below left of=3]{$\udots$};
                  \node[label=below:{1}][u](u1)[below left of=dots']{};
                  \node[label=below:{1}][u](u11)[below right of=dots]{};
                  \draw(3)--(dots');
                  \draw(4)--(dots);
                  \draw(dots)--(u11);
                  \draw(dots')--(u1);
                 \draw(1)--(4);
                 \draw(1)--(3);
                 \draw(3)--(4);
             \end{tikzpicture}
    \end{scriptsize}\end{array}$ & SU($2N+1$)$_{\mu}$ & $\sum\limits_{k=1}^N\mu_k\,\mu_{2N+1-k}\,t^{2k}$ \\ \hline
				$\begin{array}{c}
        \begin{scriptsize}
             \begin{tikzpicture}
                 \node[label=above:{1}][u](1){};
                 \node[label=left:{1}][u](2)[below left of=1]{};
                 \node[label=right:{1}][u](5)[below right of=1]{};
                 \node[label=left:{$N$}][u](3)[below of=2]{};
                 \node[label=right:{$N$}][u](4)[below of=5]{};
                  \node (dots)[below of=4]{$\vdots$};
                   \node (dots')[below of=3]{$\vdots$};
                  \node[label=below:{1}][u](u1)[below of=dots']{};
                  \node[label=below:{1}][u](u11)[below of=dots]{};
                  \draw(3)--(dots');
                  \draw(4)--(dots);
                  \draw(dots)--(u11);
                  \draw(dots')--(u1);
                 \draw(1)--(2);
                 \draw(2)--(3);
                 \draw(3)--(4);
                 \draw(4)--(5);
                 \draw(5)--(1);
             \end{tikzpicture}
    \end{scriptsize}    \end{array}$ & SU($2N+1$)$_{\mu}$ $\times$ SU(2)$_{\nu}$ $\times$ U(1)$_{q}$ & $\sum\limits_{k=1}^N\mu_k\,\mu_{2N+1-k}\,t^{2k} + (\nu^2+1)t^2 + \nu(\mu_N q+\mu_{N+1} q^{-1})t^{N+1}-\nu^2 \mu_N \mu_{N+1}t^{2N+2}$\\ \hline 
		$\begin{array}{c}
         \begin{scriptsize}
         \begin{tikzpicture}
         \node[label=above:{1}][u](u1){};
         \node[label=left:{$N-1$}][u](N-1)[above left of=u1]{};
         \node[label=right:{$N-1$}][u](N-1')[above right of=u1]{};
         \node[label=above:{$N-1$}][u](N-1'')[above of=N-1]{};
         \node[label=above:{$N-1$}][u](N-1''')[above of=N-1']{};
         \node[label=below:{1}][u](u11)[below of=u1]{};
         \node (dots)[below left of=N-1]{$\udots$};
         \node[label=below:{1}][u](1)[below left of=dots]{};
         \node (dots')[below right of=N-1']{$\ddots$};
         \node[label=below:{1}][u](1')[below right of=dots']{};
         \draw(1)--(dots);
         \draw(dots)--(N-1);
         \draw(N-1''')--(N-1');
         \draw(N-1)--(N-1'');
         \draw(N-1')--(u1);
         \draw(N-1)--(u1);
         \draw(N-1'')--(N-1''');
         \draw(N-1')--(dots');
         \draw(dots')--(1');
         \draw[double distance=2pt](u1)--(u11);
         \end{tikzpicture}
         \end{scriptsize}
    \end{array}$ & SU($2N+1$)$_{\mu}$ $\times$ SU(2)$_{\nu}$ & $\sum\limits_{k=1}^{N-1} \mu_k \mu_{2N+1-k}\,t^{2k} +\nu^2 t^2$ \\ \hline
		$\begin{array}{c}
             \begin{scriptsize}
             \begin{tikzpicture}
                 \node[label=below:{1}][u](1){};
                 \node (dots)[right of=1]{$\cdots$};
                 \node[label=below:{$N+1$}][u](N)[right of=dots]{};
                 \node (dots')[right of=N]{$\cdots$};
                 \node[label=below:{1}][u](1')[right of=dots']{};
                 \node[label=above:{2}][u](2)[above of=N]{};
                 \node[label=above:{1}][u](u1)[right of=2]{};
                 \node[label=above:{1}][u](u1')[left of=2]{};
                 \draw(1)--(dots);
                 \draw(dots)--(N);
                 \draw(N)--(dots');
                 \draw(dots')--(1');
                 \draw(N)--(2);
                 \draw(2)--(u1);
                 \draw(2)--(u1');
                 
             \end{tikzpicture}
             \end{scriptsize}
        \end{array}$ & SU($2N+2$)$_{\mu}$ $\times$ SU(2)$_{\nu_1}$ $\times$ SU(2)$_{\nu_2}$ & $\sum\limits_{k=1}^{N+1} \mu_k \mu_{2N+2-k}\,t^{2k} +(\nu_1^2+\nu_2^2)t^2+t^4 + \nu_1 \nu_2 \mu_{N+1}(t^{N+1}+t^{N+3}) - \nu_1^2 \nu_2^2 \mu_{N+1}^2 t^{2N+6}$ \\ \hline
				$\begin{array}{c}
        \begin{scriptsize}
        \begin{tikzpicture}
         \node[label=below:{1}][u](1){};   
         \node (dots)[left of=1]{$\cdots$};
         \node[label=below:{$2N$}][u](2N)[left of=dots]{};
         \node[label=below:{$2N$}][u](2N')[left of=2N]{};
         \node[label=below:{$N$}][u](N)[left of=2N']{};
         \node[label=above:{$N$}][u](N+1)[above of=2N']{};
         \node[label=above:{$1$}][u](1')[above of=2N]{};
         \draw(1)--(dots);
         \draw(dots)--(2N);
         \draw(2N)--(2N');
         \draw(2N')--(N+1);
         \draw(2N')--(N);
         \draw(2N)--(1');
        \end{tikzpicture}
        \end{scriptsize}
    \end{array}$ & SO($4N+6$)$_{\mu}$& $\sum\limits_{k=1}^N \mu_{2k} t^{2k}$ \\ \hline
		$\begin{array}{c}
         \begin{scriptsize}
         \begin{tikzpicture}
             \node[label=below:{1}][u](1){};
             \node (dots)[left of=1]{$\cdots$};
             \node[label=below:{$2N+1$}][u](2N-1)[left of=dots]{};
             \node[label=below:{$N+1$}][u](N)[left of=2N-1]{};
             \node[label=right:{$N+1$}][u](N')[above of=2N-1]{};
             \node[label=below:{$1$}][u](u1)[left of=N]{};
             \node[label=right:{$1$}][u](u1')[above of=N']{};
             \draw(1)--(dots);
             \draw(dots)--(2N-1);
             \draw(2N-1)--(N);
             \draw(N)--(u1);
             \draw(2N-1)--(N');
             \draw(N')--(u1');
         \end{tikzpicture}
         \end{scriptsize}
    \end{array}$ & SO($4N+6$)$_{\mu}$ $\times$ U(1)$_{q}$ & $\sum\limits_{k=1}^N \mu_{2k} t^{2k}+\left(\mu_{2N+2}\,q+\mu _{2N+3}\,q^{-1}\right)t^{N+1}+t^2$ \\ \hline
		$\begin{array}{c}
         \begin{scriptsize}
         \begin{tikzpicture}
             \node[label=above:{1}][u](1){};
             \node[label=below:{2}][u](2)[below of=1]{};
             \node[label=below:{$N+2$}][u](N+1)[right of=2]{};
             \node[label=below:{$2N+2$}][u](2N)[right of=N+1]{};
             \node[label=above:{$N+1$}][u](N)[above of=2N]{};
             \node (dots)[right of=2N]{$\cdots$};
             \node[label=below:{$1$}][u](1')[right of=dots]{};
             \draw(1)--(2);
             \draw(2)--(N+1);
             \draw(N+1)--(2N);
             \draw(2N)--(N);
             \draw (2N)--(dots);
             \draw(dots)--(1');
         \end{tikzpicture}
         \end{scriptsize}
    \end{array}$ & SO($4N+8$)$_{\mu}$ $\times$ SU(2)$_{\nu}$ & $\sum\limits_{k=1}^{N+1} \mu_{2k} t^{2k}+t^4+\nu  \mu_{2N+4} \left(t^{N+1}+t^{N+3}\right)+\mu_{2N+4}^2 t^{2N+4}+\nu^2 t^2 - \nu^2 \mu _{2N+4}t^{2N+6}$ \\ \hline
				    \end{longtable}

\section{Unrefined Coulomb branch Hilbert series for low rank OSp quivers}\label{appendixC}
In this appendix, we quote the results for the unrefined Hilbert series computed for the orthosymplectic quivers. These results match with the perturbative results obtained from the HWG listed in the table \ref{TableHWGOSpQuivers}.

\captionsetup{width=15cm}
\begin{longtable}{|c|C{3.6cm}|C{3.2cm}|C{4.8cm}|}
        \caption{The computation of Coulomb branch Hilbert series of orthosymplectic quivers given in main sections for small values of $N$. The full Hilbert series is given by summing over the integer and half-integer sublattices of the magnetic weights.}
			\label{TableHSOSp} \\ \hline 
   & \multicolumn{3}{c|}{Coulomb branch Hilbert series} \\ \cline{2-4}
	  \multirow{-2}{*}{Quiver} & $\vec{m} \in \mathbb{Z}$ & $\vec{m} \in \mathbb{Z}+1/2$ & HS  \\ \hline
		$\MQ_{1,1}\rvert_{N=2}$  & \footnotesize{$1+30 t^2+433 t^4+4070 t^6+28384 t^8+158174 t^{10}+\mathcal{O}(t^{12})$} & \footnotesize{not required} &  \footnotesize{$1+30 t^2+433 t^4+4070 t^6+28384 t^8+158174 t^{10}+\mathcal{O}(t^{12})$} \\ \hline
		$\MQ_{1,1}\rvert_{N=3}$  & \footnotesize{$1+70 t^2+2413 t^4+54670 t^6+917244 t^8+12178110 t^{10}+\mathcal{O}(t^{12})$} & \footnotesize{not required} &  \footnotesize{$1+70 t^2+2413 t^4+54670 t^6+917244 t^8+12178110 t^{10}+\mathcal{O}(t^{12})$} \\ \hline
		
		$\MQ_{1,3}^{(\text{I})}\rvert_{N=2}$  & \footnotesize{$1+31t^2+464t^4+4574t^6+33908t^8+203160t^{10}+\mathcal{O}(t^{12})$} & \footnotesize{$12t^3+320t^5+4188t^7+36488t^9+\mathcal{O}(t^{11})$} &  \footnotesize{$1+31t^2+12t^3+464t^4+320t^5+4574t^6+4188t^7+33908t^8+36488t^9+203160t^{10}+\mathcal{O}(t^{11})$} \\ \hline
		$\MQ_{1,3}^{(\text{I})}\rvert_{N=3}$  & \footnotesize{$1+71t^2+2484t^4+57154t^6+974748t^8+13173046t^{10}+\mathcal{O}(t^{12})$} & \footnotesize{$40t^4+2520t^6+78584t^8+1619760t^{10}+\mathcal{O}(t^{12})$} &  \footnotesize{$1+71t^2+2524t^4+59674t^6+1053332t^8+14792806t^{10}+\mathcal{O}(t^{12})$} \\ \hline
		$\MQ_{1,3}^{(\text{II})}\rvert_{N=2}$  & \footnotesize{$1+31t^2+444t^4+4059t^6+27344t^8+147137t^{10}+\mathcal{O}(t^{12})$} & \footnotesize{$8t^3+200t^5+2432t^7+19560t^9+\mathcal{O}(t^{11})$} &  \footnotesize{$1+31t^2+8t^3+444t^4+200t^5+4059t^6+2432t^7+27344t^8+19560t^9+147137t^{10}+\mathcal{O}(t^{11})$} \\ \hline
		$\MQ_{1,3}^{(\text{II})}\rvert_{N=3}$  & \footnotesize{$1+71t^2+2484t^4+56979t^6+964339t^8+12865508t^{10}+\mathcal{O}(t^{12})$} & \footnotesize{$30 t^4 + 1848 t^6 + 56250 t^8 + 1129770 t^{10}+\mathcal{O}(t^{12})$} &  \footnotesize{$1 + 71 t^2 + 2514 t^4 + 58827 t^6 + 1020589 t^8 + 13995278 t^{10}+\mathcal{O}(t^{12})$} \\ \hline
		$\MQ_{3,3}^{(\text{I})}\rvert_{N=1}$  & \footnotesize{$1 + 8 t^2 + 62 t^4 + 280 t^6 + 1011 t^8 + 2944 t^{10}+\mathcal{O}(t^{12})$} & \footnotesize{$8 t^2 + 56 t^4 + 280 t^6 + 992 t^8 + 2944 t^{10}+\mathcal{O}(t^{12})$} &  \footnotesize{$1 + 16 t^2 + 118 t^4 + 560 t^6 + 2003 t^8 + 5888 t^{10}+\mathcal{O}(t^{12})$} \\ \hline
		$\MQ_{3,3}^{(\text{I})}\rvert_{N=2}$  & \footnotesize{$1 + 32 t^2 + 496 t^4 + 5254 t^6 + 43368 t^8 + 294996 t^{10}+\mathcal{O}(t^{12})$} & \footnotesize{$24 t^3 + 664 t^5 + 9040 t^7 + 82976 t^9+\mathcal{O}(t^{11})$} &  \footnotesize{$1 + 32 t^2 + 24 t^3 + 496 t^4 + 664 t^5 + 5254 t^6 + 9040 t^7 + 
 43368 t^8 + 82976 t^9 + 294996 t^{10}+\mathcal{O}(t^{11})$} \\ \hline
$\MQ_{3,3}^{(\text{II})}\rvert_{N=1}$  & \footnotesize{$1 + 2 t^2 + 11 t^4 + 20 t^6 + 45 t^8 + 70 t^{10}+\mathcal{O}(t^{12})$} & \footnotesize{$4 t^2 + 8 t^4 + 24 t^6 + 40 t^8 + 76 t^{10}+\mathcal{O}(t^{12})$} &  \footnotesize{$1 + 6 t^2 + 19 t^4 + 44 t^6 + 85 t^8 + 146 t^{10}+\mathcal{O}(t^{12})$} \\ \hline
$\MQ_{3,3}^{(\text{II})}\rvert_{N=2}$  & \footnotesize{$1 + 32 t^2 + 456 t^4 + 4104 t^6 + 27490 t^8 + 148792 t^{10}+\mathcal{O}(t^{12})$} & \footnotesize{$16 t^3 + 416 t^5 + 4960 t^7 + 38400 t^9+\mathcal{O}(t^{11})$} &  \footnotesize{$1 + 32 t^2 + 16 t^3 + 456 t^4 + 416 t^5 + 4104 t^6 + 4960 t^7 + 
 27490 t^8 + 38400 t^9 + 148792 t^{10}+\mathcal{O}(t^{11})$} \\ \hline
$\MQ_{3,3}^{(\text{III})}\rvert_{N=1}$  & \footnotesize{$1 + 5 t^2 + 30 t^4 + 94 t^6 + 263 t^8 + 587 t^{10}+\mathcal{O}(t^{12})$} & \footnotesize{$6 t^2 + 26 t^4 + 98 t^6 + 254 t^8 + 596 t^{10}+\mathcal{O}(t^{12})$} &  \footnotesize{$1 + 11 t^2 + 56 t^4 + 192 t^6 + 517 t^8 + 1183 t^{10}+\mathcal{O}(t^{12})$} \\ \hline
$\MQ_{3,3}^{(\text{III})}\rvert_{N=2}$  & \footnotesize{$1 + 32 t^2 + 476 t^4 + 4671 t^6 + 34989 t^8 + 214034 t^{10}+\mathcal{O}(t^{12})$} & \footnotesize{$20 t^3 + 540 t^5 + 6920 t^7 + 58628 t^9+\mathcal{O}(t^{11})$} &  \footnotesize{$1 + 32 t^2 + 20 t^3 + 476 t^4 + 540 t^5 + 4671 t^6 + 6920 t^7 + 
 34989 t^8 + 58628 t^9 + 214034 t^{10}+\mathcal{O}(t^{11})$} \\ \hline
$\MQ_{3',3'}^{(\text{I})}\rvert_{N=1}$  & \footnotesize{$1 + 16 t^2 + 118 t^4 + 560 t^6 + 2003 t^8 + 5888 t^{10}+\mathcal{O}(t^{12})$} & \footnotesize{not required} &  \footnotesize{$1 + 16 t^2 + 118 t^4 + 560 t^6 + 2003 t^8 + 5888 t^{10}+\mathcal{O}(t^{12})$} \\ \hline
$\MQ_{3',3'}^{(\text{I})}\rvert_{N=2}$  & \footnotesize{$1 + 48 t^2 + 1126 t^4 + 17248 t^6 + 194729 t^8 + 1735152 t^{10}+\mathcal{O}(t^{12})$} & \footnotesize{not required} &  \footnotesize{$1 + 48 t^2 + 1126 t^4 + 17248 t^6 + 194729 t^8 + 1735152 t^{10}+\mathcal{O}(t^{12})$} \\ \hline
$\MQ_{3',3'}^{(\text{II})}\rvert_{N=1}$  & \footnotesize{$1 + 6 t^2 + 19 t^4 + 44 t^6 + 85 t^8 + 146 t^{10}+\mathcal{O}(t^{12})$} & \footnotesize{not required} &  \footnotesize{$1 + 6 t^2 + 19 t^4 + 44 t^6 + 85 t^8 + 146 t^{10}+\mathcal{O}(t^{12})$} \\ \hline
$\MQ_{3',3'}^{(\text{II})}\rvert_{N=2}$  & \footnotesize{$1 + 54 t^2 + 1283 t^4 + 18412 t^6 + 185691 t^8 + 1438022t^{10}+\mathcal{O}(t^{12})$} & \footnotesize{not required} &  \footnotesize{$1 + 54 t^2 + 1283 t^4 + 18412 t^6 + 185691 t^8 + 1438022t^{10}+\mathcal{O}(t^{12})$} \\ \hline
$\MQ_{3',3'}^{(\text{III})}\rvert_{N=1}$  & \footnotesize{$1 + 11 t^2 + 56 t^4 + 192 t^6 + 517 t^8 + 1183 t^{10}+\mathcal{O}(t^{12})$} & \footnotesize{not required} &  \footnotesize{$1 + 11 t^2 + 56 t^4 + 192 t^6 + 517 t^8 + 1183 t^{10}+\mathcal{O}(t^{12})$} \\ \hline
$\MQ_{3',3'}^{(\text{III})}\rvert_{N=2}$  & \footnotesize{$1 + 51 t^2 + 1200 t^4 + 17824 t^6 + 191099 t^8 + 1596553t^{10}+\mathcal{O}(t^{12})$} & \footnotesize{not required} &  \footnotesize{$1 + 51 t^2 + 1200 t^4 + 17824 t^6 + 191099 t^8 + 1596553t^{10}+\mathcal{O}(t^{12})$} \\ \hline
$\MQ_{3',4}^{(\text{I})}\rvert_{N=1}$  & \footnotesize{$1 + 20 t^2 + 227 t^4 + 1720 t^6 + 9552 t^8 + 42168 t^{10}+\mathcal{O}(t^{12})$} & \footnotesize{$12 t^2 + 192 t^4 + 1592 t^6 + 9184 t^8 + 41224 t^{10}+\mathcal{O}(t^{12})$} &  \footnotesize{$1 + 32 t^2 + 419 t^4 + 3312 t^6 + 18736 t^8 + 83392 t^{10}+\mathcal{O}(t^{12})$} \\ \hline
$\MQ_{3',4}^{(\text{I})}\rvert_{N=2}$  & \footnotesize{$1 + 52 t^2 + 1327 t^4 + 22500 t^6 + 286968 t^8 + 2939292t^{10}+\mathcal{O}(t^{12})$} & \footnotesize{$40 t^3 + 1780 t^5 + 39140 t^7 + 570120 t^9+\mathcal{O}(t^{11})$} &  \footnotesize{$1 + 52 t^2 + 40 t^3 + 1327 t^4 + 1780 t^5 + 22500 t^6 + 39140 t^7 + 
 286968 t^8 + 570120 t^9 + 2939292 t^{10}+\mathcal{O}(t^{11})$} \\ \hline
$\MQ_{3',4}^{(\text{II})}\rvert_{N=1}$  & \footnotesize{$1 + 15 t^2 + 145 t^4 + 879 t^6 + 3964 t^8 + 14388t^{10}+\mathcal{O}(t^{12})$} & \footnotesize{$12 t^2 + 132 t^4 + 848 t^6 + 3888 t^8 + 14240 t^{10}+\mathcal{O}(t^{12})$} &  \footnotesize{$1 + 27 t^2 + 277 t^4 + 1727 t^6 + 7852 t^8 + 28628t^{10}+\mathcal{O}(t^{12})$} \\ \hline
$\MQ_{3',4}^{(\text{II})}\rvert_{N=2}$  & \footnotesize{$1 + 55 t^2 + 1413 t^4 + 23399 t^6 + 287256 t^8 + 2808576 t^{10}+\mathcal{O}(t^{12})$} & \footnotesize{$40 t^3 + 1900 t^5 + 41680 t^7 + 585260 t^9+\mathcal{O}(t^{11})$} &  \footnotesize{$1 + 55 t^2 + 40 t^3 + 1413 t^4 + 1900 t^5 + 23399 t^6 + 41680 t^7 + 
 287256 t^8 + 585260 t^9 + 2808576 t^{10}+\mathcal{O}(t^{11})$} \\ \hline
$\MQ_{4,4}\rvert_{N=1}$  & \footnotesize{$1 + 24 t^2 + 496 t^4 + 5800 t^6 + 47734 t^8 + 299176t^{10}+\mathcal{O}(t^{12})$} & \footnotesize{$24 t^2 + 480 t^4 + 5800 t^6 + 47616 t^8 + 299176 t^{10}+\mathcal{O}(t^{12})$} &  \footnotesize{$1 + 48 t^2 + 976 t^4 + 11600 t^6 + 95350 t^8 + 598352 t^{10}+\mathcal{O}(t^{12})$} \\ \hline
$\MQ_{4,4}\rvert_{N=2}$  & \footnotesize{$1 + 56 t^2 + 1544 t^4 + 30192 t^6 + 468888 t^8 + 5964152t^{10}+\mathcal{O}(t^{12})$} & \footnotesize{$80 t^3 + 3880 t^5 + 93240 t^7 + 1510680 t^9+\mathcal{O}(t^{11})$} &  \footnotesize{$1 + 56 t^2 + 80 t^3 + 1544 t^4 + 3880 t^5 + 30192 t^6 + 93240 t^7 + 
 468888 t^8 + 1510680 t^9 + 5964152 t^{10}+\mathcal{O}(t^{11})$} \\ \hline
$\MQ_{5,5}\rvert_{N=1}$  & \footnotesize{$1 + 42 t^2 + 1805 t^4 + 42204 t^6 + 693740 t^8 + 8548816 t^{10}+\mathcal{O}(t^{12})$} & \footnotesize{$48 t^2 + 1760 t^4 + 42384 t^6 + 692960 t^8 + 8551232 t^{10}+\mathcal{O}(t^{12})$} &  \footnotesize{$1 + 90 t^2 + 3565 t^4 + 84588 t^6 + 1386700 t^8 + 17100048 t^{10}+\mathcal{O}(t^{12})$} \\ \hline
$\MQ_{5,5}\rvert_{N=2}$  & \footnotesize{$1 + 82 t^2 + 3329 t^4+\mathcal{O}(t^{6})$} & \footnotesize{$160t^3+\mathcal{O}(t^{5})$} &  \footnotesize{$1 + 82 t^2 + 160 t^3 + 3329 t^4+\mathcal{O}(t^{5})$} \\ \hline
$\MQ_{5',5'}\rvert_{N=1}$  & \footnotesize{$1 + 90 t^2 + 3565 t^4 + 84588 t^6 + 1386700 t^8 + 17100048 t^{10}+\mathcal{O}(t^{12})$} & \footnotesize{not required} &  \footnotesize{$1 + 90 t^2 + 3565 t^4 + 84588 t^6 + 1386700 t^8 + 17100048 t^{10}+\mathcal{O}(t^{12})$} \\ \hline
$\MQ_{5',5'}\rvert_{N=2}$  & \footnotesize{$1 + 182 t^2 + 16443 t^4 + 977366 t^6+\mathcal{O}(t^{8})$} & \footnotesize{not required} &  \footnotesize{$1 + 182 t^2 + 16443 t^4 + 977366 t^6+\mathcal{O}(t^{8})$} \\ \hline
$\MQ_{5',6}\rvert_{N=1}$  & \footnotesize{$1 + 91 t^2 + 4118 t^4 + 122828 t^6 + 2660832 t^8 + 44299317t^{10}+\mathcal{O}(t^{12})$} & \footnotesize{$32 t^2 + 2592 t^4 + 97984 t^6 + 2366496 t^8 + 41557344 t^{10}+\mathcal{O}(t^{12})$} &  \footnotesize{$1 + 123 t^2 + 6710 t^4 + 220812 t^6 + 5027328 t^8 + 85856661 t^{10}+\mathcal{O}(t^{12})$} \\ \hline
$\MQ_{5',6}\rvert_{N=2}$  & \footnotesize{$1 + 183 t^2 + 16626 t^4 + 1000427 t^6+\mathcal{O}(t^{8})$} & \footnotesize{$128 t^3 + 21632 t^5+\mathcal{O}(t^{7})$} &  \footnotesize{$1 + 183 t^2 + 128 t^3 + 16626 t^4 + 21632 t^5 + 1000427 t^6+\mathcal{O}(t^{8})$} \\ \hline
$\MQ_{6,6}\rvert_{N=1}$  & \footnotesize{$1 + 92 t^2 + 5696 t^4 + 235812 t^6 + 6925608 t^8 + 153422070 t^{10}+\mathcal{O}(t^{12})$} & \footnotesize{$64 t^2 + 5248 t^4 + 230784 t^6 + 6881472 t^8 + 153102720 t^{10}+\mathcal{O}(t^{12})$} &  \footnotesize{$1 + 156 t^2 + 10944 t^4 + 466596 t^6 + 13807080 t^8 + 306524790 t^{10}+\mathcal{O}(t^{12})$} \\ \hline
$\MQ_{7,7}\rvert_{N=1}$  & \footnotesize{$1 + 138 t^2 + 16303 t^4+\mathcal{O}(t^{6})$} & \footnotesize{$128 t^2 + 16128 t^4+\mathcal{O}(t^{6})$} &  \footnotesize{$1 + 266 t^2 + 32431 t^4+\mathcal{O}(t^{6})$} \\ \hline
	 \end{longtable}
\section{Central charges and 3d mirrors for D-type class-S theories}\label{appendixD}
In this appendix we briefly review two properties of class-S theories of $D_N$ type. We restrict ourself to the case in which the $6$d $(2,0)$ $D_N$ theory is compactified on the sphere, with regular (untwisted) punctures only. We discuss three instances of theories of this type in the main text, in sections \ref{E6E6section}, \ref{E7E7section} and  \ref{E8E8section}. There, we consider examples in which a single three-punctured sphere describes the 4-dimensional version of the $E_n\times E_n$ theory ($n=6,7,8)$, i.e. two copies of $E_n$ Minahan-Nemeschansky. 

The first property that we review in this appendix is the rule for the computation of superconformal central charges $a$ and $c$, giving the data labeling the punctures~\cite{Chacaltana:2011ze}. The second property is the prescription for finding the corresponding 3d $\mathcal{N}=4$ mirror theories~\cite{Benini:2010uu}.

\subsection{Central charges}
\label{appendixC1}

The central charges $a$ and $c$ of a 4d $\mathcal{N}=2$ SCFT are defined via the trace anomaly in a curved background,
\begin{equation}
T_{\mu}^{\ \mu}=\dfrac{c}{16\pi^2}(\mbox{Weyl})^2-\dfrac{a}{16\pi^2}(\mbox{Euler})^2\ .
\end{equation}
For Lagrangian theories, the central charges are related to the number of vector multiples and hypermultiplets by
\begin{equation}
a=\dfrac{5n_v+n_h}{24}, \quad c=\dfrac{2n_v+n_h}{12} \ .
\label{eq:cchargesnhnv}
\end{equation}
For non-Lagrangian theories, formula (\ref{eq:cchargesnhnv}) still holds, but now $n_h$ and $n_v$ are interpreted as an effective number of hypermultiplets and vectormultiplets.
For the subset of theories of our current interest, it holds that
\begin{equation}
\begin{aligned}
n_v=-\dfrac{1}{3}N(16N^2-24N+11)+\sum_{\alpha}\delta n_v^{(\alpha)} \ \ \mbox{,}\\
n_h=-\dfrac{8}{3}N(N-1)(2N-1)+\sum_{\alpha}\delta n_h^{(\alpha)}  \ \ \mbox{,}
\end{aligned}
\label{eq::acformulae}
\end{equation}
where $g$ is the genus of the Riemann surface and $\alpha$ runs over the set of punctures. $\delta n_v^{(\alpha)}$ and $\delta n_h^{(\alpha)}$ are local contributions coming from the $\alpha$-th puncture. Both formulae (\ref{eq::acformulae}) and an algorithmic rule for the computation of $\delta n_h$ and $\delta n_v$ are derived in \cite{Chacaltana:2011ze}. In the same paper, the explicit values of $\delta n_h$ and $\delta n_v$ is listed for all the punctures up to $N=6$. We defer the reader to such paper for further details.

\subsection{3d Mirrors}
\label{appendixC2}

The procedure consists in associating an orthosymplectic linear quiver to each puncture of the four dimensional theory. For $4$d theories of $D_N$ type, each of those $3$d $\mathcal{N}=4$ quiver tails ends with a flavor symmetry node SO($2N$). As a second step, one glues the flavour nodes together, producing a star-shaped quiver. 

The rule for associating the quiver tails to the punctures is the following. Consider a regular puncture whose Nahm partition is given by $[h_1,\cdots h_J]$. The associated 3d quiver tail is
\begin{equation}
[\text{SO}(2N)]-\text{USp}(r_1)-\text{SO}(r_2)-...-\text{USp}(r_{J-1})\ ,
\end{equation}
where the quantities $r_a$ are defined as
\begin{equation}
r_a=\left[\sum_{b=a+1}^J h_b\right]_{+,-}, \qquad + : \text{SO}, \ \ \ -: \text{USp}
\end{equation}
and $[n]_{+(-)}$ is the smallest (resp. largest) even integer $\geq n$ (resp. $\leq n$). When $r_{J-1}=0$, we remove the last group $\text{USp}(0)$. We defer the reader to \cite{Benini:2010uu} for further details.

\section{5-brane webs for SO(4) theory with different discrete theta angles}\label{appendixE}

\begin{figure}[H]
	\centering
	\begin{scriptsize}
		\begin{tikzpicture}
		\draw[thick](0,0)--(-1,-1);
		\draw[thick](0,0)--(0,1);
		\draw[thick](-1,2)--(0,1);
		\draw[thick](1,1)--(0,1);
		\draw[thick](1,1)--(2,2);
		\draw[thick](1,1)--(1,0);
		\draw[thick](1,1)--(1,0);
		\draw[thick](0,0)--(1,0);
		\draw[thick](2,-1)--(1,0);
		\end{tikzpicture}
	\end{scriptsize}
\qquad\qquad\qquad
	\begin{scriptsize}
		\begin{tikzpicture}
		\draw[thick](0,0)--(-1,-1);
		\draw[thick](0,0)--(0,1);
		\draw[thick](-1,2)--(0,1);
		\draw[thick](1,1)--(0,1);
		\draw[thick](1,1)--(1,1.5);
		\draw[thick](1,1)--(2,0);
		\draw[thick](1,1)--(1,2);
		\draw[thick](0,0)--(2,0);
		\draw[thick](3,-.5)--(2,0);
		\end{tikzpicture}
	\end{scriptsize}
	\caption{5-brane web diagrams for pure SU(2)$_0$ (left) and pure SU(2)$_\pi$ (right) gauge theories.}
	\label{fig:SU(2)_0 vs SU(2)_pi}
\end{figure}
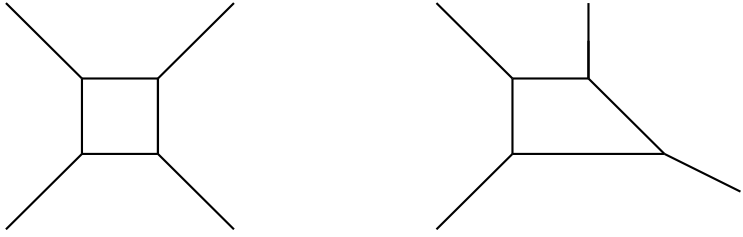

It is known that the pure SU(2) gauge theory has two choices, SU(2)$_0$ and SU(2)$_{\pi}$, depending on their discrete theta angles. 
SU(2)$_0$ and SU(2)$_{\pi}$ gauge theories are known to have different UV fixed points with different global symmetry, $E_1 =$ SU(2) and $\tilde{E}_1 =$ U(1), respectively. They also have distinct 5-brane configurations as depicted in figure \ref{fig:SU(2)_0 vs SU(2)_pi}. %
One can readily see from the 5-brane webs given in figure \ref{fig:SU(2)_0 vs SU(2)_pi} that SU(2)$_0$ gauge theory has an one-dimensional Higgs branch at infinite coupling, while SU(2)$_\pi$ gauge theory has no Higgs branch at infinite coupling. 

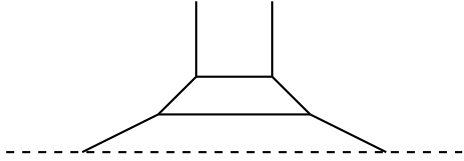
\begin{figure}[H]
    \centering
    \begin{scriptsize}
    \begin{tikzpicture}
    \draw[thick](0,0)--(1,0.5);
    \draw[thick](1,.5)--(1.5,1);
    \draw[thick](1.5,1)--(1.5,2);
    \draw[thick](1.5,1)--(2.5,1);
    \draw[thick](2.5,1)--(2.5,2);
    \draw[thick](2.5,1)--(3,.5);
    \draw[thick](1,.5)--(3,.5);
    \draw[thick](3,.5)--(4,0);
    \draw[thick,dashed](-1,0)--(5,0);
    \end{tikzpicture}
    \end{scriptsize}
    \caption{5-brane web diagram for SO(4).}
    \label{fig:SO(4)pure}
\end{figure}

\begin{figure}[H]
	\centering
	\begin{scriptsize}
		\begin{tikzpicture}
		\draw[thick](0,0)--(0,1);
		\draw[thick](0,0)--(-1,-1);
		\draw[thick](-1,2)--(0,1);
		\draw[thick](2,1)--(0,1);
		\draw[thick](2,1)--(3,2);
		\draw[thick](-.5,-1)--(2.5,2);
		\draw[thick](2,1)--(2,0);
		\draw[thick](2,0)--(0,0);
		\draw[thick](2,0)--(3,-1);
		\end{tikzpicture}
	\end{scriptsize}
\qquad\qquad\qquad
	\begin{scriptsize}
		\begin{tikzpicture}[scale=.5]
		\draw[thick](0,0)--(-1,-1);
		\draw[thick](0,0)--(0,1);
		\draw[thick](-1,2)--(0,1);
		\draw[thick](1,1)--(0,1);
		\draw[thick](1,1)--(2,2);
		\draw[thick](1,1)--(1,0);
		\draw[thick](1,1)--(1,0);
		\draw[thick](0,0)--(1,0);
		\draw[thick](2,-1)--(1,0);
		
		\draw[thick](3,-3)--(2,-4);
		\draw[thick](3,-3)--(3,-2);
		\draw[thick](2,-1)--(3,-2);
		\draw[thick](4,-2)--(3,-2);
		\draw[thick](4,-2)--(5,-1);
		\draw[thick](4,-2)--(4,-3);
		\draw[thick](4,-2)--(4,-3);
		\draw[thick](3,-3)--(4,-3);
		\draw[thick](5,-4)--(4,-3);
		\end{tikzpicture}
	\end{scriptsize}
	\caption{5-brane web for SU(2)$_0-$SU(2)$_0$ quiver. To read off the theta angles we deform the web on the left to that on the right.}
	\label{fig:SU(2)_0-SU(2)_0_quiver}
\end{figure}
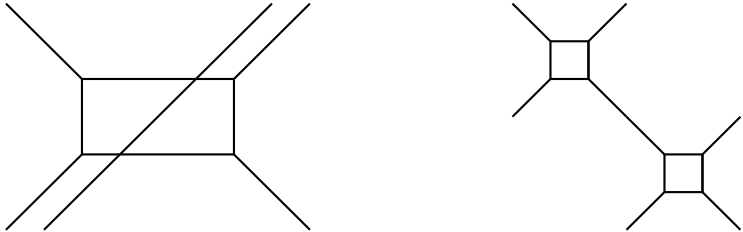

In this appendix, we discuss the 5-brane configurations for $\text{SO(4)}=\text{SU(2)}\times\text{SU(2)}$ gauge theory with different discrete theta angles. 
As each SU(2) can have discrete theta angle, we introduce the following shorthand notation to denote two discrete theta angles for SO(4),  SO(4)$_{\theta_1,\theta_2}=~$SU(2)$_{\theta_1}\times $ SU(2)$_{\theta_2}$. 
It follows that a conventional 5-brane web for SO(4) given in figure \ref{fig:SO(4)pure} corresponds to a 5-brane web for SO(4)$_{0,0}$. 
At the infinite coupling, this 5-brane web for SO(4)$_{0,0}$ is deformed to figure \ref{web diagrams EQ11 and EQ1}, from which one can read off the magnetic quiver and the corresponding Hilbert series.
In particular, we can check that the Higgs branch dimension is two, as expected. 
We also find the SU(2) global symmetry from the parallel NS5-branes, which is part of the $E_1 \times E_1$ symmetry.  
The SU(2)$_0$ $\times$ SU(2)$_0$ gauge theory can also be constructed from the 5-brane web without an O5-plane as given in figure \ref{fig:SU(2)_0-SU(2)_0_quiver}. 
We find that if the bifundamental mass is large enough, this web can be decomposed into two copies of SU(2)$_0$ webs. 

%

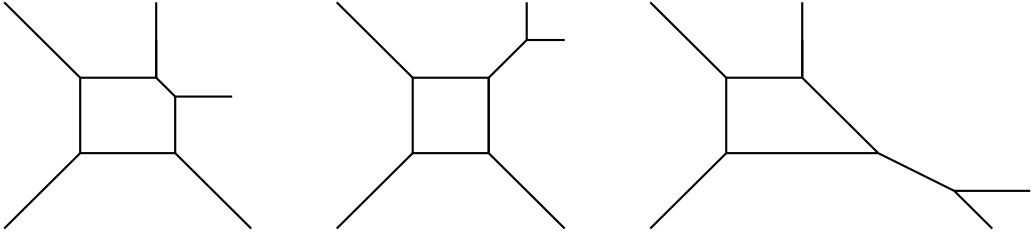
\begin{figure}[H]
	\centering
	\begin{scriptsize}
		\begin{tikzpicture}
		\draw[thick](0,0)--(-1,-1);
		\draw[thick](0,0)--(0,1);
		\draw[thick](-1,2)--(0,1);
		\draw[thick](1,1)--(0,1);
		\draw[thick](1,1)--(1,1.5);
		\draw[thick](1,1)--(1,2);
		\draw[thick](1,1)--(1.25,.75);
		\draw[thick](2,.75)--(1.25,.75);
		\draw[thick](1.25,0)--(1.25,.75);
		\draw[thick](1.25,0)--(0,0);
		\draw[thick](1.25,0)--(2.25,-1);
		\end{tikzpicture}
	\end{scriptsize}
	\qquad
	\begin{scriptsize}
		\begin{tikzpicture}
		\draw[thick](0,0)--(-1,-1);
		\draw[thick](0,0)--(0,1);
		\draw[thick](-1,2)--(0,1);
		\draw[thick](1,1)--(0,1);
		\draw[thick](1,1)--(1.5,1.5);
		\draw[thick](1.5,2)--(1.5,1.5);
		\draw[thick](2,1.5)--(1.5,1.5);
		\draw[thick](1,1)--(1,0);
		\draw[thick](1,1)--(1,0);
		\draw[thick](0,0)--(1,0);
		\draw[thick](2,-1)--(1,0);
		\end{tikzpicture}
	\end{scriptsize}
	\qquad
	\begin{scriptsize}
		\begin{tikzpicture}
		\draw[thick](0,0)--(-1,-1);
		\draw[thick](0,0)--(0,1);
		\draw[thick](-1,2)--(0,1);
		\draw[thick](1,1)--(0,1);
		\draw[thick](1,1)--(1,1.5);
		\draw[thick](1,1)--(2,0);
		\draw[thick](1,1)--(1,2);
		\draw[thick](0,0)--(2,0);
		\draw[thick](3,-.5)--(2,0);
		\draw[thick](3,-.5)--(4,-.5);
		\draw[thick](3,-.5)--(3.5,-1);
		\end{tikzpicture}
	\end{scriptsize}
	\caption{Left: 5-brane web diagram for SU(2) +1\textbf{F}. Middle: decoupling the hypermultiplet by giving it a positive mass. Taking the mass to infinity results in the web diagram for pure SU(2)$_0$ theory. Right: decoupling the hypermultiplet by giving it a negative mass. Taking the mass to negative infinity results in the web diagram for pure SU(2)$_\pi$ theory.}
	\label{fig:SU21Fweb}
\end{figure}

To construct 5-brane webs for SO(4)$_{\theta_1,\theta_2}$ gauge theories with different discrete theta angles, we first recall how 5-brane web for  SU(2)$_{0}$ and SU(2)$_{\pi}$ can be obtained from that for SU(2) gauge theory with one flavor. Pure SU(2)$_0$ theory is obtained by decoupling the flavor with a positive infinity mass, while pure SU(2)$_{\pi}$ theory is obtained by giving a negative infinity mass, as depicted in figure \ref{fig:SU21Fweb}.
%
%

\begin{figure}[H]
	\centering
\begin{scriptsize}
\begin{tikzpicture}
\draw[thick](0,0)--(1,0.5);
\draw[thick](1,.5)--(1.5,1);
\draw[thick](1.5,1)--(1.5,2);
\draw[thick](1.5,1)--(2.5,1);
\draw[thick](2.5,1)--(2.5,2);
\draw[thick](2.5,1)--(3,.5);
\draw[thick](1,.5)--(3,.5);
\draw[thick](3,.5)--(4,0);
\draw[thick](5,0)--(6,.5);
\draw[thick](7,1.5)--(6,.5);
\draw[thick](7,.5)--(6,.5);
\draw[thick,dashed](-1,0)--(8,0);

\draw[thick](10,0)--(11,0.5);
\draw[thick](11,.5)--(11.5,1);
\draw[thick](11.5,1)--(11.5,2);
\draw[thick](11.5,1)--(12.5,1);
\draw[thick](11,.5)--(12.5,.5);
\draw[thick](12.5,.5)--(12.5,1);
\draw[thick](12.5,.5)--(13,0);
\draw[thick](13,1.5)--(12.5,1);
\draw[thick](13.5,.5)--(13,0);
\draw[thick](13.5,.5)--(13.5,1.5);
\draw[thick](13.5,.5)--(14,.5);
\draw[thick](13,1.5)--(13.5,1.5);
\draw[thick](13,1.5)--(13,2);
\draw[thick](14,2)--(13.5,1.5);
\draw[thick,dashed](9,0)--(15,0);
			\end{tikzpicture}
		\end{scriptsize}
	\caption{Left: The web diagram for SO(4) gauge theory with 1\textbf{s} + 1\textbf{c}. Right: The web after flop transitions.}
	\label{fig:web-SO4-1S1C}
\end{figure}
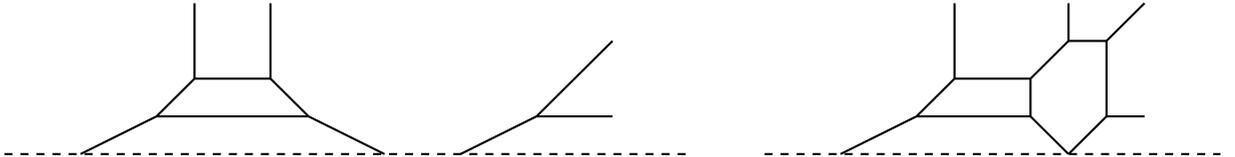

\begin{figure}[H]
\centering
    \begin{scriptsize}
    \begin{tikzpicture}
    \draw[thick](0,0)--(1,0.5);
    \draw[thick](1,.5)--(1.5,1);
    \draw[thick](1.5,1)--(1.5,2);
    \draw[thick](1.5,1)--(2.5,1);
    \draw[thick](1,.5)--(2.5,.5);
    \draw[thick](2.5,.5)--(2.5,1);
    \draw[thick](2.5,.5)--(3,0);
    \draw[thick](3.5,2)--(2.5,1);
    \draw[thick](3.5,.5)--(3,0);
    \draw[thick](3.5,.5)--(3.5,2);
    \draw[thick](4,3)--(3.5,2);
    \draw[thick](4,3)--(4,3.5);
    \draw[thick](4,3)--(4.5,3.5);
    \draw[thick](3.5,.5)--(4,.5);
    \draw[thick,dashed](-1,0)--(5,0);

    \draw[thick](7,0)--(8,0.5);
\draw[thick](8,.5)--(8.5,1);
\draw[thick](8.5,1)--(8.5,2);
\draw[thick](8.5,1)--(9.5,1);
\draw[thick](8,.5)--(9.5,.5);
\draw[thick](9.5,.5)--(9.5,1);
\draw[thick](9.5,.5)--(10,0);
\draw[thick](10,1.5)--(9.5,1);
\draw[thick](10,0)--(10,1.5);
\draw[thick](10.5,2.5)--(10,1.5);
\draw[thick,dashed](6,0)--(10,0);
\draw[thick,dashed](11,0)--(12,0);
\draw[thick](10,0)--(11,0);
\node[label=above:{(1,2)}][7brane] at (10.5,2.5){};
\node[7brane] at (11,0){};
\node[7brane] at (8.5,2){};
    \end{tikzpicture}
    \end{scriptsize}
\caption{Left: The web diagram for SO(4) gauge theory with 1\textbf{s} + 1\textbf{c} with negative large masses. Right: 5-brane web diagram for SO(4)$_{\pi, \pi}$.}
\label{fig:webSO4pipi}
\end{figure}
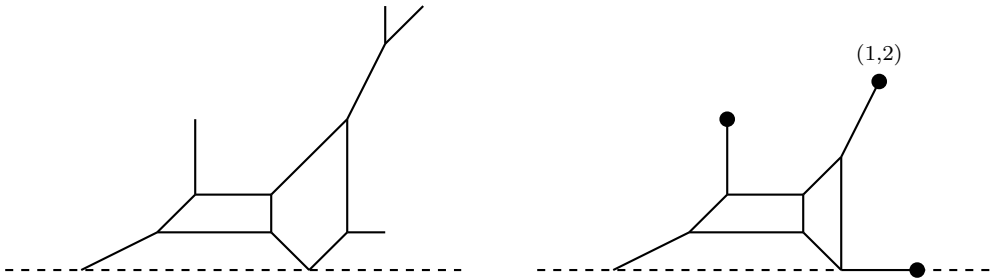

We then consider an analogous situation with 5-brane web with an O5-plane. For instance, consider a 5-brane web with spinor matter as in figure \ref{fig:web-SO4-1S1C}. The left of figure \ref{fig:web-SO4-1S1C} is the web for SO(4) gauge theory with one spinor and one conjugate spinor,
which corresponds to two copies of SU(2) gauge theories with one flavor. If we take all of their masses to be positive infinity, which is to move the branes corresponding to the spinor matters to the right, then the resulting 5-brane web trivially goes back to the pure SO(4)$_{0,0}$ gauge theory in figure \ref{fig:SO(4)pure}. On the other hand, one may instead bring the spinor matter closer to the pure SO(4)$_{0,0}$ followed by 
a sequence of flop transitions including the ``generalized flop transition'' \cite{Hayashi:2017btw}, which yields the right of figure \ref{fig:web-SO4-1S1C}. 
When we consider their masses to be negatively large enough, further flop transition is induced to yield the left of figure \ref{fig:webSO4pipi}. 
Taking their masses to be identical and further tuning this mass to be negative infinity, we find the SO(4)$_{\pi, \pi}$ gauge theory in the right of figure \ref{fig:webSO4pipi}. We can check from this web that there is no Higgs branch. Also, we find no non-Abelian global symmetry, as expected. 

\begin{figure}[H]
    \centering
    \begin{scriptsize}
    \begin{tikzpicture}
    \draw[thick](1.5,1)--(1,1.5);
    \draw[thick](1.5,1)--(2.5,1);
    \draw[thick](1.5,.5)--(2.5,.5);
    \draw[thick](1.5,.5)--(1.5,1);
    \draw[thick](1.5,.5)--(1,0);
    \draw[thick](1,1.5)--(1,0);
    \draw[thick](1,1.5)--(.5,2.5);
    \draw[thick](0,0)--(1,0);
    \draw[thick](2.5,.5)--(2.5,1);
    \draw[thick](2.5,.5)--(3,0);
    \draw[thick](3,1.5)--(2.5,1);
    \draw[thick](3,0)--(3,1.5);
    \draw[thick](3.5,2.5)--(3,1.5);
    \draw[thick,dashed](-1,0)--(3,0);
    \draw[thick,dashed](4,0)--(5,0);
    \draw[thick](3,0)--(4,0);
    \node[label=above:{(1,2)}][7brane] at (3.5,2.5){};
    \node[label=above:{(1,-2)}][7brane] at (.5,2.5){};
    \node[7brane] at (0,0){};
    \node[7brane] at (4,0){};
    \end{tikzpicture}
    \end{scriptsize}
    \caption{5-brane web diagram for SO(4)$_{2\pi,2\pi}$}
    \label{fig:SO(4) 2pi-2pi}
\end{figure}
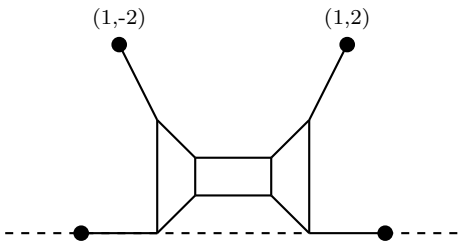

We can even construct a 5-brane web for SO(4)$_{2 \pi, 2 \pi}$ gauge theory, by repeating the procedure above. Given a 5-brane web for SO(4)$_{\pi, \pi}$ as in figure \ref{fig:webSO4pipi}, we can introduce another set of spinor and conjugate spinor on the left. By taking their masses to be negatively large, we obtain the 5-brane web in figure \ref{fig:SO(4) 2pi-2pi}. This is a 5-brane web for SO(4)$_{2 \pi, 2 \pi}$ gauge theory, whose Higgs branch should be identical to SO(4)$_{0,0}$ gauge theory. From this 5-brane web diagram, we can explicitly check using \cite{Akhond:2020vhc} that it has Higgs branch with dimension two at the infinite coupling, as expected. 
%
We note that it is also interesting to observe that the S-dual of this web diagram looks like SU(2) $\times$ SU(2) gauge theory as a special case of the D-type Dynkin quiver gauge theory constructed with ON$^0$-plane, where the gauge coupling constants are tuned to be identical. The symmetric shape indicates that the corresponding SU(2) theories have discrete theta angle 0 rather than $\pi$.


All these web diagrams for SO(4)$_{\theta_1, \theta_2}$ gauge theories with different discrete theta angles and their consistency checks give further support for the correspondence between \eqref{E1xE1 electric OSp} and \eqref{E1xE1 electric unitary}.

\bibliographystyle{JHEP}

\bibliography{ref}
\end{document}